\documentclass[11pt,a4paper]{article}
\usepackage[T1]{fontenc}
\usepackage{lmodern}
\usepackage{amsmath,amssymb,amsthm,mathtools}
\usepackage[a4paper,margin=1.05in,includefoot]{geometry}
\usepackage{enumitem,booktabs,float}\usepackage[noend]{algpseudocode}
\usepackage[protrusion=false]{microtype}
\usepackage[colorlinks=true,allcolors=blue]{hyperref}
\newtheorem{theorem}{Theorem}[section]
\newtheorem{lemma}[theorem]{Lemma}
\newtheorem{proposition}[theorem]{Proposition}
\newtheorem{corollary}[theorem]{Corollary}
\theoremstyle{definition}

\newcommand{\sig}{\sigma}
\newcommand{\calR}{\mathcal R}
\newcommand{\Forest}{\operatorname{Forest}}

\newcommand{\Insert}{\operatorname{Insert}}
\newcommand{\Pull}{\operatorname{Pull}}
\newcommand{\Merge}{\operatorname{Merge}}
\newcommand{\Init}{\operatorname{Init}}
\newcommand{\Delete}{\operatorname{Delete}}
\newcommand{\dist}{\operatorname{dist}}

\newcommand{\J}{\operatorname{J}}
\newfloat{algorithm}{htbp}{loa}
\floatname{algorithm}{Algorithm}
\algrenewcommand\algorithmicrequire{\textbf{Input:}}
\algrenewcommand\algorithmicensure{\textbf{Output:}}
\algrenewcommand\alglinenumber[1]{\normalsize #1:}
\makeatletter
\newlength\AlgRuleTop
\newlength\AlgRuleMid
\newlength\AlgRuleBot
\newlength\AlgBarWidth
\newlength\AlgInset
\newcommand\floatc@algruled[2]{\setbox\@tempboxa\hbox{\kern\AlgInset{\@fs@cfont#1:}\enspace}%
  \hangindent\wd\@tempboxa\hangafter\@ne\noindent\strut\box\@tempboxa#2\strut\par}
\newcommand\fs@algruled{\def\@fs@cfont{\bfseries}\let\@fs@capt\floatc@algruled
  \def\@fs@pre{\hrule\@height\AlgRuleTop\@depth\z@\kern2pt}%
  \def\@fs@mid{\kern1pt\hrule\@height\AlgRuleMid\@depth\z@\kern3pt}%
  \def\@fs@post{\kern2pt\hrule\@height\AlgRuleBot\@depth\z@}%
  \let\@fs@iftopcapt\iftrue}
\floatstyle{algruled}\restylefloat{algorithm}\floatstyle{plain}
\newcommand\AlgMark{\ensuremath{\triangleright}}
\newcommand\AlgPhase[1]{\rule{0pt}{\dimexpr\ht\strutbox+2.5pt\relax}\AlgMark~\textbf{#1}}
\algrenewcommand\algorithmiccomment[1]{\unskip\nobreak\hfil\penalty50\hskip1.5em
  \hbox{}\nobreak\hfil\mbox{\AlgMark~#1}\parfillskip\z@\finalhyphendemerits\z@}
\newlength\Alg@numwd\newlength\Alg@numsep\newlength\Alg@ind
\newlength\Alg@barpos\newlength\Alg@hang
\newcount\Alg@depth\newcount\Alg@cnt
\newif\ifAlg@open
\def\Alg@bars{\ifnum\Alg@cnt>\z@\kern\Alg@barpos\vrule\@width\AlgBarWidth
  \kern\dimexpr\Alg@ind-\Alg@barpos-\AlgBarWidth\relax
  \advance\Alg@cnt\m@ne\expandafter\Alg@bars\fi}
\def\Alg@open#1#2#3{\nointerlineskip\hbox to\linewidth\bgroup\kern\AlgInset#1%
  \vtop\bgroup\hsize\dimexpr#2\relax\linewidth\hsize\parindent\z@\parskip\z@skip
  \leftskip\z@skip\rightskip\z@skip\parfillskip\@flushglue\everypar{}%
  \hangindent\dimexpr#3\relax\hangafter\@ne\noindent\strut\global\Alg@opentrue}
\def\Alg@shut{\unskip\strut\par\egroup\hss\kern\AlgInset\egroup}
\def\Alg@close{\ifAlg@open\global\Alg@openfalse\expandafter\Alg@shut\fi}
\def\Alg@line#1#2{\Alg@depth#2\relax\Alg@cnt\Alg@depth
  \Alg@open{\hbox to\Alg@numwd{\hss#1}\kern\Alg@numsep\Alg@bars}%
    {\linewidth-2\AlgInset-\Alg@numwd-\Alg@numsep-\Alg@ind*\Alg@depth}{\Alg@hang}}
\def\Alg@head#1{\Alg@open{}{\linewidth-2\AlgInset}{\Alg@numwd+\Alg@numsep}#1\enspace}
\def\Alg@stateblock{\csname ALG@bl@\ALG@L @STATE\endcsname}
\def\Statex{\Alg@close\Alg@depth\value{ALG@nested}%
  \ifnum\csname ALG@currentblock@\theALG@nested\endcsname=\Alg@stateblock\relax
    \advance\Alg@depth\m@ne\fi
  \Alg@line{}\Alg@depth}
\algrenewcommand\Require{\Alg@close\Alg@head\algorithmicrequire}
\algrenewcommand\Ensure{\Alg@close\Alg@head\algorithmicensure}
\let\Alg@makeentity\ALG@makeentity
\def\ALG@makeentity{\Alg@close\Alg@makeentity}
\def\ALG@doentity{%
  \edef\ALG@thisblock{\csname ALG@currentblock@\theALG@nested\endcsname}%
  \expandafter\ifx\csname ALG@b@\ALG@L @\ALG@thisentity @\ALG@thisblock\endcsname\relax
    \def\ALG@thisblock{0}\fi
  \ALG@getentitytext
  \ifnum\ALG@thisblock=0 \else\ALG@doend\fi
  \Alg@depth\value{ALG@nested}%
  \expandafter\ifnum0=\csname ALG@b@\ALG@L @\ALG@thisentity @\ALG@thisblock\endcsname
  \else\ALG@dobegin\fi
  \def\ALG@entitiecommand{\ALG@displayentity}%
  \ifx\ALG@text\ALG@x@notext\else\Alg@line\ALG@step\Alg@depth\fi}
\renewenvironment{algorithmic}[1][0]{\par
  \edef\ALG@numberfreq{#1}\def\@currentlabel{\theALG@line}%
  \setcounter{ALG@line}{0}\setcounter{ALG@rem}{0}\setcounter{ALG@nested}{0}%
  \expandafter\edef\csname ALG@currentblock@\theALG@nested\endcsname{0}%
  \expandafter\let\csname ALG@currentlifetime@\theALG@nested\endcsname\relax
  \ALG@tlm\z@\settowidth\Alg@numwd{\alglinenumber{99}}%
  \setlength\Alg@numsep{0.5em}\setlength\Alg@ind{\algorithmicindent}%
  \setlength\Alg@barpos{0.5em}\setlength\Alg@hang{1em}%
  \global\Alg@openfalse\ALG@beginalgorithmic}
 {\Alg@close\ALG@closeloops
  \expandafter\ifnum\csname ALG@currentblock@\theALG@nested\endcsname=0 \else
    \PackageError{algorithmicx}{Some blocks are not closed!!!}{}\fi
  \ALG@endalgorithmic\par}
\newcommand{\AlgWrap}[1]{#1}%
\makeatother
\newcommand{\DocumentTitle}{Improved Bounds for Directed SSSP in Sparse and Intermediate-Density Graphs}
\hypersetup{
  pdftitle={\DocumentTitle},
  pdfauthor={Valery (Binyamin) Grishin, Aryeh Lev Zabokritskiy (Yohananov)},
  pdflang={en}
}
\title{\DocumentTitle}
\author{
  \normalsize Valery (Binyamin) Grishin\\[0.5em]
  \small Department of Computer Science, Tel-Hai University of Kiryat Shmona\\
  \small and the Galilee, Kiryat Shmona, Israel\\[0.25em]
  \small\href{mailto:valerig.tech@gmail.com}{\texttt{valerig.tech@gmail.com}}\\[2em]
  \normalsize Aryeh Lev Zabokritskiy (Yohananov)\\[0.5em]
  \small Department of Computer Science, MIGAL--Galilee Research Institute,\\
  \small Kiryat Shmona, Israel, and Department of Computer Science, Tel-Hai\\
  \small University of Kiryat Shmona and the Galilee, Kiryat Shmona, Israel\\[0.25em]
  \small\href{mailto:yuhanalev@telhai.ac.il}{\texttt{yuhanalev@telhai.ac.il}}\\[0.5em]
  \small ORCID: \href{https://orcid.org/0000-0003-3151-6192}{0000-0003-3151-6192}
}
\date{}
\begin{document}
{\hypersetup{urlcolor=black}\maketitle}
\begin{abstract}
We give a deterministic algorithm for exact single-source shortest paths (SSSP) in directed graphs with $n$ vertices, $m$ edges, and nonnegative real edge weights. When $m=O(n)$, it runs in $O(n\sqrt{\log n}(\log\log n)^{1/4})$ time. For $m=\Theta(n)$, this lowers the exponent of $\log\log n$ in the bound of Duan, Mao, Shu and Yin from $1/2$ to $1/4$. At intermediate densities, writing $d=m/n$, the same algorithm improves on the bound of Duan et al.\ when $d\geq 3$ and $d\log(2+d)=o(\log\log n)$. For example, when $m=\Theta(n\sqrt{\log\log n})$, it runs in $O(n\sqrt{\log n}(\log\log n)^{5/8})$ time, lowering their exponent from $3/4$ to $5/8$.
\end{abstract}

\section{Introduction}
Let $n$ and $m$ denote the numbers of vertices and edges in a directed graph with nonnegative real edge weights. Dijkstra's algorithm computes exact single-source shortest paths (SSSP) in $O(m+n\log n)$ time using Fibonacci heaps~\cite{dijkstra1959,FredmanTarjan1987}. In the comparison-addition model, Duan, Mao, Shu and Yin~\cite{DMSY2026} obtain the deterministic bound
\[
 O\!\left(n+m\sqrt{\log n}+\sqrt{mn\log n\log\log n}\right).
\]
We work in this model as stated by Duan et al.~\cite[Section 2.2]{DMSY2026}: real weights and path lengths are only added and compared, each operation at unit cost; that section does not restrict integer operations. For integers, such as indices and pointers, we assume $O(\log(n+m+2))$-bit words with unit-cost elementary operations; integer products, division, logarithms and roots are computed from these operations and charged.

We improve the worst-case bound of Duan et al.\ over graphs with $m=O(n)$ and obtain improvements at intermediate densities. When $m=O(n)$, our running time is
\[
 O\!\left(n\sqrt{\log n}(\log\log n)^{1/4}\right),
\]
compared with $O(n\sqrt{\log n}(\log\log n)^{1/2})$ for Duan et al.\ at $m=\Theta(n)$. When $m/n=O((\log\log n)^{-1/2})$, the bound of Duan et al.\ is no larger than ours, up to a constant factor. Write $d=m/n$ for the average out-degree. Our uniform intermediate-density bound covers $3\leq d\leq C\log\log n$ for each fixed $C>0$; strict improvement holds on a smaller range. At $d=\Theta(\sqrt{\log\log n})$, the exponent of $\log\log n$ decreases from $3/4$ to $5/8$.

The main theorem of the public B1 draft~\cite[Section 2]{B1PublicDraft2026}, hosted as an unverified historical draft in the C-HD proof package~\cite{CHDGeneric2026}, states the sparse bound
\[
 O\!\left(n+m\sqrt{\log n}(\log\log n\,\log\log\log n)^{1/4}\right).
\]
For $m=\Theta(n)$, our bound removes the factor $(\log\log\log n)^{1/4}$ from B1's bound. For smaller $m$, B1's displayed bound, which retains its additive $n$ and factor $m$, can be smaller than our uniform sparse bound. The draft also gives the general-degree expression~\cite[Section 2, Remark~(iii), and Section S4.9]{B1PublicDraft2026}
\[
 O\!\left(n+m\sqrt{\log n}(\log\log n)^{1/4}
       (1+(\log\log\log n)/d)^{1/4}\right)
\]
as an explicitly unclaimed remark outside its main theorem. The draft states strict improvement over Duan et al.\ for $d=o(\sqrt{\log\log n})$. We compare with this expression on that range when $d\geq3$.

We adapt the capped recursion of B1~\cite{B1PublicDraft2026} and the worklist and counting methods of the public C-HD candidate~\cite{CHDPublic2026}.\footnote{\fontsize{10.2pt}{12pt}\selectfont Both sources contain substantial related proofs.} The analysis of the lazy worklist uses operation bounds from the C-HD proof package~\cite{CHDGeneric2026}.

For sufficiently large $n$, write $\lambda_0\coloneqq\log\log n$ and
\begin{equation}\label{eq:primary-coefficient}
 h(d,z)\coloneqq\sqrt{dz}+d\log(2+d)+d+1.
\end{equation}
\begin{theorem}\label{thm:main}
There is a single deterministic algorithm for exact SSSP in directed graphs with nonnegative real edge weights, with the following guarantees.
\begin{enumerate}
\item For every fixed $C>0$, uniformly on $m\leq Cn$, its time is
\begin{equation}\label{eq:main-sparse}
 O\!\left(n\sqrt{\log n}(\log\log n)^{1/4}\right).
\end{equation}
\item For every fixed $C>0$, uniformly on $3\leq d\leq C\lambda_0$, its time is
\begin{equation}\label{eq:main}
 O\!\left(n+m+n\sqrt{d\log n\,h(d,\lambda_0)}\right).
\end{equation}
\end{enumerate}
The constants and thresholds may depend on $C$. The algorithm and its fixed constants do not. It uses unit-cost comparisons and additions on exact real weights. On $O(\log(n+m+2))$-bit integer words, comparisons, addition, subtraction, constant-size copying, shifts, bit access, and direct-address and pointer reads and writes cost one step. Integer products, division, logarithms and roots are computed from these primitives and charged accordingly. It returns all distances and a shortest-path tree rooted at the source, given by one original predecessor edge per reachable vertex other than the source, and handles zero weights, loops, parallel arcs and unreachable vertices. Its storage, counted in real-weight cells and integer words, is bounded by a constant multiple of the corresponding time bound. It remains exact on every input through parameter tests and an exact fallback.
\end{theorem}

The improvements are asymptotic. The accelerated branch of the algorithm can be selected only if $\log_2(N+2)>2^{8211}$, where $N$ is the number of vertices of the degree-reduced graph that this branch searches. A test can fail only on inputs with fewer than an absolute number of vertices; after a failed test, the algorithm uses the exact Dijkstra fallback. The algorithm also returns exact distances outside the theorem's density ranges.

Theorem~\ref{thm:main} has the following fixed-power consequence.
\begin{corollary}\label{cor:fixed-powers}
For every fixed $0\leq a\leq b<1$ and $c_1,c_2>0$, uniformly for $\alpha\in[a,b]$ and $c_1\lambda_0^\alpha\leq d\leq c_2\lambda_0^\alpha$, the algorithm of Theorem~\ref{thm:main} takes
\begin{equation}\label{eq:main-packed}
 O\!\left(n+m+n\sqrt{\log n}(\log\log n)^{(3\alpha+1)/4}\right)
\end{equation}
time. The constants and threshold may depend on the displayed fixed interval and density constants.
\end{corollary}
For fixed $0\leq\alpha<1$, the bound of Duan et al.\ has exponent $(\alpha+1)/2$, exceeding ours by $(1-\alpha)/4$. For $0<\alpha<1/2$, we compare with B1's unclaimed general-degree expression, which is smaller than theirs, with exponent $\alpha+1/4$, exceeding ours by $\alpha/4$. Thus at $\alpha=1/3$ the comparison is $7/12$ versus $1/2$, and at $\alpha=1/2$ it is $3/4$ versus $5/8$.

The bound \eqref{eq:main} is of smaller order than that of Duan et al.\ exactly on the strict-improvement window
\begin{equation}\label{eq:gainwindow}
 3\leq d,\qquad d\log(2+d)=o(\lambda_0).
\end{equation}
Within this window, it is also of smaller order than B1's general-degree expression when $d=o(\sqrt{\lambda_0})$. The term $d\log(2+d)$ in $h(d,\lambda_0)$ determines where the improvement over Duan et al.\ ends. At $d=\Theta(\lambda_0)$, the stated bound in \eqref{eq:main} instead exceeds that of Duan et al.\ by a factor $\sqrt{\log\lambda_0}$.

Section~\ref{sec:overview} describes the algorithm. Sections~\ref{sec:model} and~\ref{sec:framework} define its search rules and bounded recursion. Section~\ref{sec:outer} bounds the outer recursion in terms of the cost of local searches. Section~\ref{sec:invocation-cost} reduces that cost to the cost of smaller searches, implemented in Section~\ref{sec:packed}. Section~\ref{sec:packed-global} derives the running times, Section~\ref{subsec:comparisons} compares them with earlier bounds, and Section~\ref{sec:conclusion} discusses limitations and open questions.

\section{The shortest-path algorithm}\label{sec:overview}

Algorithm~\ref{alg:driver} in Section~\ref{subsec:main-driver} computes all distances from the source $s$ and an original-edge predecessor tree on reachable vertices. To limit the number of arcs scanned at a vertex, we first reduce the degrees by splitting vertices into ports joined by zero-weight edges. The distances on the new graph give the distances on the input graph. On the resulting graph we use the bounded recursion of Duan et al.~\cite[Lemmas 3 and 9]{DMSY2026}. Its first call starts at the original source's port with no upper bound. A call takes a set of starting vertices with tentative routes and an upper bound on route keys, ordered by length and then by fixed tie rules. It settles a prefix of the vertices it is asked to reach and returns a cutoff key and a worklist for the caller to continue the search.

To form a batch for a recursive child, the call explores locally from its starting vertices, one at a time. Each local search records a tree of reached vertices. These vertices are called retained. A search can retain an endpoint even when it rejects the candidate route and leaves the label unchanged. The search ends when it exhausts the eligible routes, retains a prescribed number of vertices (its \emph{retention limit}), or reaches a tree retained by an earlier search. Searches that exhaust supply the vertices they have accepted. The stopped trees form a forest. The call divides this forest into pieces, groups its remaining starting vertices by piece, and keeps one representative per group in a worklist ordered by tentative route. The call pulls a batch of the smallest stored keys, adds group members with current keys below the batch bound whenever their representative is pulled, and recurses on the resulting set. When the child returns, the call uses outgoing edges from the newly completed vertices to update its worklist before taking another batch. At level zero it processes vertices with a priority queue.

The heap searches of Duan et al.~\cite[Lemma~4]{DMSY2026} have a logarithmic term in the cost per extraction. Following B1~\cite[Sections S2--S3]{B1PublicDraft2026}, each local search of this \emph{outer recursion} runs the bounded recursion again from its single starting vertex, with the entry labels as upper limits on acceptable extensions. This \emph{capped invocation} forms its own batches with smaller \emph{inner forest searches}, the local searches of its own calls.

For comparison, consider an inner forest search that finds the minimum pending vertex by a scan, as in C-HD~\cite[Section 3.3]{CHDPublic2026}; such a search may probe one endpoint from several tails. In the packed version, we list the short walks from each possible root once and rank them by length and the route tie rules. A bit at a rank indicates that a candidate extending an accepted prefix is pending. A lookup finds the first set bit. If that candidate is within the search bound, test its route. If the invocation continues after the test, mark every rank ending at this endpoint as unavailable and clear their pending bits in one operation. If the route is accepted and the search continues, enable its eligible one-edge extensions, omitting the marked ranks. This uses constant work per retained vertex of an inner forest search, apart from outgoing-slot work charged separately for the whole invocation.

The walk order is computed by comparisons of exact real lengths and the tie rules. We add real arc lengths to form the walk records and compare real route lengths when testing candidates. The masks and the lookup indices are bounded integers representing walk ranks and outgoing slots. The mask width permits tables for the first set position and for intersections and complements. To remove ranks, intersect the pending mask with the complement of the mask of those ranks. We charge for building these tables and sorting the walks before the first search (Section~\ref{sec:packed}). The number of walks grows quickly with their length, so we use these masks only for the inner forest searches inside a capped invocation.

Packing reduces the \emph{local coefficient}, the cost of a capped invocation per retained vertex. For its per-level part, let $k$ be the retention limit of an inner forest search and $t$ the inner recursion scale, so that level-$j$ capacities in \eqref{eq:capacities} grow like $2^{jt}$. A tree piece has at least $k$ retained vertices (Section~\ref{subsec:procedure}), so one worklist insertion costing $O(t)$ contributes $O(t/k)$ per retained vertex of the piece. With an $O(k)$ search charge per retained vertex, as for a scan of the pending vertices, choosing $k$ of order $\sqrt t$ gives $O(\sqrt t)$ per level in these two terms. With packing, choosing $k$ proportional to $t$ makes both terms constant. Write $A=\lceil\log_2(P+2)\rceil$ for the logarithmic scale of the invocation's retention limit $P$. On parameter records that pass the driver's tests (Section~\ref{subsec:main-driver}), there are $O(A/t)$ levels. With routing degree bound $D$, each completed vertex pays $O(Dt)$ once, for at most one worklist insertion per outgoing edge, and the proof balances $A/t$ against $Dt$ for packing. Balancing $A/\sqrt t$ against $Dt$ instead would give the local coefficient a term $D^{1/3}A^{2/3}$; packing gives $\sqrt{DA}$, with the degree-logarithm term $D\log(2+D)$, which pays for base-heap operations and a logarithmic floor on $t$ (Sections~\ref{sec:invocation-cost}--\ref{sec:packed}). B1's inner heap searches and representative insertions~\cite[Section S4.3, Lemma~I, and Sections S4.6--S4.7]{B1PublicDraft2026} carry a logarithmic per-level charge. Our packed inner search makes the corresponding charge constant at bounded degree and removes the resulting $(\log\log\log n)^{1/4}$ factor.

\subsection{Branch selection and output}\label{subsec:main-driver}

If $n>8$ and $m<3n$, the driver adds $n$ dummy vertices whose edges never enter the input graph. This raises the density for the parameter tests and preserves the original distances. The packed driver, Algorithm~\ref{alg:driver}, tests the recursion and worklist capacities, then checks that the walk ranks fit in a mask and that their tables fit the preparation budget. If a test fails, it sets aside the preliminary routing state and returns Fibonacci-heap Dijkstra on the original input.

\begin{algorithm}[!t]
\caption{Single-source shortest paths with the packed engine}\label{alg:driver}
\small
\begin{algorithmic}[1]
\Require Directed graph $I_0$ with nonnegative real edge weights and source $s$. Edges have identifiers. Loops and parallel edges are allowed.
\Ensure Distances from $s$ in $I_0$ and a shortest-path tree rooted at $s$, given by one original predecessor edge per reachable vertex other than $s$.
\Statex \AlgPhase{Prepare the input}
\If{$|V(I_0)|\leq8$}
    \State \Return Fibonacci-heap Dijkstra on $I_0$
\EndIf
\State $I\gets I_0$
\If{$|E(I_0)|<3|V(I_0)|$} \label{alg:driver-pad}
    \State $I\gets$ the padded graph of Section~\ref{subsec:sparse}
\EndIf
\State $(n,m)\gets(|V(I)|,|E(I)|)$
\State \label{alg:driver-simplify}On a copy of $I$, remove loops and keep a lightest arc for each ordered pair, taking the first on ties
\If{the source and all remaining endpoints occupy at most two vertices}
    \State \Return the direct solution on $I_0$ \Comment{Section~\ref{subsec:routing}}
\EndIf
\State \label{alg:driver-reduce}With $D$ from \eqref{eq:packed-degree}, build the routing graph with $N$ vertices, $M$ arcs and source $s'$ \Comment{Section~\ref{subsec:routing}}
\Statex \AlgPhase{Select a search}
\State \label{alg:driver-tests}Run the following tests in order, stopping at the first failure
\State \label{alg:driver-outertests}Form and test the outer record (including depth $e$) \Comment{Section~\ref{subsec:packed-driver}}
\State \label{alg:driver-innertests}Form and test the inner record \Comment{Section~\ref{subsec:packed-driver}}
\State \label{alg:driver-pretests}Test the walk-list sizes and preparation budget in \eqref{eq:packed-globalguards}
\If{a test fails} \label{alg:driver-fallback}
    \State Set aside the preliminary routing state
    \State \Return Fibonacci-heap Dijkstra on $I_0$ \Comment{Section~\ref{subsec:packed-driver}}
\EndIf
\Statex \AlgPhase{Compute distances with the packed engine}
\State \label{alg:driver-setup}Prepare the walk lists, lookup tables and shared state \Comment{Section~\ref{sec:packed} and Appendix~\ref{sec:resources}}
\State Initialize the label of $s'$ to the empty route and all other labels to infinity
\State \label{alg:driver-run}Run the top call $\calR(\infty,\{s'\},e)$ of Algorithm~\ref{alg:recursion} with the outer record of line~\ref{alg:driver-outertests}, using the packed engine in each capped invocation \Comment{Section~\ref{subsec:outer-run}}
\Statex \AlgPhase{Recover the original output}
\State \label{alg:driver-output}Take the routing distance at one port per vertex of $I_0$
\State \label{alg:driver-tree}Traverse the tight edges of $I_0$ from $s$ and record the edge of first discovery at each vertex \Comment{Section~\ref{subsec:routing}}
\State \Return the distances and the predecessor edges
\end{algorithmic}
\end{algorithm}

For the running times, Lemma~\ref{lem:outer-cost} and the balance after it turn a local coefficient $O(g)$ into outer time $O(N\sqrt{Dg\log N})$, plus terms that the driver absorbs. On the intermediate range of Theorem~\ref{thm:main}, the packed driver has $N=O(n)$, $D\leq d$ and $A=O(\log\log n)$, so its coefficient $h(D,A)$ of \eqref{eq:primary-coefficient} gives \eqref{eq:main}, and bounded densities with padding give \eqref{eq:main-sparse} (Section~\ref{sec:packed-global}). A test fails only on inputs with fewer than an absolute number of vertices (Section~\ref{subsec:uniform-packed}).

\section{Path labels and capped searches}\label{sec:model}

A search keeps one current route to each vertex it has reached. We need a rule for comparing routes of equal length and a rule for extending a route during a local search. The local search may improve the common labels, but each new arrival must respect the label that was in place when that search began.

All logarithms are base two. Write $\J(z)\coloneqq\lceil\log_2(z+2)\rceil$ for nonnegative integers $z$. Several letters denote more than one quantity; the table below lists their meanings and where each applies.
\begin{center}
\small
\begin{tabular}{@{}lp{0.72\linewidth}@{}}
\toprule
Symbol & Meaning \\
\midrule
$G$, $g$, $G_p$ & Routing graph; outer estimate of the cost per retained vertex; the packed driver's bound on that cost, used as $g$ (Section~\ref{subsec:packed-driver}). \\
$M$, $M_j$, $\mathcal M$ & Routing arc count, pull quota at level $j$, and total retained mass. The worklist appendices use $M$ locally for a quota. \\
$L$, $L_*$, $\mathcal L$, $L_0$ & $L=\J(N)$ in the scale analysis, the lower cut $L$ in a call, the invocation's lower threshold $L_*$, the lazy live map $\mathcal L$, and $L_0=\log n$ in Section~\ref{subsec:comparisons}. \\
$A$, $A_j$, $A_*$ & $A=\J(P)$, the input capacity at level $j$, and the invocation's active set $A_*$. \\
$R$, $R_j$, $R_{\rm B}$, $R_X$ & $R=N+M+2$ and the level-specific and local bounds on explicit worklist records; $R_X$ counts placements in call $X$'s subtree (Appendices~\ref{app:lazy} and~\ref{app:counting}). \\
$C$ & A fixed density-class constant in a running-time claim, or the map of entry witnesses in a capped invocation; in Appendix~\ref{app:word-bounds}, an unnamed constant. \\
$\J$, $J$ & $\J(z)=\lceil\log_2(z+2)\rceil$; a set $J$ of outgoing slots in $\operatorname{Child}(p,J)$ (Section~\ref{sec:packed}). \\
$P$, $P_S$ & Retention limit of a capped invocation; $P_S=S\setminus Q$, the grouped sources of a call (Section~\ref{sec:framework}); in Appendix~\ref{app:word-bounds}, $P=N$ at the outer scale. \\
$r$, $r_i$ & Number of accepted output vertices of a completed run; $r_i$, the number retained by the $i$th inner forest search (Section~\ref{sec:invocation-cost}). \\
$S$, $S_0$ & Source set of a call ($S_c$ of a child, $S_X$ of call $X$); $S_0$, the bound on walk records per root (Section~\ref{sec:packed}); $S_M$, a sum in Appendix~\ref{app:lazy-primitives}. \\
$T$, $\mathcal T$, $T_P$ & The outer scale (Section~\ref{subsec:outer-run}); the target set $\mathcal T(B,S)$ of a call, with $T_P=\mathcal T(B,P_S)$ in Appendix~\ref{app:return}; $T=1$, a unit block charge in Appendix~\ref{app:lazy-primitives}. \\
$U$, $U_0$ & The output that Algorithm~\ref{alg:recursion} accumulates, and the completed output a call returns (Section~\ref{subsec:worklists-returns}); $U_0$, a per-root preparation bound (Section~\ref{subsec:packed-driver}). \\
$W$, $W_H$ & Accepted output of a call's small full searches (Section~\ref{sec:framework}); $W_H$, $W_{\rm extra}$ and $W_{\rm DS}$ measure work (Sections~\ref{sec:invocation-cost}--\ref{sec:packed}, Appendices~\ref{app:lazy}, \ref{app:counting} and~\ref{app:coupling}). \\
$w_0$, $w_y$, $w$ & The mask width $w_0$ (Section~\ref{sec:packed}); the actual witness $w_y$ of the root $y$ of a packed search; $w(f)$, the weight of an arc $f$ (Section~\ref{subsec:routing}). \\
$\lambda_0,\lambda,\lambda_n,\lambda_{\rm B}$ & The input scale $\log\log n$, the routing scale $\J(\J(N))$, the packed driver's $\J(\J(n))$ on its working input, and the explicit worklist's logarithmic factor. \\
\bottomrule
\end{tabular}
\end{center}

We first work on a simple directed routing graph with $N$ integer-indexed vertices, $M$ arcs and maximum incoming and outgoing degree at most $D$. Section~\ref{subsec:routing} obtains it by degree reduction from the working input $I$ of Algorithm~\ref{alg:driver}, whose vertex and arc counts are $n$ and $m$, with a degree bound chosen by the driver (Section~\ref{subsec:packed-driver}). Static outgoing lists are sorted once by $(\text{weight},\text{head identifier})$. This costs $O(M\log(2+D))$. Deterministic selection~\cite{bfprt1973}, balanced search trees~\cite{avl1962}, Fibonacci heaps~\cite{FredmanTarjan1987}, and indexed binary heaps are used on stored records.

Duan et al.~\cite[Section 2.2]{DMSY2026} restrict operations on real weights to addition and comparison and state no restriction on integer operations there. We use these operations on weights and route lengths, each in unit time. Graph indices, ranks, masks and pointers use $O(\log(n+m+2))$-bit integer words. The unit-cost word operations are comparisons, addition and subtraction, constant-size copying, shifts and bit access, and direct-address and pointer reads and writes. Integer products, division, logarithms and roots are computed by the charged algorithms of Appendix~\ref{sec:resources}. The packed engine's tables are indexed by integer masks of walk ranks and outgoing slots. Their construction is charged in Section~\ref{sec:packed}. All source, adjacency and record traversal orders and primitive tie choices are fixed.

\subsection{Witness order and active vertices}
A full witness is a source walk, ordered lexicographically by length, hop count and reverse vertex-identifier sequence. Its four-field signature is
\[
 \sig(\omega)\coloneqq(\operatorname{length}(\omega),\operatorname{hops}(\omega),
                 \operatorname{last}(\omega),\operatorname{penultimate}(\omega)).
\]
The zero-hop source has a fixed predecessor sentinel; infinity is greatest. This representation is inherited from Duan et al.~\cite[Section 2.3]{DMSY2026}. Full historical witnesses are proof objects. Label, cap and bound comparisons during the recursive searches use the four copied fields. The global preparation in Section~\ref{sec:packed} is performed once; it additionally compares explicitly stored short continuations and pays for traversing their reverse sequences. Different endpoints have different signatures. Appending a nonnegative edge strictly increases the pair (length, hops), including at weight zero.

The common array $\widehat d$ starts with the source's zero-hop witness and infinity elsewhere. A legitimate update extends the current tail witness by one routing edge and is accepted only if its signature is at most the old head signature. Equal signatures are accepted: a later scan may still need to insert a head into its worklist after a forest search has published the same label. A \emph{probe} inspects an outgoing edge $(u,v)$ to test its candidate $c$, the current witness at $u$ extended by that edge. Along a tail's sorted static outgoing list, candidates extending one tail witness appear in increasing signature order, since they share hop count and penultimate vertex. Stored queue keys, bounds and separators are immutable copies of signatures.

A \emph{legitimate snapshot} is a common-label state obtained from the initialization above by accepted updates. A \emph{cap context} starts at seed $x$ with such a snapshot $C$ at entry (in Theorem~\ref{thm:main}, $C$ denotes a density constant) and the fixed inherited witness $C(x)$. A feasible local route appends a routing continuation from $x$ to this witness; at each newly appended arrival $v$, its prefix must be at most $C(v)$ in full witness order. Earlier arrivals in the inherited witness are not retested against $C$.

Canonical routing paths, prefixes and ancestry refer to this local continuation starting at $x$, with the inherited prefix fixed. The canonical optimum $\pi_C(v)$ is the minimum full witness among these feasible local routes to $v$, or infinity if none exists. A vertex lies below a signature bound $B$ when $\sig(\pi_C(v))<B$; a label or candidate lies below $B$ when its own signature does.

Only $x$ starts active. On an accepted relaxation from an active tail, the head becomes active, and the common label is replaced by the candidate. The effective local label is $\widehat d(v)$ at active vertices and infinity elsewhere. Before activation, the common label remains $C(v)$. Consequently, for a probe of $(u,v)$ with candidate $c$ obtained by appending that arc to the current witness at $u$, the runtime guard

\begin{equation}\label{eq:guard}
 \sig(c)\leq\sig(\widehat d(v))
\end{equation}
is exactly the conjunction of the local-current guard (the candidate is at most the effective local label of $v$) and the entry-cap guard $c\leq C(v)$, both in full witness order; on legitimate histories, comparing signatures gives the same answer as these full-order comparisons (Appendix~\ref{app:witness}).

The containing outer context, the cap context of the source whose snapshot is the initial labelling (Section~\ref{sec:framework}), is paused while a capped invocation runs in its own cap context. Publication of an accepted update writes the common label and records accepted (active) membership in the invocation's cap context and its containing outer context. Thus invocation writes belong to the outer context's accepted history as well, while its worklist and groups remain paused.

A vertex is \emph{complete} relative to $C$ when the full witness of its current common label $\widehat d(v)$ equals $\pi_C(v)$.

\begin{lemma}\label{lem:semantics}
Accepted full witnesses are feasible in every cap context in which they are accepted, and nonincreasing in full order. Prefixes of canonical routing paths are canonical. While every write is accepted in a given cap context, completeness relative to it persists, and every vertex on a complete vertex's canonical routing path is complete relative to it. If a capped invocation or an inner forest search starts at a complete vertex of its containing context, the canonical witnesses of the search's context (for an inner forest search, a proof-only cap context at its root with its entry snapshot) and of its containing context are equal on every canonical descendant of that seed in the containing context that lies below the search's upper bound there.
\end{lemma}
By the last assertion, a local search that completes the canonical descendants of a complete seed below its bound completes them in the containing context as well; Sections~\ref{subsec:outer-run} and~\ref{subsec:packed-coverage} obtain property (F) of Section~\ref{sec:framework} in this way. The chronological proof, including the equal-signature and inherited-prefix cases, is in Appendix~\ref{app:witness}; it gives the equality of the last assertion on every canonical descendant of the seed, and only those below the bound are used.

\subsection{Retained endpoints and stopping}\label{subsec:stopping}
A \emph{capped invocation} is a run of the recursion of Section~\ref{sec:framework} from a single seed $x$ in its own cap context (Section~\ref{subsec:core}). The outer recursion uses it as its local search (Section~\ref{subsec:outer-run}): it either completes ordinarily and returns its output with complete witnesses, or stops and exports the tree of vertices it has retained. A local search can spend work probing a vertex even when its label stays unchanged. The size and contact tests of a capped invocation count the endpoints of \emph{eligible} probes, those that meet the retention test~\eqref{eq:retain} below, including rejected relaxations. At a stop the caller uses the edges that first retained these endpoints to form groups. The work bounds of Section~\ref{sec:invocation-cost} are stated in terms of the same count.

The capped invocation receives a retention limit $P$, an upper signature bound $B_*\!>\sig(C(x))$, a lower threshold $L_*\leq\sig(C(x))$, and a fixed external forest $F_*$ excluding $x$. In the outer recursion, the call that starts the invocation supplies these: its retention limit, upper bound, lower threshold (the least current signature of its starting vertices) and forest of earlier stopped trees (Section~\ref{subsec:procedure}). The invocation's \emph{retained} set $K_*$ and \emph{active} set $A_*$ both initially contain only $x$.

The invocation has two \emph{outward stops}: a size stop on its $P$th retained vertex and a contact stop on its first retained contact with $F_*$. These differ from the size and contact stops of an inner forest search, which fire on the $k$th vertex of its own retained tree and on its first contact with the forest retained by its call. Such a search tests retention against its call's bound $B\leq B_*$ and lower threshold, at least $\sig(C(x))$, whereas the invocation uses $B_*$ and $L_*$ (Section~\ref{subsec:core}). An outward stop exports the actual tree of retained vertices and leaves accepted label updates in place; it does not certify them as complete. Write $q\coloneqq|K_*|$ for the number of retained vertices. Each probe introduces at most one new retained head, so $q\leq P$, counting the triggering endpoint.

An ordinary completion before either outward stop returns exactly the feasible vertices whose canonical local witnesses from $x$ have signature below $B_*$, with complete witnesses; Section~\ref{sec:invocation-cost} derives this \emph{ordinary-completion claim} from Theorem~\ref{thm:return}.

On every graph probe $(u,v)$, before its internal tests or the deletion of its edge (Algorithm~\ref{alg:packed-search}), the invocation's \emph{retention observer} forms its candidate $c$ and retains $v$ if

\begin{equation}\label{eq:retain}
 \sig(c)<B_*\quad\hbox{and}\quad\sig(\widehat d(v))\geq L_*.
\end{equation}

On ordinary completion, the number $r$ of accepted output vertices can be smaller than $q$. For example, with lower threshold $0$ and upper bound $3$, a seed of label $0$ with an edge of weight $2$ to a vertex capped at $1$ retains that vertex and rejects the relaxation.

If the candidate is below $B_*$ but the current head signature is below $L_*$, the observer does not retain the head on this probe. The candidate extends an active route from $x$, so it is above $L_*$ and cannot improve that head. A head retained by \eqref{eq:retain} remains in $K_*$ even if the relaxation fails; the comparison of Section~\ref{subsec:packed-projection} runs on the subgraph induced by $K_*$.

On first retention the observer records the endpoint's first-retention parent edge from the already active tail; probing tails are active because inner forest searches probe only from their roots and accepted vertices, and complete-tail scans (Section~\ref{subsec:core}) only from call outputs, which are call sources or accepted vertices. Every accepted candidate lies below a bound at most $B_*$ and strictly above the seed signature $\sig(C(x))\geq L_*$. Its old head label is at least that candidate. Hence it meets \eqref{eq:retain}; with the seed initialized in both sets, $A_*\subseteq K_*$. Each first-retention parent is therefore already retained, so the first-retention edges form a tree rooted at $x$. A rejected endpoint has only this metadata until a later accepted relaxation promotes it.

Each graph probe is one atomic transaction: the retention observer acts first, then any permitted relaxation and its accepted publication, then all membership, count and stop-flag updates; a firing outward stop is taken before any pending queue action or the next probe. A \emph{sentinel probe}, on the first edge whose candidate is at least the current scan's upper bound, performs only the retention observation; the observer acts on every graph probe, so $K_*$ consists of $x$ and the endpoints of all eligible probes. A sentinel probe can still have its candidate below the invocation's upper bound $B_*$, so it can retain a vertex and trigger an outward stop. On a stop the invocation exports its actual retained tree with its first-retention edges and destroys its owned state; accepted labels remain. Appendix~\ref{app:transactions} fixes the remaining transaction details, which the comparison of Section~\ref{subsec:packed-projection} uses.

Within a capped invocation, the inner forest searches are those of the \emph{packed engine} (Section~\ref{sec:packed}); Section~\ref{sec:invocation-cost} states what each inner forest search must supply to its caller.

\section{The bounded recursive search}\label{sec:framework}

A call starts from a set of vertices with known routes and an upper bound on the routes it should consider. It determines the correct labels below a returned cut and leaves a worklist for continuing above that cut. The outer recursion runs this procedure from the routing vertex representing the input source with an infinite bound. Each capped invocation uses the same procedure from its own seed with its own bound and cap context.

\subsection{Targets and the forest contract}

The \emph{targets} of a call are the vertices whose canonical routes pass through at least one of its input sources and fall below its bound. Fix a cap context and abbreviate its canonical witness by $\pi(v)$. The outer recursion of Section~\ref{sec:outer} uses the \emph{outer context}, the cap context of the source whose snapshot is the initial labelling; its caps at all other vertices are infinite. Every accepted write is published in the outer context, so a vertex is active in it exactly when its common label is finite. A capped invocation uses its own cap context (Section~\ref{subsec:core}). For an upper signature bound $B$ and input source set $S$, set
\[
 \mathcal T(B,S)\coloneqq\{v:\pi(v)\text{ is finite},\ \sig(\pi(v))<B,
                   \text{ and its canonical routing path visits }S\}.
\]
A call $X\coloneqq\calR(B,S,j)$ has a nonempty deduplicated input source set $S$ of active vertices, finite current source signatures below $B$, and two input conditions:
\begin{enumerate}[label=(I\arabic*)]
\item Every target has a complete canonical ancestor in $S$, possibly itself.
\item Every feasible vertex below $B$ outside $\mathcal T(B,S)$ is complete.
\end{enumerate}
Condition (I2) controls endpoints outside the target: with it, Lemma~\ref{lem:foreign} shows that retained ones keep fixed labels and that accepted extensions from target tails stay in the target. The single-seed root satisfies (I1) and (I2), since every canonical routing path of its context starts at its complete seed (Section~\ref{sec:model}). A source of minimum canonical signature has no proper source ancestor, so it is complete by (I1). Thus
\[
 a_X\coloneqq\min_{s\in S}\sig(\pi(s))=\min_{s\in S}\sig(\widehat d(s)).
\]

The recursion level decreases from a call at level $j$ to its children at level $j-1$; level zero is the base case. Fix integers $t\geq1$ and $k\geq2$ for one realization of the shared recursion. The parameter $t$ controls growth between levels, and $k$ is its local retention limit. Inside a capped invocation, $(t,k)$ is the inner record of Section~\ref{subsec:packed-cost}; in the outer recursion it is $(T,P)$ of Section~\ref{subsec:outer-run}. The capacities and guards below use the pair fixed for that realization.

The worklist represents vertices with stored keys that are immutable copies of signatures. The current key of a vertex $v$ is $\sig(\widehat d(v))$; its stored key may be larger.

To form a batch, the call first searches from individual sources. Searches that exhaust below the retention limit contribute their output directly. Trees from searches that stop serve to group the remaining sources. Write $\Forest(B,S)$ for this operation. Its output consists of an actual retained edge forest $F$, a persistent set $Q$ of roots whose searches exhaust with fewer than $k$ retained vertices, and the accepted output $W$ of those \emph{small full searches}. Put $P_S\coloneqq S\setminus Q$. The sources in $P_S$ are assigned to disjoint \emph{groups}, each lying in one connected tree piece and having fewer than $3k$ members. A group enters the worklist through one representative; its remaining members are expanded only when that representative is pulled.

The accepted output of each small full completion is complete relative to its own cap context; $W$ need not be complete in the containing context. Completeness in the containing context comes from the three forest properties below, obtained from complete starting sources.

The forest is required to satisfy three semantic properties:
\begin{align}
 &\text{each target has a complete $P_S$ ancestor or is complete in $W$;}\label{eq:forest1}\\
 &\mathcal T(B,S)\setminus\mathcal T(B,P_S)\text{ is complete in }W;\label{eq:forest2}\\
 &\text{each target of $\mathcal T(B,P_S)$ has a complete $P_S$ ancestor.}\label{eq:forest3}
\end{align}
They follow from (I1) and one property of the local search, which Sections~\ref{subsec:outer-run} and~\ref{sec:packed} verify for their searches:
\begin{enumerate}[label=(F)]
\item a local search that starts at a complete source $s$ and exhausts with fewer than $k$ retained vertices leaves every target whose canonical path visits $s$ complete in this cap context, not only relative to its own search, and puts it in $W$.
\end{enumerate}
Indeed, by (I1) a target has a complete source ancestor $s$. If the search from $s$ exhausts with fewer than $k$ retained vertices, (F) puts the target, complete, in $W$; otherwise $s\notin Q$, so $s$ is a complete $P_S$ ancestor. This proves \eqref{eq:forest1}. A target outside $\mathcal T(B,P_S)$ has no $P_S$ ancestor, so \eqref{eq:forest1} gives \eqref{eq:forest2}. A complete target of $\mathcal T(B,P_S)$ has a complete $P_S$ canonical prefix, because complete vertices have complete canonical prefixes (Lemma~\ref{lem:semantics}); so \eqref{eq:forest1} also gives \eqref{eq:forest3}.

The forest construction below also uses the following structural requirement of a local search started at a root $s\notin F$ of the call:
\begin{enumerate}[label=(F$'$)]
\item every vertex that the search retains after $s$ is retained from an active tail of the search, with candidate signature below $B$ and current head signature at least $a_X$, and is attached by that first edge, so its retained vertices form a tree; every vertex it accepts is retained and lies in $\mathcal T(B,S)$; it stops on its $k$th retained vertex or on its first retained contact with $F$; and a stopped tree meets $F$ only at its contact endpoint.
\end{enumerate}
Retained vertices outside the call's target are called foreign leaves: they are rejected endpoints and hence are not expanded. By (F$'$) they satisfy the hypotheses of Lemma~\ref{lem:foreign}, which keeps their labels fixed, with signatures in $[a_X,B)$, for the counting of Section~\ref{subsec:foreign}. The component sizes used in the forest construction come from the size and contact stops in (F$'$), and the group sizes from Lemma~\ref{lem:partition}.

Each small full search starts at a target source and accepts only target vertices by Lemma~\ref{lem:foreign}, so $W\subseteq\mathcal T(B,S)$. Each accepts fewer than $k$ vertices, and there is one such search per root in $Q$, giving $|W|\leq k|Q|$. Group lists change only by permanent deletion. Each group has a designated representative, chosen from its members with minimum current key when the group is formed or when its previous representative is returned. The latter event marks the group dirty; dirty lists avoid scanning all groups after each child.

\noindent\textbf{Capacities for batches and output.} For integers $j\geq0$ define

\begin{equation}\label{eq:capacities}
 A_j\coloneqq t^2 2^{jt},\quad H_j\coloneqq t^3 2^{jt},\quad
 M_0\coloneqq1,\quad M_j\coloneqq16t2^{(j-1)t}\quad(j\geq1).
\end{equation}
Each call is required to have $|S|\leq A_j$; the child-input argument in Appendix~\ref{app:return} preserves this premise. Below, a level-$j$ call pulls at most $M_j$ entries at a time and stops pulling or extracting once its output reaches $H_j$. At nonbase levels $j\geq1$ require
\begin{equation}\label{eq:capacityguards}
 (3k+1)M_j\leq A_{j-1},\qquad
 H_j+2H_{j-1}+kA_j<2H_j.
\end{equation}
The first inequality permits expansion of pulled group representatives; the second bounds the final output after the last child and the small full searches. The outer and inner realizations below enforce $t\geq4096$ and $16(3k+1)\leq t$, which imply $k\leq t/16$ and both guards: for $j\geq1$,
\[
 \frac{(3k+1)M_j}{A_{j-1}}=\frac{16(3k+1)}{t}\leq1,\qquad
 \frac{2H_{j-1}+kA_j}{H_j}=2\,2^{-t}+k/t<1.
\]
They also give $3M_{j-1}\leq M_j$, which the lazy merge of Appendix~\ref{app:lazy} needs: $M_1=16t\geq3M_0$, and at higher levels $M_j/M_{j-1}=2^t\geq3$.

\subsection{Worklists and return values}\label{subsec:worklists-returns}

A parent pulls candidate sources from its worklist and merges the residual worklist returned by each child. The worklist supports insertion keeping the smaller key for a repeated identifier, deletion, a smallest-key pull, and ordered merge. A pull removes all logical entries if there are at most $M_j$, and otherwise exactly the $M_j$ smallest; it returns their vertex identifiers and a cut $b$: their stored keys are below $b$, and all remaining stored keys are at least $b$; exhaustion gives $b=B$. A child structure keeps its input bound $b$ as its final bound until it is merged into its parent; the ordered merge uses this bound (Appendix~\ref{app:lazy}).

A call that no outward stop interrupts is \emph{ordinary}. Write $(\beta,U,\mathcal D_X)$ for its return: $\beta$ is its cut, $U$ its completed output (in a nonbase call of Algorithm~\ref{alg:recursion}, the returned set $U\cup W_{\rm new}$, whose $U$ collects the child outputs), and $\mathcal D_X$ the residual worklist. In a nonbase call, $L$ is the lower cut for the remaining recursive work; the procedure below specifies its initialization and updates.

An ordinary return is \emph{full} if $\beta=B$ and \emph{partial} if $\beta<B$.

\subsection{The recursive procedure}\label{subsec:procedure}

Level zero is a direct priority-queue search; a nonbase call forms its forest and then processes pulled batches through children. Algorithm~\ref{alg:recursion} states both cases; its nonbase case uses the forest constructed next.

\begin{algorithm}[!t]
\caption{Bounded recursion $\calR(B,S,j)$ in a fixed cap context}\label{alg:recursion}
\begin{algorithmic}[1]
\Require Upper bound $B$, level $j$, active sources $S$ with finite current signatures below $B$ satisfying (I1)--(I2), $0<|S|\leq A_j$; fixed parameters $t,k$ and a worklist representation
\Ensure Returned cut $\beta$, completed prefix $U$, residual worklist $\mathcal D_X$; in a capped invocation, an outward stop instead interrupts the procedure and exports the retained tree \Comment{Section~\ref{subsec:stopping}}
\State $U\gets\varnothing$
\If{$j=0$}
  \State Put $S$ in a minimum queue, keyed by current signatures.
  \While{the queue is nonempty and $|U|<H_0$}
    \State Extract the minimum and add its vertex to $U$.
    \State Scan its outgoing prefix below $B$. Insert or update every accepted head.
  \EndWhile
  \If{the queue is empty}
    \State \Return $(B,U,\varnothing)$
  \EndIf
  \State \AlgWrap{\Return the queue's minimum key, $U$, and the residual queue in the required representation.}
\EndIf
\State Form $F,Q,W$ and the groups using $\Forest(B,S)$.
\State Choose a minimum-current-key representative from each group and insert it into the initially empty worklist $\mathcal D_X$.
\State $L\gets$ the smallest inserted key, or $B$ if there is none.
\While{$\mathcal D_X$ has logical entries and $|U|<H_j$}
  \State Pull the least entries and their cut $b$.
  \State Discard pulled identifiers whose current keys are below $L$.
  \State \AlgWrap{Expand pulled group representatives with remaining group members whose current keys are below $b$. Deduplicate the surviving pulled identifiers and the added members into $S_c$.}
  \If{$S_c\ne\varnothing$}
    \State $(\beta_c,U_c,\mathcal D_c)\gets\calR(b,S_c,j-1)$. Merge $\mathcal D_c$ into $\mathcal D_X$.
    \State \AlgWrap{Delete $U_c$ from the lazy worklist if using it. Remove $U_c$ from the groups, marking groups that have lost their representative.}
    \State Scan outgoing edges of $U_c$ in $[b,B)$ and insert accepted heads.\label{line:scan-child}
    \State Reselect and insert a current minimum for each marked nonempty group.
    \State $U\gets U\cup U_c$ and $L\gets\beta_c$.\label{line:next-cut}
  \EndIf
\EndWhile
\If{$\mathcal D_X$ has no logical entries}
  \State $L\gets B$
\EndIf
\State Reinsert input sources with current keys in $[L,B)$.
\State $W_{\rm new}\gets\{v\in W\setminus U:\sig(\widehat d(v))<L\}$
\State Scan outgoing edges of $W_{\rm new}$ in $[L,B)$ and insert accepted heads.
\State \Return $(L,U\cup W_{\rm new},\mathcal D_X)$
\end{algorithmic}
\end{algorithm}

\noindent\textbf{Constructing the forest.} Process $S$ once in its fixed order, maintaining $F,Q,W$ as specified above. A root already in $F$ is skipped. Every other root $s$ starts a local search with inherited witness $\widehat d(s)$ and the then-current common snapshot, upper bound $B$, lower threshold $a_X$, retention limit $k$, and the current $F$ held fixed for contact. In the outer recursion this local search is a capped invocation with retention limit $k$ and $(L_*,B_*,F_*)=(a_X,B,F)$, so that its retention test \eqref{eq:retain} imposes the candidate and head bounds of Lemma~\ref{lem:foreign}. The outer parameter scheme \eqref{eq:outer-scheme} chooses this retention limit as $P$. In the inner recursion it is the packed search of Section~\ref{sec:packed}; its entry snapshot is used only in the proof and creates no additional publication context. The local search stops on $k$ retained vertices or first contact with $F$. A stopped tree is appended, joining at its sole contact endpoint if present. A small full completion puts its accepted output into $W$ and its root in $Q$, where it remains until this call returns.

Every root outside $Q$ lies in $F$, and every component of $F$ has at least $k$ vertices, since its first tree stopped at size $k$ and each later tree joins it at one contact endpoint. Partition each component of $F$ into connected edge-disjoint pieces, each with at least $k$ and fewer than $3k$ vertices, assign each call input source of $P_S$ to one containing piece, and call the sources assigned to a piece its \emph{group}. Drop empty groups. Lemma~\ref{lem:partition} in Appendix~\ref{sec:structures} proves the partition and its linear total endpoint-list cost.

A scan processes edges under \eqref{eq:guard}, including equal-signature updates, and publishes accepted updates. An insertion keeps the smaller key and stores a copy of the current signature. In a capped invocation every probe is a transaction of Section~\ref{subsec:stopping}. Each capped invocation uses the lazy worklist, with the base queue conversion and scans of Section~\ref{subsec:core}; the outer recursion uses the explicit worklist, with the base search tree and scans of Section~\ref{subsec:outer-run}.

All sets in Algorithm~\ref{alg:recursion} use distinct identifiers. Forest updates publish labels without inserting worklist entries. Appendix~\ref{app:return} proves that $W_{\rm new}$ is complete in the containing context. The explicit worklist may keep records for vertices of $U_c$; their current keys lie below the updated cut and the filter discards them when pulled.

For example, if a child runs with a bound whose length field is $2$ and completes a vertex $u$ at distance $1$, an edge of weight $2$ out of $u$ lies beyond that child's search. If the parent's bound has length field greater than $3$, the parent's scan in step~\ref{line:scan-child} tests this edge at candidate distance $3$. The same scan works after a partial child return because it starts at the child's input bound. The child's residual worklist carries the pending work below that bound.

\subsection{Complete-tail scans and lazy worklists}\label{subsec:core}
The outer recursion uses an explicit worklist, whose per-entry merges are paid by the outer operation count. Inside a capped invocation the recursion must neither rescan complete tails at every level nor merge returned worklists entry by entry, since either would cost about $D$ per returned vertex at every inner level. Instead, static complete-tail cursors pass each static arc at most once, and the lazy worklist has an amortized merge charge independent of the number of returned physical entries, with physical work accounted for by the family potential and placement counts of Appendices~\ref{app:lazy} and~\ref{app:counting}.

For a capped invocation, fix its parameters $t,k,\ell$, the packed record of Section~\ref{subsec:packed-cost}. It uses the capacities \eqref{eq:capacities} and the procedure above with the lazy ordered structure of Appendix~\ref{app:lazy}. A capped invocation with seed $x$ runs the root call $\calR(B_*,\{x\},\ell)$ at initial level $\ell$ in its cap context. Its target consists of the cap-feasible vertices whose canonical local witness from $x$ has signature below $B_*$. Its $a_X$ is $\sig(C(x))\geq L_*$; the invocation's lower threshold $L_*$ enters only the retention observer's test \eqref{eq:retain} and the packed slot exclusions \eqref{eq:rootlow}, and every inner call's lower threshold $a_X$ is at least $\sig(C(x))$.

The lazy live map $\mathcal L$ (Appendix~\ref{app:lazy}), which records each vertex's live entry, is shared by all inner recursive levels in this one capped invocation. Complete-tail scans are the scans performed by base extraction, after a child returns, and for $W_{\rm new}$; they exclude inner forest exploration, which uses slot masks. A base call uses an indexed binary heap; before returning, it inserts its remaining keys into a fresh level-zero lazy structure with parameter $M_0=1$ and final bound $B$. The heap has no entries in the lazy live map. Conversion finishes before the parent resumes.

Each complete-tail scan reads the sorted static outgoing list of its complete tail through one cursor kept for the whole invocation, a static complete-tail cursor as in C-HD~\cite[Sections 3.2--3.3 and 4.4]{CHDPublic2026}. At its first base extraction, traverse the sorted static prefix below $B$, leaving the cursor at the first candidate at least $B$ if one exists, and at list end otherwise. If the tail is instead first output in $W_{\rm new}$, start from the static beginning: pass candidates below $L$ only to the retention observer, process $[L,B)$, and leave the cursor at that first candidate at least $B$, or at list end if there is none. At successive ancestors, continue from that cursor across the band $[b,B)$, where $b$ was the child input bound. The tail witness is fixed after completeness, so these are ascending disjoint candidate bands. The cursor advances past each static edge at most once; at most one upper sentinel probe occurs per scan episode. Each execution of a tail scan is a scan episode and has a charged header, even if its list is empty.

The lazy family keeps at most one live physical entry per vertex, skips an insertion when a no-larger live value exists, and merges a returned child in constant amortized time. For the lazy worklist, we use the generic operation specifications, amortized work bounds, and block-count transition lemmas proved in the C-HD proof package~\cite{CHDGeneric2026} at the cited revision. Appendix~\ref{app:lazy} proves their applicability to our recursion and bounds the work of our implementation in our cost model. Appendix~\ref{app:lazy-use} shows that the recursion supplies the merge premises \eqref{eq:mergepremises} and the skip condition, and states the survival condition under ancestor invalidation, which the recursive induction of Appendix~\ref{app:return} proves.

\subsection{Correctness of the recursive return}\label{subsec:return-correctness}

When a child is called while its parent's current lower cut is $L$, and the child returns cut $\beta_c$, we call $[L,\beta_c)$ the child's \emph{handled interval}; the root call's handled interval is $[a_X,\beta)$ for its returned cut $\beta$. The canonical signatures of a child's returned output lie in its handled interval, and Section~\ref{subsec:foreign} counts retained vertices through the disjointness of these intervals at each level; the minimum canonical signature of the child's sources is at least its entering parent cut but need not equal it (the child-target argument in Appendix~\ref{app:return}).

A partial return must complete at least $t$ times as many vertices as there were input sources; this pays for source work in the cost analysis. Residual sources and canonical boundary heads must remain represented so that the parent can continue beyond the returned cut. For a vertex set $U$, write $N^+(U)$ for the heads of routing arcs with tails in $U$.

Algorithm~\ref{alg:recursion} meets the following contract. It adapts the bounded recursion of Duan et al.~\cite[Lemmas 3 and 9]{DMSY2026} and B1~\cite[Sections S2--S3]{B1PublicDraft2026}, using C-HD's treatment of retained foreign leaves~\cite[Sections 4.2 and 5.1--5.3]{CHDPublic2026}; Appendix~\ref{app:return} proves the contract under the present cap conditions for both worklist representations.

\begin{theorem}\label{thm:return}
For a call as defined above, assume $|S|\leq A_j$, (I1)--(I2), properties (F) and (F$'$) of the local search, $k\geq2$, the capacity guards \eqref{eq:capacityguards}, and one of the two worklist representations: the explicit structure of Appendix~\ref{sec:structures} with a base search tree and full outgoing scans, or the lazy structure of Appendix~\ref{app:lazy} with the base, conversion and scan rules of Section~\ref{subsec:core}, under the parameter conditions stated with Proposition~\ref{prop:ds} and in Appendix~\ref{app:lazy}, respectively. Every ordinary call terminates and returns $(\beta,U,\mathcal D_X)$ such that:
\begin{enumerate}
\item $a_X<\beta\leq B$, and $U$ is exactly the target prefix with canonical signature below $\beta$. Its vertices are complete and $1\leq|U|<2H_j$.
\item If $\beta<B$, then $|U|\geq H_j\geq t|S|$. If $\beta=B$, the entire target and all sources are returned and the logical worklist is empty.
\item Vertices represented in the logical residual worklist lie in $\mathcal T(B,S)\cap(S\cup N^+(U))$, with stored keys in $[\beta,B)$ and at least their current keys.
\item Every unreturned call input source is represented at its final current key. Every canonical edge from $U$ to a residual target has its complete head represented at its optimal key.
\item Immediate child outputs are disjoint. Outputs at a fixed level are disjoint subsets of the root output. Child handled intervals nest in their parent's interval, and handled intervals at one level are disjoint.
\end{enumerate}
\end{theorem}
The full proof is in Appendix~\ref{app:return}, jointly with the legality of every worklist operation. The numerical requirements are those of Proposition~\ref{prop:ds} for the explicit representation, ensured by the outer tests and explicit allowance of Section~\ref{subsec:outer-run}, and those of Appendix~\ref{app:lazy} for the lazy one, ensured by the engine guards \eqref{eq:packed-innerguards}. Section~\ref{subsec:packed-cost} checks the shared semantic and worklist premises for the packed engine, and Appendix~\ref{app:return} verifies the operation-specific semantic premises. For the explicit outer structure, each vertex represented in the logical residual worklist has one record. For the lazy inner structure, stale physical entries are additional storage, charged by placement counts.

\subsection{Fixed ranks for foreign leaves}\label{subsec:foreign}
Fixed ranks ensure that each vertex is retained by at most one full call per level, even when it lies outside the target or is cap-infeasible. C-HD's foreign-leaf argument~\cite[Sections 4.2 and 5.1--5.3]{CHDPublic2026} counts endpoints using disjoint handled intervals. Caps introduce infeasible vertices, which have no finite local optimum. The following rank extends that counting argument to them:
\begin{equation}\label{eq:mu}
 \mu(v)\coloneqq\begin{cases}\pi_C(v),&v\text{ cap-feasible},\\ C(v),&v\text{ cap-infeasible}.\end{cases}
\end{equation}
The context has no outside writer. An infeasible vertex never accepts a route, so its common label is always $C(v)$. At a feasible vertex, $\pi_C(v)\leq\widehat d(v)\leq C(v)$ in full order.

The mutable edges below are the surviving entries of a tail's slot mask (Section~\ref{sec:packed}); their deletion leaves the static outgoing lists used in Section~\ref{subsec:core} unchanged.
\begin{lemma}\label{lem:foreign}
In a call $X\coloneqq\calR(B,S,j)$, suppose an endpoint $v$ outside its target is retained from an active target tail with candidate signature below $B$ and current head signature at least $a_X$. Then
\[
 \widehat d(v)=\mu(v),\qquad a_X\leq\sig(\mu(v))<B,
\]
and this equality persists in the context. Accepted below-$B$ extensions from target tails stay in the target. A mutable edge from a target tail, deleted when its head signature is below $a_X$, remains ineligible for accepted relaxation within this cap context.
\end{lemma}
\begin{proof}
Let $c$ be the candidate witness of the probe that retained $v$. If $c$ respects $C(v)$, it is a feasible route and $\pi_C(v)\leq c$. If it violates the cap, then $C(v)<c$; in either case $\mu(v)\leq c$. Thus its fixed rank is below $B$. A feasible endpoint outside the target is complete by (I2); an infeasible endpoint keeps its cap as its common label. The retention test gives the lower bound, and completeness or infeasibility gives persistence.

If an accepted head were outside the target, it would be feasible and complete. Acceptance then forces its full witness to be canonical; its canonical tail prefix visits $S$, a contradiction. Finally, a tail in this target has canonical rank at least $a_X$. After deletion, every later tail label is at least that rank, while the head stays below the deletion-time threshold by nonincrease. Nonnegative extension with increased hops can never pass the runtime guard again. A canonical edge from a target tail has head signature at least $a_X$, so these deletions remove no such edge.
\end{proof}

For a full call, target vertices are returned and foreign vertices satisfy Lemma~\ref{lem:foreign}; every retained vertex has fixed signature $\sig(\mu(v))$ in $[a_X,B)$. At a fixed level, these intervals are disjoint, since they lie in the disjoint handled intervals of full calls (the interval argument in the proof of Theorem~\ref{thm:return}, Appendix~\ref{app:return}). Thus the retained vertices are distinct across those calls, and their retained forests contain at most as many vertices in total as the routing graph on which the recursion runs. A local search stopped on contact contributes only one previously retained endpoint, so the sum of exported tree sizes is at most twice this total. Exhausted searches with fewer than $k$ retained vertices are charged separately through $Q$.

Full-call group reselections are paid by one charge per piece plus at most one charge per routing arc in the context. In the adapted home-color argument of C-HD~\cite[Section 5.3]{CHDPublic2026}, a retained vertex is colored by the child's handled interval containing its fixed rank $\mu$, or by an auxiliary color if no interval contains it. A group member can be removed only by a child of its color, and edges joining different colors pay extra reselections (Appendix~\ref{app:counting}). For partial calls, the bound $t|S|\leq|U|$ of Theorem~\ref{thm:return} pays for their group reselections.

\section{The routing graph and the outer recursion}\label{sec:outer}

The outer recursion in Algorithm~\ref{alg:driver} works on a bounded-degree routing graph. We construct that graph, recover the input-graph solution from its distances, reduce the exactness of the recursion to rules for its local searches, and bound its cost in terms of those searches.

\subsection{Routing graph and original output}\label{subsec:routing}
To limit the number of arcs scanned at a vertex, replace a vertex with many incident arcs by a cycle of ports. The counts $n,m$ remain those of the working input $I$ of Algorithm~\ref{alg:driver}, including loops and parallel arcs. Remove nonnegative self-loops and keep a minimum-weight original edge record for each ordered pair, choosing the first on ties. One destination-stamp array, reused across tails, performs this in $O(n+m)$ time; let $m_0$ be the number of surviving arcs. If the set consisting of the source and all endpoints of surviving arcs has at most two vertices, solve it directly by inspecting its surviving arcs; every other vertex is unreachable. No later construction, test or bound relies on this shortcut. Otherwise keep all original vertices, and let $D$ be the integer degree bound chosen by the driver, with $3\leq D\leq\min\{\lfloor m/n\rfloor,n\}$ (Section~\ref{subsec:packed-driver}).

For each original vertex, concatenate its incoming and outgoing incidences and divide this combined list into segments of at most $D-2$. Make one port per segment, one port for an isolated vertex, and connect its ports by a directed zero-weight cycle, omitting a singleton loop. Attach surviving original arcs at their assigned ports. Choose the original source's first port $s'$ as the routing source. This is the degree reduction of Duan et al.~\cite[Section 2.1]{DMSY2026}, with the rounding of combined incidence lists fixed here; the driver chooses its degree bound from these counts.

The routing graph is simple with both degrees at most $D$, so $M\leq ND$. Every original path lifts through zero-weight cycles, and every routing path projects with the same weight. All ports of an original vertex have its original distance. A vertex with $i\geq1$ incidences receives at most $1+i/(D-2)$ ports, so $n\leq N\leq n+2m_0/(D-2)$. Also $M\leq m_0+N$, and $D\leq m/n$ and $D/(D-2)\leq3$ give $ND\leq nD+2m_0D/(D-2)\leq m+6m_0\leq7m$. Since $N\leq n+2m$ and $M\leq n+3m$, the reduction costs $O(n+m)$ even when $N$ exceeds a constant multiple of $n$.

After the routing distances are computed, read one port distance per original vertex; when padding was used, the original vertices are the unpadded ones (Section~\ref{subsec:sparse}). Traverse the \emph{original} tight-edge graph from the source, which starts discovered: for each original arc $f=(u,v)$ of weight $w(f)$ leaving a discovered vertex $u$, test finite labels and $\dist(u)+w(f)=\dist(v)$, and assign a predecessor only on first discovery. Every reachable vertex has a tight shortest path; first discovery gives an acyclic tree even with zero-weight cycles. This costs $O(n+m)$ and returns original predecessor edges. Unreachable vertices keep distance infinity.

\subsection{Outer recursion with capped invocations}\label{subsec:outer-run}
The recursion starts from $s'$ with no upper bound. Its local searches start from individual seeds and must return either completed vertices or a tree that the caller uses to form batches. The packed engine of Section~\ref{sec:packed} implements this interface with its inner parameter record.

Let $g\geq D$ be the driver's integer estimate of the cost per retained vertex of a capped invocation; it is fixed before $P$ is chosen, although that cost itself depends on $P$ (Section~\ref{subsec:packed-driver}). With $L\coloneqq\J(N)=\lceil\log_2(N+2)\rceil$ (a logarithmic scale, not the lower cut of Algorithm~\ref{alg:recursion}), choose the outer scale $T$, retention limit $P$ and recursion depth $e$ by
\begin{equation}\label{eq:outer-scheme}
 T\coloneqq\min\{s\in\mathbb Z_{>0}:Ds^2\geq Lg\},\qquad
 P\coloneqq\left\lfloor\frac{T}{1024g}\right\rfloor,\qquad
 e\coloneqq\lceil\log_2(N)/T\rceil.
\end{equation}
The scale $T$ balances the level work $NLg/T$ against the edge-insertion work $NDT$ (Section~\ref{subsec:outer-cost}); $P$ is small enough that a small full search costs $O(T)$ and, once $P\geq2$, large enough that $T/P=O(g)$. The choice $P=\lfloor T/(1024g)\rfloor$ gives $Pg\leq T/1024$. The capacity formulas \eqref{eq:capacities} use $(t,k)=(T,P)$. Before constructing capacity tables, the driver tests
\[
 4096\leq T\leq L,\quad T^2\leq N,\quad2\leq P\leq N,\quad
 16(3P+1)\leq T,\quad P\leq T/16;
\]
these are part of \eqref{eq:packed-outerguards}. The inequalities $16(3P+1)\leq T$ and $P\leq T/16$ imply \eqref{eq:capacityguards} for $(t,k)=(T,P)$. The bound $T\leq L$ and the coarse parameter bounds tested by the driver control the finite record and setup; $T^2\leq N$ already follows from $4096\leq T\leq L$.

To bound the amortized insertion charge of the outer worklist by $O(T)$, we test an explicit allowance. For the explicit structure at each nonbase level, whose record bound $R_{\rm B}$ in Proposition~\ref{prop:ds} is at most $R_j$ (Appendix~\ref{app:return}), require
\begin{equation}\label{eq:outerds}
 R_j\coloneqq A_j+2DH_j,\qquad 2M_j+R_j\leq M_j2^{2T}.
\end{equation}
The test is equivalent to $\log_2(2+R_j/M_j)\leq2T$, so it bounds the insertion charge $\lambda_{\rm B}$ of Proposition~\ref{prop:ds} by $2T$, which supplies the factor $T$ in the outer insertion costs below. Since $M_j\geq16T\geq256$, it also gives $\max\{1,\log_2(2+R_j/M_j)\}\leq2T\leq M_j/8$, so Proposition~\ref{prop:ds}, whose insertion charge is this maximum, applies with $c_0=1/8$; its per-entry merge cost is unchanged. After $T\leq L$ has passed, \eqref{eq:outerds} is tested in a cancelled scalar form, without constructing its tables (Appendix~\ref{app:word-bounds}).

In the top call $\calR(\infty,\{s'\},e)$, use the explicit worklists with per-entry merge and separate per-level arrays (Appendix~\ref{sec:structures}), the base balanced search tree, the full outgoing scans, and the representatives of Section~\ref{sec:framework}. Whenever a call starts a local search from $x$, pass the retention limit $P$, the lower threshold $a_X$, the upper bound $B$ and its current forest $F$. The capped invocation runs its own copy of the recursion from $x$ in the cap context of Section~\ref{subsec:core}. It publishes accepted labels and the specified outer accepted-membership records. The outer worklist and its maps, groups, forest and control state wait for the search to stop or complete.

The search stops when it retains its $P$th vertex or first retains a vertex already in $F$. It stops at the end of the probe transaction (Section~\ref{subsec:stopping}) and returns the tree of vertices it retained, including any rejected endpoints. Append this tree to $F$, joining at the contact if there is one. If the search exhausts ordinarily, discard the tree, put $x$ in $Q$, and add its accepted output to $W$.

Exactness of the outer recursion follows from Theorem~\ref{thm:return} applied to the top call, whose premises we now check. Those on the local search, properties (F) and (F$'$), need only that every capped invocation obey the rules of Section~\ref{subsec:stopping}, including the ordinary-completion claim derived in Section~\ref{sec:invocation-cost}. The packed engine does so under \eqref{eq:walkguard} and \eqref{eq:packed-innerguards} (Theorem~\ref{thm:packed-cap}). A complete outer source's canonical descendants below $B$ are cap-feasible for its entry snapshot, with the same canonical witnesses, by Lemma~\ref{lem:semantics}. An invocation that exhausts with fewer than $P$ retained vertices completes ordinarily and therefore, by the ordinary-completion claim of Section~\ref{subsec:stopping}, returns those descendants as accepted output, complete in the outer context and not only relative to the invocation, and they enter $W$; this is property (F). The retention test \eqref{eq:retain} with $(L_*,B_*,F_*)=(a_X,B,F)$, the actual first-retention tree that the invocation hands back (Section~\ref{subsec:stopping}) and the two outward stops give (F$'$), and $W$ contains only accepted targets of the outer call.

The top call has $|S|=1\leq A_e$ and, as a single-seed root, satisfies (I1)--(I2) (Section~\ref{sec:framework}); the tests above give $k=P\geq2$ and \eqref{eq:capacityguards}, and with \eqref{eq:outerds} they give the numerical requirements of Proposition~\ref{prop:ds}. For its merge condition, a nonbase child of a level-$j$ call has quota $M_{j-1}=2^{-T}M_j<M_j/3$, while a base child supplies its search tree. Hence Theorem~\ref{thm:return} applies to the top call, including the retained foreign leaves of its forests; foreign metadata is counted separately. Since $H_e>N$, the top call returns fully and computes all routing distances.

\subsection{Outer cost}\label{subsec:outer-cost}
At scale $T$, there are $e+1$ recursion levels. Increasing $T$ reduces the number of levels but increases the work charged to each edge insertion. The next bound counts both costs.

\begin{lemma}\label{lem:outer-cost}
Suppose the outer tests and the explicit allowance \eqref{eq:outerds} of Section~\ref{subsec:outer-run} pass, every capped invocation of the outer recursion obeys the rules of Section~\ref{subsec:stopping}, including the ordinary-completion claim, and every such invocation that retains $q$ vertices, whether it stops or completes, has per-invocation work (Section~\ref{sec:invocation-cost}) at most $c\,qh_{\rm loc}$, for a number $h_{\rm loc}\geq1$ and a constant $c$. Then the outer recursion, including the sorting of static outgoing lists and the once-only array setup of Appendix~\ref{sec:resources}, takes time
\begin{equation}\label{eq:outercost}
 \begin{split}
 O\big(&N+M\log(2+D)+N(e+1)(h_{\rm loc}+D+T/P+1)\\
       &+Ph_{\rm loc}(N+M+Ne/T)+MT+ND\log(2+TD)\big),
 \end{split}
\end{equation}
with an implicit constant depending only on $c$. The original reduction and output add $O(n+m)$; any global preparation of the engine is paid separately.
\end{lemma}
\begin{proof}
At each level the retained forests of full calls contain at most $N$ vertices in total by their fixed global ranks and disjoint handled intervals (Section~\ref{subsec:foreign}). First contacts and piece endpoint occurrences (Lemma~\ref{lem:partition}) multiply this by only a constant. The sum of exported tree sizes in a partial forest is at most $P|S|$, paid by $T|S|\leq|U|$. Invocations that exhaust with fewer than $P$ retained vertices are charged by $Ph_{\rm loc}|Q|$. Write $Q_X$ for the set $Q$ of an outer call $X$, the roots of searches that exhaust with fewer than $P$ retained vertices. The source-origin proof in Appendix~\ref{app:counting}, applied to the outer context, gives
\[
 \sum_X|Q_X|\leq N+M+Ne/T.
\]
Hence all capped invocations and forest processing cost
\[
 O\!\left(N(e+1)(h_{\rm loc}+1)+Ph_{\rm loc}(N+M+Ne/T)\right).
\]
Initial group representatives and per-piece reselections cost $O(N(e+1)T/P)$. The coloring by child intervals in Section~\ref{subsec:foreign} charges all other full-call reselections to $O(M)$ edges. Each charge pays for an $O(P)$ group scan and an explicit insertion of amortized cost $O(T)$, inside the $MT$ term of \eqref{eq:outercost}. Partial-call group work and source restorations use the bound $T|S|\leq|U|$ of Theorem~\ref{thm:return}. Edge-insertion events occur at most once per physical edge across the disjoint output bands and cost $O(MT)$ (Appendix~\ref{app:counting}).

Outer full outgoing scans can revisit a tail at different levels; they cost $O(N(e+1)D)$. A full child returns an empty explicit structure; a partial child has at most $|S_c|+D|U_c|$ distinct vertex/key records. Summing over child outputs bounds the explicit per-entry merges, the pulled records with outdated stored keys and the conversions into the explicit arrays by the same $O(N(e+1)D)$ term. Base search trees have support $O(T^2+DT^3)$, and their distinct output and edge events cost $O(ND\log(2+TD))$. Including sorting and the once-only array setup gives \eqref{eq:outercost}. Appendix~\ref{app:counting} gives the source-origin, group and explicit record counts used here, including empty filtered batches and input-source restorations.
\end{proof}

Suppose in addition that $h_{\rm loc}\leq c_1g$ for a constant $c_1$. The definition of $P$ and $P\geq2$ give $Pg\leq T/1024$ and $T/P\leq2048g$; also $g\geq D$, $e+1\leq L/T+2$ and $M\leq ND$. The level term $N(e+1)(h_{\rm loc}+D+T/P+1)$ and the small-search term $Ph_{\rm loc}(N+M+Ne/T)$ of \eqref{eq:outercost}, together with $MT$, are therefore
\[
 O\bigl(NLg/T+Ng+NDT+Ne\bigr),
\]
with a constant depending only on $c$ and $c_1$. The terms $NLg/T$ and $NDT$ balance at $N\sqrt{DLg}$: the least-integer choice of $T$ gives $DT^2\geq Lg$, hence $NLg/T\leq NDT$, and $Lg\geq D$ gives $T\leq\sqrt{Lg/D}+1\leq2\sqrt{Lg/D}$, hence $NDT\leq2N\sqrt{DLg}$. Also $Ne=O(NL/T+N)$, which is $O(NLg/T+N)$ since $g\geq1$. Thus the bound \eqref{eq:outercost} is $O(N\sqrt{DLg}+Ng+N+M\log(2+D)+ND\log(2+TD))$; the driver absorbs the terms after the first (Section~\ref{subsec:packed-branch}).

For the driver estimates of $P$, note that $DT^2\geq Lg$ also gives
\[
 T/g\geq\sqrt{L/(Dg)}.
\]
Hence, once $\sqrt{L/(Dg)}\geq2048$, which holds eventually whenever $Dg=o(L)$, rounding down in the definition of $P$ loses at most a factor $2$: $T/(2048g)\leq P\leq T/(1024g)$. Together with $\sqrt{Lg/D}\leq T\leq2\sqrt{Lg/D}$ this gives $P=\Theta(\sqrt{L/(Dg)})$.

\section{The cost of a capped invocation}\label{sec:invocation-cost}
The outer bound charges a capped invocation $O(qg)$ when it retains $q$ vertices. The invocation may stop early, and some of the vertices it retains may never receive an update. For an inner forest search with the properties listed below, we first bound the work of a completed run on those $q$ vertices. The packed analysis then uses that bound for the \emph{actual run}, the invocation as executed.

\noindent\textbf{Completed comparison runs.} A \emph{completed comparison run} is a proof-only execution on the subgraph induced by the actual retained vertices, started from the invocation's entry state (entry caps as labels, only the seed active), with the same parameter record and outward stops masked. B1~\cite[Section S1.C, Lemmas D1--D3]{B1PublicDraft2026} analyzes a local search using its entry labels as caps and matches the search to an execution on the induced region reachable by admissible walks. In B1, an admissible walk avoids the current sinks and, at each nonempty prefix, its label is at most its endpoint's initial label and strictly below the call bound; the induced region contains the root and the admissibly reachable endpoints. Here every active tail and accepted head is retained, as are eligible rejected endpoints. On the retained graph, the proof matches the actual and comparison runs' labels, accepted membership and recursive worklists up to the stopping probe, with the separately charged exceptions in Appendix~\ref{app:coupling}. The match holds because an actual probe to an unretained head either is its scan's terminating sentinel probe, paid by the scan header, or meets a head whose current label is below $L_*$; such an edge never carries an accepted relaxation in the invocation. Differences in the comparison's retention tree and outward-stop flags do not affect the matched state. Proposition~\ref{prop:packed-projection} bounds the actual work by a constant multiple of the completed comparison work plus $O(Dq+q)$.

\noindent\textbf{Work convention.} Per-invocation work includes probes and scan headers, state touches and list copying, concrete worklist operations, allocation, publication to the two contexts, return or export copying, restoration and owned cleanup. Once-prepared static sorting, shared arrays and parameter tables are excluded, as are original-input reduction, final global output and the global preparation of the packed engine; Section~\ref{sec:outer}, Section~\ref{subsec:packed-driver} and Appendix~\ref{sec:resources} pay them separately. Each invocation initializes and clears state only as it is touched (Appendix~\ref{sec:resources}).

\noindent\textbf{The fixed record.} Every invocation with retention limit $P$ uses one parameter record $(t,k,\ell)$, prepared from $P$ and $D$ before the first invocation (Section~\ref{subsec:packed-cost}), and its completed comparison run on $q\leq P$ vertices uses the same record. Put $A\coloneqq\J(P)$ and $\ell\coloneqq\lceil\log_2(P)/t\rceil$, and use the capacities \eqref{eq:capacities}. The packed guards \eqref{eq:packed-innerguards} contain
\begin{equation}\label{eq:shared-guards}
 \begin{gathered}
 1\leq D\leq P,\qquad4096\leq t\leq A,\qquad t^2\leq P,\\
 16(3k+1)\leq t,\qquad k\leq t/16,\qquad\J(D+\ell+2)+6\J(t+2)\leq t/4.
 \end{gathered}
\end{equation}
The last guard bounds the amortized lazy insertion charge above the base by $O(t)$ (Appendix~\ref{app:counting}). These guards imply \eqref{eq:capacityguards}, the condition $3M_{j-1}\leq M_j$ that the lazy merge of Appendix~\ref{app:lazy} needs at nonbase levels $j\geq1$ (Section~\ref{sec:framework}), and $H_\ell=t^32^{\ell t}\geq t^3P>P$.

\noindent\textbf{Ordinary completion.} The inequality $H_\ell>P$ gives the ordinary-completion claim of Section~\ref{subsec:stopping}: once the inner forest search has properties (F) and (F$'$) of Section~\ref{sec:framework} in every inner call, Theorem~\ref{thm:return} applies to the root call $\calR(B_*,\{x\},\ell)$, a single-seed root, so $|S|=1\leq A_\ell$ and (I1)--(I2) hold (Section~\ref{sec:framework}); it uses the lazy representation of Section~\ref{subsec:core}, and the packed record has $k\geq64$ (Section~\ref{subsec:packed-cost}). An ordinary invocation has fewer than $P$ retained vertices, whereas a partial root return would contain at least $H_\ell$ output vertices, all active and hence retained. Hence the root returns fully and, by Theorem~\ref{thm:return}, outputs its entire target (Section~\ref{subsec:core}) with complete witnesses.

\noindent\textbf{Requirements on the inner forest search.} Lemma~\ref{lem:completed-shared} assumes three properties of the inner forest search. First, in every inner call it has properties (F) and (F$'$) of Section~\ref{sec:framework}. Second, it creates no keyed placement (a new worklist entry with a stored key), recursive source or group membership by itself, so that only accepted complete-tail relaxations followed by insertion create the edge-insertion events of Appendix~\ref{app:counting}. Third, in a completed comparison run, the inner forest searches, the $i$th of which retains $r_i$ vertices, have total work
\begin{equation}\label{eq:engine-work}
 O\Bigl(c_s\sum_i(1+r_i)+W_{\rm extra}\Bigr),
\end{equation}
where $c_s\geq1$ is fixed by the record and $W_{\rm extra}\geq0$ is engine-specific work per run that is not charged to retained vertices, such as slot masks and permanent slot deletions. The packed engine has $c_s=1$ and $W_{\rm extra}=O(Dq)$ (Lemma~\ref{lem:packed-forest}).

\begin{lemma}\label{lem:completed-shared}
Fix the capped-invocation inputs of Section~\ref{subsec:stopping}: a legitimate ambient snapshot $C$, inherited seed witness $C(x)$, bounds $L_*,B_*$ and external forest $F_*$, together with the retention limit $P$, the routing degree bound $D$ and integers $t,k$ with $k\geq2$ satisfying \eqref{eq:shared-guards}. Let the inner forest search of the engine have the three properties above. Run the engine on the subgraph induced by any $q\leq P$ routing vertices containing $x$, preserving their ambient identifiers, as a completed comparison run: only new routing continuations are restricted to this graph, the inherited witness and cap witnesses may leave it, the invocation's retention observer records retention, its outward size and contact stops are masked, and inner forest stops remain enabled. Then the run returns fully, with $1\leq r\leq q$ feasible output vertices, and the number $E$ of outgoing arcs at its output vertices satisfies $E\leq Dr\leq Dq$. Its inner forest searches have total retained mass
\begin{equation}\label{eq:packed-mass}
 \mathcal M\coloneqq\sum_i(1+r_i)=O\bigl((\ell+1)q+k(r+E+\ell r/t)\bigr),
\end{equation}
and, with $a_0\coloneqq1+\log(2+tD)$, its total work is
\begin{equation}\label{eq:shared-coefficient}
 O\Bigl(c_s\mathcal M+W_{\rm extra}+q\bigl[(\ell+1)(t/k+1)+D(t+a_0)+a_0+1\bigr]\Bigr).
\end{equation}
The implicit constant in \eqref{eq:packed-mass} is absolute, and that in \eqref{eq:shared-coefficient} depends only on the constant in \eqref{eq:engine-work}.
\end{lemma}
\noindent\emph{Proof idea.} The single-seed root meets the premises of Theorem~\ref{thm:return}, and $H_\ell>P\geq q$ forces a full return. Because the inner search creates no source or placement by itself, the source sums and the source-origin bound of Appendix~\ref{app:counting} apply. The fixed ranks of Section~\ref{subsec:foreign} let the full calls of one level retain only $O(q)$ vertices, which gives \eqref{eq:packed-mass}. Every remaining operation belongs to a class that does not depend on the inner forest search and is paid at amortized insertion charge $O(t)$ above the base and $O(a_0)$ at the base. Appendix~\ref{app:completed-shared-proof} gives the complete proof.

By \eqref{eq:packed-mass} and $r\leq q$, $E\leq Dq$, the mass contributes $c_s\mathcal M=O(q[c_s(\ell+1)+c_sk(1+D+\ell/t)])$ to \eqref{eq:shared-coefficient}. The engine therefore controls only the per-level term $(\ell+1)(c_s+t/k)$: the packed engine has $c_s=1$ and $k\leq t$, so $c_sk(1+D)=O(tD)$ lies within the edge term $Dt$.

\section{Prepared walks and the packed engine}\label{sec:packed}
The inner forest search needs to select the least pending continuation and to remove all continuations to an encountered endpoint. We implement those operations with the walk ranks of Section~\ref{sec:overview}. The preparation below sorts the walks once and stores their masks. During a search, a rank enters the pending set only if it extends an accepted prefix.

The packed search supplies the same local-search properties (F) and (F$'$) to the recursive procedure. It costs $O(1)$ per search and per retained endpoint, plus $O(Dq)$ for slots of accepted tails during an invocation (Lemma~\ref{lem:packed-forest}). The smaller search cost allows $k$ proportional to $t$ in Section~\ref{subsec:packed-cost}, which gives the term $\sqrt{D\J(P)}$ of the packed local coefficient. Lemma~\ref{lem:outer-cost} then turns the reduced local coefficient into the global time bound.

\subsection{Prepared walks and their order}
Fix the routing graph $G$, its counts $N,M$, its degree bound $D$, and one integer inner retention limit $k\geq2$ for every packed invocation. The preparation uses up to $S_0$ walk records per root and up to $w_0$ bits per mask, where
\[
 R\coloneqq N+M+2,\qquad w_0\coloneqq\lfloor\log_2R/8\rfloor,\qquad
 S_0\coloneqq(D+1)^{k+1}.
\]
To fit the ranks from one root in a single mask, we require
\begin{equation}\label{eq:walkguard}
 S_0\leq w_0.
\end{equation}
The further tests of Section~\ref{subsec:packed-driver} keep this preparation within the final running time. All walks and slots below belong to $G$, even in the induced comparison used to analyze a stopped invocation.

For each vertex $y$, enumerate the directed walks from $y$ of at most $k$ edges as a prefix tree. Its nodes, the \emph{walk records}, number at most $S_0$. A walk record stores its parent, endpoint, final arc and its outgoing slot number, relative real length, and number of edges. Its children correspond to the outgoing slots of its endpoint. The empty walk is the root record and never a candidate. Sort the nonempty records by relative length, number of edges, and reverse vertex-identifier sequence, and assign them distinct walk ranks starting at zero.

\begin{lemma}\label{lem:relative-order}
The walk-rank order of the nonempty walk records of $y$ agrees with the full witness order obtained by appending each record's walk to any one fixed legitimate witness ending at $y$. Every child is strictly later than its parent. Walk ranks can be computed by comparisons taking $O(k)$ time each; subsequent rank comparisons take constant time.
\end{lemma}
\begin{proof}
The common inherited witness contributes the same length and number of hops to both concatenations. If the resulting first two coordinates tie, the two continuations have equally many vertices. Their reverse sequences either first differ within these suffixes, independently of the inherited prefix, or the suffixes coincide. In a simple graph coincident suffixes give the same walk record. Nonnegative weight and one additional hop make every child strictly later than its parent. Parent links reveal a reverse suffix in at most $k$ steps during sorting. The computed rank thereafter supplies the order directly.
\end{proof}

\noindent\textbf{Frontier operations.} Represent sets of ranks and outgoing slots by integer masks with at most $w_0$ bits. These masks contain identifiers only; relative lengths remain exact real fields. For each endpoint $v$ occurring in $y$'s prefix tree, use a mask $\operatorname{End}(y,v)$ containing all walk records that end at $v$. Each such record stores a direct pointer to this mask, and $\operatorname{End}(y,y)$ includes all walks that return to $y$. For a prefix $p$ of fewer than $k$ edges and a subset $J$ of its endpoint's outgoing slots, use

\[
 \operatorname{Child}(p,J)
 \coloneqq\{\text{ranks of children of }p\text{ whose slots lie in }J\};
\]
depth-$k$ records have empty child masks. The search adds children using mask union and removes endpoints using mask difference. Lemma~\ref{lem:relative-order} lets it select the minimum candidate by finding the least set position. Slot masks have at most $D$ bits, and \eqref{eq:walkguard} gives $D\leq w_0$, so they use the same operations.

\noindent\textbf{Paid preparation.} Appendix~\ref{app:mask-tables} realizes these operations in the integer model of Section~\ref{sec:model} by standard small-universe tabulation~\cite{ADKF1970}. After a construction taking
\begin{equation}\label{eq:mask-setup}
 O(2^{2w_0}+2^{w_0}w_0^2+\log R)=O(R)
\end{equation}
time and cells, each union, difference or least set position takes a constant number of table lookups, additions and comparisons, without a machine bitwise operation, and each $\operatorname{End}$ or $\operatorname{Child}$ answer is one stored mask or one lookup. Walk generation, exact real additions, sorting, endpoint masks and child masks together take
\begin{equation}\label{eq:walk-setup}
 O\bigl(NS_0[k\log(2+S_0)+D+2^D]+N+M\bigr)
\end{equation}
time and $O(NS_0(1+D+2^D)+N+M)$ cells. This preparation, together with the lookup tables for mask operations, is the \emph{global preparation}; it is performed once for $G$ and fixed $k$ and paid by the packed driver (Section~\ref{subsec:packed-driver}).

\subsection{The packed forest search}
\begin{algorithm}[!t]
\caption{Packed forest search from a root $y$ in a call $X$}\label{alg:packed-search}
\small
\begin{algorithmic}[1]
\Require Root $y\notin F$ with actual witness $w_y$; the call's bound $B\leq B_*$, lower threshold $a_X$, fixed old forest $F$ and retention limit $k$; the walks, ranks and masks prepared for $y$
\Ensure A small full completion, or the retained tree stopped at its $k$th retained vertex or at its first contact with $F$; an outward stop of the invocation ends the search earlier
\Statex $0_y$: empty prefix. $\operatorname{Live}(u)$: available outgoing slots at $u$.
\Statex $Z$: occurrences of encountered endpoints. $V$: pending walk ranks.
\State Initialize the retained and accepted sets of the search to $\{y\}$
\State \label{alg:packed-init}$Z\gets\operatorname{End}(y,y)$;\enspace $V\gets\operatorname{Child}(0_y,\operatorname{Live}(y))\setminus Z$, preparing $\operatorname{Live}(y)$ on first use
\Loop
  \If{$V=\varnothing$}
    \State \Return a small full completion
  \EndIf
  \State $p\gets$ the walk record whose rank is least in $V$; form the copied candidate $c$ from $w_y$ and $p$
  \If{$\sig(c)\geq B$} \label{alg:packed-range}
    \State \Return a small full completion
  \EndIf
  \If{the final arc $(u,v)$ of $p$ is not in the graph being searched} \label{alg:packed-member}
    \State Clear $p$ in $V$ and delete slot $(u,v)$ from $\operatorname{Live}(u)$
  \Else
    \State \label{alg:packed-probe}Probe $(u,v)$ from the actual accepted witness of $u$: apply the retention test \eqref{eq:retain}
    \If{the pre-write label of $v$ is not below $a_X$}
      \State Make the update permitted by \eqref{eq:guard}, with both accepted publications
    \EndIf
    \State \label{alg:packed-flags}Finish the membership and count flags
    \If{this transaction triggers an outward stop}
      \State Take the outward stop \Comment{Section~\ref{subsec:stopping}}
    \ElsIf{the pre-write label of $v$ was below $a_X$}
      \State $Z\gets Z\cup\operatorname{End}(y,v)$;\enspace $V\gets V\setminus\operatorname{End}(y,v)$; delete slot $(u,v)$ from $\operatorname{Live}(u)$
    \Else
      \State \label{alg:packed-record}Record $v$ as retained, with first edge $(u,v)$ and whether it accepted; if it did, $p_v\gets p$
      \State $Z\gets Z\cup\operatorname{End}(y,v)$;\enspace $V\gets V\setminus\operatorname{End}(y,v)$
      \If{$v\in F$}
        \State \Return the stopped tree \Comment{first contact with $F$}
      \ElsIf{$v$ is the $k$th retained vertex}
        \State \Return the stopped tree \Comment{size $k$}
      \ElsIf{$v$ accepted}
        \State \label{alg:packed-children}$V\gets V\cup\bigl(\operatorname{Child}(p_v,\operatorname{Live}(v))\setminus Z\bigr)$, preparing $\operatorname{Live}(v)$ on first use
      \EndIf
    \EndIf
  \EndIf
\EndLoop
\end{algorithmic}
\end{algorithm}
Fix a capped invocation with entry snapshot $C$, lower threshold $L_*$ and upper bound $B_*$. A tail's slot mask $\operatorname{Live}(u)$ records which of its outgoing edges the search may still use. The first time an accepted tail needs the mask, scan its static outgoing list and exclude the slots $(u,v)$ with
\begin{equation}\label{eq:rootlow}
 \sig(C(v))<L_*.
\end{equation}
The entry cap is read from a saved copy if the invocation has already written at $v$, and from the unchanged common label otherwise. Appendix~\ref{sec:resources} supplies this constant-time access. By Lemma~\ref{lem:packed-seen}, exclusions by \eqref{eq:rootlow} suppress no feasible canonical edge and no retention by the invocation's retention observer. Mask preparation tests no candidate, so it is not a probe: it performs no retention observation or activation at the scanned heads. The mask persists through all inner levels of this invocation. Its total preparation and destruction cost is $O(Dq)$, because every tail whose mask is prepared is accepted and hence belongs to the invocation's retained set of size $q$. Complete-tail cursors remain separate and use the immutable static outgoing lists.

An inner search in a call $X$ has bound $B\leq B_*$, lower threshold $a_X\geq L_*$, fixed old forest $F$, and root $y\notin F$. Store its actual root witness $w_y$ once. Let $Z$ mark every occurrence of endpoints already encountered, and let the frontier $V$, the pending set, contain the ranks of candidate walk records. Write $0_y$ for the empty prefix. Every accepted vertex $u$ stores the walk record $p_u$ for its accepted prefix in $y$'s prefix tree. Algorithm~\ref{alg:packed-search} states the search; it uses the prepared lookups for all mask operations.

The test in line~\ref{alg:packed-range} is a \emph{pure range test}: it reads only the selected walk record, whose copied candidate combines $w_y$ with the record's relative length, hops, endpoint and final predecessor, and it does not probe an edge or invoke the retention observer. In the actual execution the stored arc pointer of the selected record answers the test in line~\ref{alg:packed-member} immediately. A missing arc occurs only in the induced comparison of Section~\ref{subsec:packed-projection}; such a missing-edge selection reads no current head label and performs no retention, acceptance or endpoint marking. Lines~\ref{alg:packed-probe}--\ref{alg:packed-flags} form the probe transaction of Section~\ref{subsec:stopping}. A probed head whose pre-write label is below $a_X$ is \emph{low}; the search makes no update at a low head and, unless an outward stop is taken, marks its endpoint and deletes its slot. Before the first accepted write at the head, including an equal-signature write, the update saves the entry cap and, for the comparison of Section~\ref{subsec:packed-projection}, the containing context's accepted-membership bit, as in Appendix~\ref{sec:resources}. An outward stop triggered by the transaction is honored before any pending local insertion or child-mask operation. For the endpoint that triggers it, the recording of that endpoint and its first edge in the inner search tree, line~\ref{alg:packed-record}, may remain pending; the invocation's retention record and any accepted publications have already finished.

Lines~\ref{alg:packed-init} and~\ref{alg:packed-children} prepare slot masks with the exclusions \eqref{eq:rootlow}. These exclusions ensure that every selected candidate in the actual packed search whose signature is below the bound $B$ of its inner search has its head retained, so its arc lies in the retained induced graph used in Section~\ref{subsec:packed-projection}. Appendix~\ref{app:coupling-packed} records how the comparison uses these exclusions and the later slot deletions. A root in $F$ is skipped before the search is constructed. Static complete-tail scans continue to probe their sentinel edges as specified in Section~\ref{subsec:core}.

\subsection{Endpoint removal and complete-source coverage}\label{subsec:packed-coverage}
We need to show that processing the first candidate at an endpoint suffices, including when an equal-signature update would trigger a publication. The proof uses the order of the prepared walks and the rule that a vertex adds children only from its accepted prefix.

\begin{lemma}\label{lem:packed-seen}
During a packed search, each accepted endpoint has one fixed accepted prefix walk. Removing all occurrences of an encountered endpoint suppresses no later permissible accepted update, including an equal-signature update. Each selected low-head slot can remain deleted for the entire capped invocation. Exclusions by \eqref{eq:rootlow} suppress no feasible canonical edge and no retention by the invocation's retention observer.
\end{lemma}
\begin{proof}
The first two assertions are proved together by induction over selected records; the inductive claim is that every accepted vertex has one fixed prefix. At every continuation, $V$ contains children of accepted prefixes, minus marked endpoints and slots already deleted. A newly added child has a larger walk rank than its selected parent by Lemma~\ref{lem:relative-order}. Selection takes the least rank in $V$. Thus selected records are nondecreasing in rank, also when a record enters $V$ after earlier selections. Removing records or missing edges cannot decrease the next rank.

Consider a removed occurrence at $v$. If its parent never accepts, it cannot be supplied by an accepted prefix. Otherwise its candidate is no earlier in full order than the first candidate at $v$. Its copied signature cannot be smaller. If the copied signatures were equal, their final predecessor, length and hops would agree. The simple graph has one arc from that predecessor to $v$, and the predecessor has one accepted prefix and adds its children once. There is therefore only one possible occurrence with that final arc supplied by an accepted prefix. Distinct later occurrences supplied by accepted prefixes have strictly larger copied signatures. This also closes the induction: after its first accepted candidate every later candidate supplied by an accepted prefix in this search is strictly worse, while descendants added from that prefix reproduce exactly its actual-tail extensions. Full ties between arbitrary prepared walks were resolved once during sorting; the argument does not identify arbitrary full histories merely because their copied signatures agree.

It remains to show that each later occurrence at $v$ from an accepted prefix would be rejected. If the first candidate accepted, its common value is thereafter no larger. If the first candidate failed the entry-cap or local-current guard, that obstruction persists because caps are fixed and labels do not increase. A first encounter below $a_X$ also remains a strict rejection during this search. A return walk to $y$ has nonnegative added length and positive added hops and is strictly worse than $w_y$, so its initial endpoint marking is valid.

For a slot $(u,v)$ deleted as low in a call $X$, Lemma~\ref{lem:foreign} gives, at every later time in this cap context,
\[
 \sig(\widehat d_{\rm later}(u))\geq\sig(\pi_C(u))\geq a_X
 >\sig(\widehat d_{\rm deletion}(v))
 \geq\sig(\widehat d_{\rm later}(v)).
\]
Every extension is strictly later than its tail. No later relaxation along the slot is accepted, regardless of later calls' lower thresholds. The head $v$ of an initially excluded slot instead has $\sig(C(v))<L_*$, while every actual tail witness is at least $L_*$. The head is therefore cap-infeasible, and an observation by the invocation's retention observer would fail its lower test. These deletions are private to this cap context and leave static outgoing lists intact.
\end{proof}

\begin{lemma}\label{lem:packed-forest}
The packed search supplies property (F), containment of accepted extensions in the current target, and stopped trees that meet $F$ only at their contact endpoint, if any, as required in Section~\ref{sec:framework}. Its fixed ranks for foreign leaves are those of Lemma~\ref{lem:foreign}. If $r_i$ counts the vertices retained by search $i$ and $q$ counts those retained by the invocation, the total work of all packed searches in a capped invocation is
\begin{equation}\label{eq:packed-local}
 O\left(\sum_i(1+r_i)+Dq\right),
\end{equation}
provided the arc-membership test in line~\ref{alg:packed-member} of Algorithm~\ref{alg:packed-search} takes constant time. This includes the probe transactions, both publications for every accepted update (including an equal-signature update), and local allocation and cleanup.
\end{lemma}
\begin{proof}
\emph{Distinct endpoints and depth.} Accepted prefixes have distinct endpoints, since marked endpoints never activate again. Before a continuing child addition there are fewer than $k$ retained vertices; its prefix has at most $k-2$ edges, and its children have at most $k-1$. The $k$th retained endpoint is not expanded. Thus the prepared depth $k$ never truncates a continuing search.

\emph{Property (F).} First suppose the root is complete at forest entry. A canonical descendant below $B$ has a cap-feasible canonical suffix from that root by Lemma~\ref{lem:semantics}, whose agreement argument in Appendix~\ref{app:witness} covers an inner search with its proof-local entry snapshot. At each continuation we maintain the following boundary property for every such suffix: its accepted vertices form a canonical initial segment, and, unless the entire suffix has accepted, the next canonical edge contributes an active candidate to $V$. Here ``accepted'' refers to this search, including the initialized root. The property holds initially. The next edge fits the prepared depth by the preceding paragraph, and neither an entry-cap exclusion in \eqref{eq:rootlow} nor a permanently low slot can remove it by Lemma~\ref{lem:packed-seen}.

To preserve the property, consider a selected candidate whose endpoint lies on a required canonical suffix and has not yet accepted. Let $v$ be the first unaccepted vertex of the canonical path to that endpoint. The active boundary candidate at $v$ is no later than the canonical candidate at the selected endpoint, and is strictly earlier if $v$ is a proper predecessor. The selected candidate is least in $V$, whereas every feasible candidate at that endpoint is no earlier than its canonical optimum. If the selected candidate is infeasible, it is strictly later than the canonical candidate, since the latter respects the entry cap and the current common label. Thus least selection forces the endpoint to be $v$ and its witness to be canonical. Its equal or improving update is accepted. On continuation its children supply the next boundary. A selection outside these required paths cannot mark their boundary endpoints, and deleted low or missing slots cannot remove their canonical edges. This proves the induction, including selections that were activated only after earlier acceptances. Consequently a search that ends by exhaustion or by the pure range test has completed every required descendant below $B$; an unfinished canonical boundary would still be present below $B$. A path requiring $k$ retained vertices causes a size stop first. This proves (F) for sources complete at entry. Applying the same argument to the search's own entry snapshot, in which its inherited root witness is the canonical seed, proves completeness of the accepted output of a small full search relative to that snapshot. Only the first application asserts completeness in the containing context.

\emph{Containment, contact and fixed ranks.} Every accepted extension remains in the current target by Lemma~\ref{lem:foreign}. Only accepted vertices enter $W$ when a small full search ends. Every new nonlow retained endpoint, including a rejected one, attaches by its first actual edge to an already accepted and retained parent. The first contact ends the search before another probe. A previously marked nonlow contact would already have ended it, and a low endpoint remains low during this search. Thus endpoint marking hides no first contact. Rejected leaves are never expanded or made sources. The retained trees, persistent $Q$, groups of call input sources and the fixed-rank argument consequently have the properties used in Section~\ref{sec:framework}. In particular an infeasible foreign leaf keeps $C(v)$, while a feasible foreign leaf is complete by (I2); retention gives $a_X\leq\sig(\mu(v))<B$ in both cases.

\emph{Work.} Each continuing nonlow selection adds a new retained endpoint. Each low or missing-edge selection deletes a slot from its accepted tail's mask, so there are at most $Dq$ such selections and at most $Dq$ slot inspections to build the masks. Each selection, termination and accepted prefix takes constant mask work. A mask difference clears all occurrences of an endpoint, and later child additions exclude that endpoint. Each probe and its two publications take constant work. The costs of creating lists, first edges, witness records and touch records include their destruction. An outward stop adds $O(1)$ work for the last selection. This gives \eqref{eq:packed-local}.
\end{proof}

After the range test in line~\ref{alg:packed-range}, Algorithm~\ref{alg:packed-search} processes only candidates below $B$; line~\ref{alg:packed-record} records a retained endpoint, with its first edge, only when its head is at least $a_X$, and the search stops on first contact or on the $k$th retained vertex. Together with Lemma~\ref{lem:packed-forest}, these properties give (F) and (F$'$), so the argument of Section~\ref{sec:framework} gives \eqref{eq:forest1}--\eqref{eq:forest3} and the same return induction applies. A packed write changes no worklist entry or group membership of the surrounding recursion, so the source-origin counts also apply.

\subsection{The packed record and the return contract}\label{subsec:packed-cost}
Apart from the separate $O(Dq)$ cost of slot masks, the packed forest-search work is linear in the retained search sizes. This allows the inner retention limit $k$ to be proportional to the recursion scale $t$; we fix the resulting record and check the premises of Theorem~\ref{thm:return}.

For a fixed invocation retention limit $P$ and degree bound $D$ of the routing graph, put
\begin{equation}\label{eq:packed-inner}
 \begin{gathered}
 A\coloneqq\J(P),\qquad \gamma\coloneqq\J(A+D+2),\qquad
 t_{\rm sq}\coloneqq\min\{s\in\mathbb Z_{>0}:Ds^2\geq A\},\\
 t\coloneqq\max\{4096,256\gamma,t_{\rm sq}\},\qquad
 k\coloneqq\lfloor t/64\rfloor,\qquad
 \ell\coloneqq\lceil\log_2(P)/t\rceil.
 \end{gathered}
\end{equation}
With $c_s=1$ and $k=\Theta(t)$, the per-level term $(\ell+1)(c_s+t/k)$ identified after Lemma~\ref{lem:completed-shared} is $O(A/t+1)$; balancing $A/t$ against the edge term $Dt$ gives the square-root scale $t_{\rm sq}$. The logarithmic floor $256\gamma$ supplies the logarithmic guard on records with $t\leq A$, even when $t_{\rm sq}$ is smaller. The proof of Theorem~\ref{thm:packed-cap} bounds its cost. The shared capacities are \eqref{eq:capacities}. The record must satisfy \eqref{eq:packed-innerguards} in Appendix~\ref{app:record-guards}: the shared guards \eqref{eq:shared-guards} together with $64\leq k\leq P$ and $H_\ell>P$. The driver checks those conditions before enumerating walks or constructing tables.

\begin{theorem}\label{thm:packed-cap}
With the parameter record \eqref{eq:packed-inner}, under the global preparation of Section~\ref{sec:packed} and the guards \eqref{eq:walkguard} and \eqref{eq:packed-innerguards}, the packed engine obeys the capped-invocation rules of Section~\ref{sec:model} and the ordinary-return contract of Theorem~\ref{thm:return}. An invocation retaining $q\leq P$ vertices and ending at ordinary completion or an outward stop takes
\begin{equation}\label{eq:packed-capcost}
 O\bigl(q[\sqrt{D\J(P)}+D\log(2+D)+D+1]\bigr)
\end{equation}
time, including owned state, publication, export and cleanup. The global preparation is paid once, separately.
\end{theorem}

The coefficient in \eqref{eq:packed-capcost} is $h(D,\J(P))$ from \eqref{eq:primary-coefficient}. With $H_\ell>P$, the premise check below gives the ordinary-completion claim as in Section~\ref{sec:invocation-cost}; Section~\ref{subsec:packed-projection} then transfers completed costs to stopped prefixes, and Section~\ref{sec:packed-cost}, with Appendix~\ref{app:record-guards}, bounds the coefficient and completes the proof of Theorem~\ref{thm:packed-cap}.

By Appendix~\ref{app:record-guards}, every record passing $t\leq A$ satisfies the logarithmic guard, through \eqref{eq:packed-logguard}, and the scale inequalities $k\geq64$, $64\leq t/k<65$, $16(3k+1)\leq t$ and $k\leq t/16$. The common calculation following \eqref{eq:capacityguards} gives the capacity and merge conditions. Also $H_j=tA_j$. The partition still gives groups of fewer than $3k$ call input sources, and their $O(k)$ expansion and reselection costs are paid using $k\leq t$. 
By Lemma~\ref{lem:packed-forest} and the paragraph following it, the packed search has properties (F) and (F$'$). A packed relaxation introduces no recursive source or keyed placement by itself, and the last paragraph of Appendix~\ref{app:return} checks that every other operation premise of the recursion is unchanged for the packed engine. This completes the premise check of Theorem~\ref{thm:return} under \eqref{eq:packed-innerguards}.

\subsection{Stopped packed invocations}\label{subsec:packed-projection}
The packed forest search meets its upper bound by a pure range test rather than by a sentinel probe, so its stopped prefix needs its own comparison. That comparison keeps the walks prepared on $G$, although it runs on the smaller graph $\mathcal H=G[K_*]$ induced by the retained set $K_*$ of the actual prefix, for two reasons. (i) Below the bound $B$ of an inner search, the entry-cap exclusions \eqref{eq:rootlow} ensure that the head of every candidate selected by the actual prefix is retained, so its arc survives in $\mathcal H$. (ii) At or above $B$, a globally prepared rank may refer to an arc outside $\mathcal H$, but the common pure range test ends the search before any membership test or label access. Hence the comparison tests the range first and arc membership second, so the walks prepared on $G$ suffice. Accepted and retained sets are tracked separately, and the actual prefix keeps the transaction and stop semantics of Section~\ref{subsec:stopping}. Appendix~\ref{app:coupling-packed} proves (i) and (ii), specifies the comparison state, whose two vertex-indexed arrays the algorithm allocates in its paid global preparation, and proves the following proposition.

\begin{proposition}\label{prop:packed-projection}
Consider an actual packed invocation with data prepared on $G$ and parameter record \eqref{eq:packed-inner} satisfying all guards \eqref{eq:packed-innerguards}, including $H_\ell>P$. Freeze it at ordinary completion or after the final transaction that triggers an outward stop, and let $K_*$ be its retained set, $q=|K_*|\leq P$, and $\mathcal H=G[K_*]$. Let $W_H$ denote the work of the completed comparison run on $\mathcal H$, started as in Section~\ref{sec:invocation-cost} from the invocation's entry state with the comparison arrays of Appendix~\ref{app:coupling-packed}, using the same data prepared on $G$ and the same parameter record. In that comparison, both outward stops are masked, all inner forest size and contact stops remain enabled, and the invocation's accepted-update and publication rules, including equal-signature updates, apply to the comparison state. Then the actual prefix work is
\[
 O(W_H+Dq+q).
\]
The implicit constant is absolute once the primitive realization and transaction convention are fixed. The global preparation is shared and excluded from both sides. The actual prefix's allocation, publication in both contexts, export of its actual retained tree, and cleanup are included. The completed comparison is used only in the proof.
\end{proposition}
In the proof the coupled packed transactions agree from the entry state, a selected missing arc in the comparison deletes its slot once, and the complete-tail scans couple through their induced sorted lists (Appendix~\ref{app:coupling-packed}).

\subsection{The packed coefficient}\label{sec:packed-cost}
With $c_s=1$ and $O(Dq)$ slot work, Lemma~\ref{lem:completed-shared} gives the packed coefficient. To bound $W_H$ in Proposition~\ref{prop:packed-projection}, consider the completed comparison run of a packed invocation on $q\leq P$ vertices, returning $r$ feasible vertices, and let $E$ be the number of induced arcs with tail in that output. The guards \eqref{eq:packed-innerguards} contain \eqref{eq:shared-guards}, and the packed search has properties (F) and (F$'$) (Section~\ref{subsec:packed-coverage}). It preserves the generation and counting rules for recursive sources and keyed placements: only accepted complete-tail relaxations followed by insertion create the edge events used in Appendix~\ref{app:counting}, and packed mask operations create no keyed placement. Lemma~\ref{lem:packed-forest}, with the constant-time arc-membership test of Appendix~\ref{app:coupling-packed}, makes all local forest work $O(\mathcal M+Dq)$. Preparing slot masks and deleting low slots and slots with missing edges have their separate $O(Dq)$ charge; these operations need not have complete tails and are not charged to $E$. Thus \eqref{eq:engine-work} holds with $c_s=1$ and $W_{\rm extra}=O(Dq)$.

With $c_s=1$, the mass contribution $c_s\mathcal M=O(q[c_s(\ell+1)+c_sk(1+D+\ell/t)])$ noted after Lemma~\ref{lem:completed-shared} becomes $O(q[(\ell+1)+k(1+D+\ell/t)])$, and $W_{\rm extra}=O(Dq)$ adds the slot work $D$ per retained vertex. Adding the remaining terms of \eqref{eq:shared-coefficient}, Lemma~\ref{lem:completed-shared} gives a completed bound $O(qc_p)$, where $a_0=1+\log(2+tD)$ and
\begin{equation}\label{eq:packed-coefficient}
 c_p\coloneqq(\ell+1)(t/k+1)+D(t+a_0)+a_0+1
             +(\ell+1)+k(1+D+\ell/t)+D.
\end{equation}
The same bound absorbs the separate $Dq+q$ in Proposition~\ref{prop:packed-projection}.

\begin{proof}[Proof of Theorem~\ref{thm:packed-cap}]
Correctness and the completed cost follow from the preceding return proof and operation counts, and Proposition~\ref{prop:packed-projection} transfers them to a stopped prefix. It remains to bound \eqref{eq:packed-coefficient} on every parameter record passing the guards, including records whose $t$ is set by the logarithmic floor. Appendix~\ref{app:record-guards} shows that \eqref{eq:packed-coefficient} is $O(\sqrt{DA}+D\log(2+D)+D+1)$ with an absolute constant on every passing record: the least-integer choice of $t_{\rm sq}$ gives $A/t\leq\sqrt{DA}$, and $Dt$ and $Da_0$ are both $O(\sqrt{DA}+D\log(2+D)+D)$, also when the floor $256\gamma$ sets $t$.
\end{proof}

\section{The packed driver: intermediate-density and sparse bounds}\label{sec:packed-global}

The packed driver prepares short walks once. Its tables include a factor $2^D$, so the driver limits $D$ when it builds the routing graph. The smaller degree saves table space but can create more ports. We bound the resulting graph size and preparation cost, then show which inputs pass the tests in Algorithm~\ref{alg:driver}. Together with Theorem~\ref{thm:packed-cap} and Lemma~\ref{lem:outer-cost}, this proves the intermediate-density bound \eqref{eq:main} and the sparse bound \eqref{eq:main-sparse} of Theorem~\ref{thm:main} (Sections~\ref{subsec:uniform-packed} and~\ref{subsec:sparse}).

\subsection{Parameter tests and branch selection}\label{subsec:packed-driver}

Let $n,m$ count the working input $I$ of line~\ref{alg:driver-pad}, including padding when used. Then $m\geq3n$. The degree scale $\Theta(\min\{m/n,\log\log n\})$ already appears in the parameter choice of Duan et al.~\cite[Section 3.6]{DMSY2026} and in B1's general-degree calculation~\cite[Section S4.9]{B1PublicDraft2026}. With $m_0$ surviving simple arcs, we use the following rounded choice to meet the packed preparation tests:
\begin{equation}\label{eq:packed-degree}
 \lambda_n\coloneqq\J(\J(n)),\qquad
 D_{\rm lim}\coloneqq\max\{3,\lfloor\lambda_n/1024\rfloor\},\qquad
 D\coloneqq\min\{\lfloor m/n\rfloor,n,D_{\rm lim}\}.
\end{equation}
The degree reduction of Section~\ref{subsec:routing}, applied with this $D$ in line~\ref{alg:driver-reduce}, needs only $3\leq D\leq\min\{\lfloor m/n\rfloor,n\}$. The routing counts satisfy
\begin{equation}\label{eq:packed-routing}
 \begin{gathered}
 n\leq N\leq n+\frac{2m_0}{D-2}\leq n+\frac{6m}{D},\qquad
 M\leq m_0+N,\qquad ND\leq7m,\\
 R\coloneqq N+M+2\leq2n+5m+2.
 \end{gathered}
\end{equation}
The general counts of Section~\ref{subsec:routing} give the first line, since $2m_0/(D-2)\leq6m/D$ for $D\geq3$. Also $N\leq n+2m_0\leq n+2m$, which gives the bound for $R$. Thus the degree reduction costs $O(n+m)$ even when $N$ exceeds a constant multiple of $n$, outside the density range of Theorem~\ref{thm:main}. The graph remains simple, with both degrees at most $D$ and $M\leq ND$; its adjacency is still unsorted. Put $L\coloneqq\J(N)$, $\lambda\coloneqq\J(L)$ and define
\begin{equation}\label{eq:packed-outer}
 \begin{aligned}
 G_p&\coloneqq\lceil\sqrt{D\lambda}\rceil+D\J(D+2)+D+1,
 &T&\coloneqq\min\{s\in\mathbb Z_{>0}:Ds^2\geq LG_p\},\\
 P&\coloneqq\left\lfloor\frac{T}{1024G_p}\right\rfloor,
 &e&\coloneqq\lceil\log_2(N)/T\rceil.
 \end{aligned}
\end{equation}
Here $G_p$ bounds the local coefficient of a packed invocation, $T$ balances the outer level cost against edge work, and $P$ is the retention limit of an outer local search; thus $(T,P,e)$ is the outer scheme \eqref{eq:outer-scheme} with $g=G_p$. The packed outer capacities are \eqref{eq:capacities} with $(t,k)=(T,P)$. Appendix~\ref{app:branch-tests} lists the scalar tests and worklist capacity checks. The driver checks the outer record first. Once $P\geq2$ passes, it forms the inner record \eqref{eq:packed-inner}, checks \eqref{eq:packed-innerguards}, and verifies that $D,P,T,t,k\leq N$ and $M\leq N^2$. It tests the walk-list sizes and the preparation budget last.

Line~\ref{alg:driver-pretests} forms, with arithmetic stopped at the tested bounds,
\[
 S_0\coloneqq(D+1)^{k+1},\qquad
 U_0\coloneqq S_0^2\,2^D(D+k+2)^4,
\]
and requires
\begin{equation}\label{eq:packed-globalguards}
 S_0\leq\lfloor\log_2(N+M+2)/8\rfloor,
 \qquad U_0^8\leq L.
\end{equation}
The first condition is \eqref{eq:walkguard}; the second bounds the cost of the global preparation in \eqref{eq:packed-pre} and implies $D\leq\lambda$. All powers and scalar computations are evaluated and charged as in Appendix~\ref{app:word-bounds}, so a failed test never enumerates its nominal value, and all scalar tests cost $O(N+M)$ on every input reaching them.

A failed test takes the branch at line~\ref{alg:driver-fallback} before the packed tables and arrays of line~\ref{alg:driver-setup} are built, and before any label is initialized. The branch sets aside the padded graph when padding was used, the simplified copy, the routing graph with its original-edge records, $D$, $\lambda_n$ and $D_{\rm lim}$, the outer and inner records and the test scalars. Preprocessing, the tests and disposal together cost $O(n+m)$. The branch then returns Fibonacci-heap Dijkstra on the preserved original input $I_0$, which costs $O(m+n\log n)$; Algorithm~\ref{alg:driver} runs the same routine on inputs with at most eight vertices.

In both uses, on a strict improvement, the routine records the edge whose relaxation improved the unsettled head as that head's predecessor edge; on ties it leaves predecessor edges unchanged, and the source has none. With nonnegative weights every settled label is final and equals the distance: zero weights are allowed, a loop never improves a label, parallel arcs are relaxed separately, and unreachable vertices keep distance infinity. A settled vertex is never strictly improved, so a recorded predecessor was settled before its head. The predecessor edges therefore go forward in settle order and are tight, and every reachable vertex other than the source has one; they form a shortest-path tree of original edges on the reachable vertices, even when zero-weight cycles exist. The routine uses only comparisons and additions on the real weights and unit-cost word operations; its root-degree table has $|V(I_0)|+1$ entries, so it needs no charged product or logarithm. The fallback path holds $O(n+m)$ cells.

When all tests pass, let $\operatorname{PRE}$ denote the cost of the global preparation. Equations~\eqref{eq:walk-setup} and~\eqref{eq:mask-setup} give
\begin{equation}\label{eq:packed-pre}
 \operatorname{PRE}=O(NU_0+N+M)
                =O(NL^{1/8}+N+M).
\end{equation}
Indeed $S_0[k\log(2+S_0)+D+2^D]=O(U_0)$. Static outgoing-list sorting separately costs $O(M\log(2+D))$. The cap, role and comparison arrays are a constant number of arrays indexed by routing vertices and edges; the outer per-level arrays still cost $O(N(e+1))$ to initialize. Every invocation uses this same $P,D,t,k,\ell$ and the same walks prepared on the routing graph.

\subsection{Cost of the packed branch}\label{subsec:packed-branch}
Theorem~\ref{thm:packed-cap} shows that packed invocations meet the stopping and completion conditions of the outer recursion. Thus line~\ref{alg:driver-run} of Algorithm~\ref{alg:driver} computes exact routing distances by Section~\ref{subsec:outer-run}.

The local coefficient in \eqref{eq:packed-capcost} is $O(G_p)$ on every parameter record that selects the packed branch. Indeed, $P\leq T\leq L$ gives $\J(P)\leq\lambda$, and the remaining terms of the local coefficient are included in $G_p$. Lemma~\ref{lem:outer-cost} with $h_{\rm loc}=G_p$ therefore gives the shared outer bound \eqref{eq:outercost}, which with the global preparation becomes
\begin{equation}\label{eq:packed-outercost}
 \begin{split}
 O\big(&N+M\log(2+D)+N(e+1)(G_p+D+T/P+1)\\
       &+PG_p(N+M+Ne/T)+MT+ND\log(2+TD)\big)
       +\operatorname{PRE}.
 \end{split}
\end{equation}
This bound still pays for the outer per-entry merges, full outgoing scans and per-level arrays. Only the inner lazy family has constant-amortized ordered merge.

The balance in Section~\ref{subsec:outer-cost} bounds the level and small-search terms together with $MT$ by $O(NDT+NG_p+N)$. The inequalities $LG_p\leq DT^2$ and $T\leq L$ give $G_p\leq DT$. The outer logarithmic guard of \eqref{eq:packed-outerguards} gives $\log(2+TD)=O(T)$, so sorting and base searches cost $O(NDT)$ using $M\leq ND$. Since $T\geq\sqrt L$, the preparation cost \eqref{eq:packed-pre} is $O(NDT+N+M)$. With $T=O(\sqrt{LG_p/D})$ from \eqref{eq:outer-scheme}, the time on every input that selects the packed branch, including the original reduction and output, is
\begin{equation}\label{eq:packed-finite}
 O(n+m+N\sqrt{DLG_p}).
\end{equation}
Its constant is absolute. This estimate is expressed in the routing size $N$; its $n+m$ term uses $R=O(n+m)$ from \eqref{eq:packed-routing}, including when $N$ exceeds a constant multiple of $n$. For densities $d\geq\lambda_0$, including those above the ranges of Theorem~\ref{thm:main}, \eqref{eq:packed-finite} gives an upper bound for the same algorithm: for all sufficiently large $n$, every such input is solved in time $O(m\sqrt{\log n\,\log\log\log n})$ with an absolute constant, and for $d\geq\sqrt{\log n/\log\log\log n}$ this bound does not improve on Dijkstra's $O(m+n\log n)$. Indeed, such an input is not padded and either takes the direct case of Section~\ref{subsec:routing} in $O(n+m)$ time or has $D=D_{\rm lim}=\lfloor\lambda_n/1024\rfloor$ by \eqref{eq:packed-degree}, since $\lambda_n=\lambda_0+O(1)$; then $N\geq n$ gives $D\leq\lambda/1024$, and by Lemma~\ref{lem:bounded-degree-guards} the driver selects the packed branch. By \eqref{eq:packed-routing}, $ND\leq7m$ and $n\leq N\leq n+2m_0<2n^2$, as the $m_0$ surviving arcs join distinct ordered pairs. Hence $L=\Theta(\log n)$, $\lambda=\lambda_0+O(1)$, $D=\Theta(\lambda_0)$ and $G_p=O(D\log(2+D))$ by \eqref{eq:packed-outer}, so $N\sqrt{DLG_p}\leq7m\sqrt{LG_p/D}=O(m\sqrt{\log n\,\log\log\log n})$, while $n+m\leq2m$.

\subsection{Uniform branch selection and the intermediate-density bound}\label{subsec:uniform-packed}
For the degree choice in \eqref{eq:packed-degree}, the following lemma verifies all packed tests uniformly in its stated routing-degree range. The argument that follows, in the original counts, applies it throughout each fixed intermediate-density range to sufficiently large inputs not already handled directly.

\begin{lemma}\label{lem:bounded-degree-guards}
For all sufficiently large $N$, every packed parameter and preparation test passes uniformly for $3\leq D\leq\lambda/1024$ and $M\leq ND$ on a simple routing graph, where $L=\J(N)$ and $\lambda=\J(L)$. These tests are \eqref{eq:packed-outerguards}, the positive-level explicit allowance in Appendix~\ref{app:branch-tests}, \eqref{eq:packed-innerguards}, \eqref{eq:packed-globalguards}, and $D,P,T,t,k\leq N$, $M\leq N^2$. The remaining resource bounds of Appendix~\ref{sec:resources} follow as well.
\end{lemma}
The proof, in Appendix~\ref{app:test-passage}, estimates $G_p$, $T$, $P$ and the inner record uniformly on this range, using the rounding observation of Section~\ref{subsec:outer-cost}, and checks each test in turn. The preparation tests pass because $\log_2S_0=O(\sqrt\lambda+(\log(2+\lambda))^2)=o(\lambda)$ while $D\leq\lambda/1024$ keeps the factor $2^D$ in $U_0$ below $L^{1/8}$. The explicit constants require very large inputs before the packed tests pass (Appendix~\ref{app:finite-thresholds}).

Lemma~\ref{lem:bounded-degree-guards} also confines the branch at line~\ref{alg:driver-fallback} to bounded inputs. Let $N_*$ be such that the conclusion of the lemma holds for all $N\geq N_*$, and put
\[
 n_0\coloneqq\max\bigl\{N_*,\ \min\{n:\J(\J(n))\geq3072\}\bigr\}.
\]
A working input that reaches line~\ref{alg:driver-tests} has $m\geq3n$ and $n\geq|V(I_0)|\geq9$, so \eqref{eq:packed-degree} gives $D\geq3$. Its routing graph is simple with $M\leq ND$, and $N\geq n$ gives $\lambda\geq\lambda_n$, because $\J$ is nondecreasing. If $n\geq n_0$, then $\lambda_n\geq3072$, so $D\leq D_{\rm lim}=\lfloor\lambda_n/1024\rfloor\leq\lambda/1024$, and $N\geq N_*$; hence every test passes. A failed test therefore implies $n<n_0$. The number $n_0$ is absolute; in particular it does not depend on the density constant $C$ of Theorem~\ref{thm:main}.

Now fix $C>0$ and consider original inputs with $3\leq d\leq C\lambda_0$. This range is contained in $3\leq d\leq C'\lambda_n$ for a fixed constant $C'$. The direct cases already take $O(n+m)$. For large $n$, the truncation at $n$ in \eqref{eq:packed-degree} is inactive and
\[
 \lambda_n/2048\leq D_{\rm lim}\leq\lambda_n/1024.
\]
If $D=\lfloor d\rfloor$, then $d/D\leq4/3$; if $D=D_{\rm lim}$, then $d/D\leq2048C'$. The routing bound \eqref{eq:packed-routing} therefore implies
\begin{equation}\label{eq:packed-original-band}
 n\leq N\leq c_C n,\qquad L=\Theta_C(\log n),\qquad
 \lambda=\lambda_n+O_C(1)=\lambda_0+O_C(1).
\end{equation}
Moreover $N\geq n$ gives $\lambda\geq\lambda_n$, and hence $D\leq\lambda_n/1024\leq\lambda/1024$. By Lemma~\ref{lem:bounded-degree-guards}, for $n$ above a uniform threshold, every input in this range that is not handled directly passes all packed tests and selects the packed branch.

Recall $h(d,z)=\sqrt{dz}+d\log(2+d)+d+1$ from \eqref{eq:primary-coefficient}; it is increasing in both arguments. From $D\leq d$ and \eqref{eq:packed-original-band},
\[
 G_p=O_C(h(d,\lambda_0)),\qquad
 N\sqrt{DLG_p}=O_C\!\left(n\sqrt{d\log n\,h(d,\lambda_0)}\right).
\]
Substitution in \eqref{eq:packed-finite} proves \eqref{eq:main}. Below the resulting uniform threshold, both $n$ and $m$ are bounded on this range; in particular, an input of this range that takes the fallback has $n<n_0$ and $m<Cn_0\log\log n_0$. The setup, the packed branch and the fallback path, with its $O(n+m)$ preliminary work and $O(m+n\log n)$ for Dijkstra on $I_0$, have operation counts bounded independently of the weights, so this bounded initial range is absorbed in the constant. This proves the uniform intermediate-density assertion of Theorem~\ref{thm:main}, on the $O(\log(n+m+2))$-bit words fixed in Appendix~\ref{sec:resources} before any optimized exponent is used.

\subsection{Sparse inputs}\label{subsec:sparse}
Bounded densities are a special case of the intermediate range, and smaller densities are padded. For every fixed density bound $C_0\geq3$, inputs with $3\leq d\leq C_0$ lie in the range of \eqref{eq:main} for sufficiently large $n$. There $D_{\rm lim}$ eventually exceeds $C_0$, so $D=\lfloor d\rfloor$ is bounded. Lemma~\ref{lem:bounded-degree-guards} therefore shows that all inputs reaching the packed tests pass them, and $h(d,\lambda_0)=\Theta_{C_0}(\sqrt{\lambda_0})$ on this range gives
\begin{equation}\label{eq:bounded-density}
 O\!\left(n+m+n\sqrt{\log n}(\log\log n)^{1/4}\right).
\end{equation}
An input of this range that fails a packed test has $n<n_0$ and $m\leq C_0n<C_0n_0$, so its fallback path is absorbed as above.

Let the unpadded input have $n>8$ and $m<3n$. Keep every original arc and add $n$ new vertices $z_0,\ldots,z_{n-1}$ and the zero-weight arcs
\[
 z_i\longrightarrow z_{(i+r)\bmod n}\quad(0\leq i<n,\ 1\leq r\leq8),
 \qquad s\longrightarrow z_0.
\]
The added part is simple and source-reachable, and its arcs never enter an original vertex. The augmented counts satisfy
\[
 n'=2n,\qquad m'=m+8n+1,\qquad4\leq m'/n'<6.
\]
Each added index uses addition and at most one wrap. Construction costs $O(n+m)$ time and cells. Every path from $s$ to an original vertex stays in the original graph, so its distance and reachability are unchanged.

Algorithm~\ref{alg:driver} uses the augmented counts $n',m'$ to choose the degree bound. Eventually $D_{\rm lim}>6$, while $4\leq m'/n'<6$, so $D=\lfloor m'/n'\rfloor$ is bounded and the preceding proof applies uniformly. The packed branch computes the augmented distances exactly, and line~\ref{alg:driver-tree} constructs the tree directly from the tight edges of the original graph. The fallback runs on $I_0$ itself; a padded input that takes it has $2|V(I_0)|<n_0$ and $|E(I_0)|<3|V(I_0)|$, so it is absorbed as above.

Apply \eqref{eq:bounded-density} to $n',m'$. Their sum is $O(n+m)$ and $n'=2n$, so the padding, driver and original-output pass satisfy \eqref{eq:main-sparse}, with $O(\log(n+m+2))$-bit counters, identifiers and addresses by \eqref{eq:packed-routing}. For each fixed $C>0$, inputs with $3n\leq m\leq Cn$ use \eqref{eq:bounded-density} directly and those with $m<3n$ use padding. Algorithm~\ref{alg:driver} solves the inputs with $n\leq8$ directly by the Fibonacci-heap Dijkstra routine of Section~\ref{subsec:packed-driver}; they lie below a size threshold, which Theorem~\ref{thm:main} permits. This proves the sparse assertion of Theorem~\ref{thm:main}, including $m=o(n)$. With the original-output pass of Section~\ref{subsec:routing}, the Dijkstra fallback of Section~\ref{subsec:packed-driver}, and the storage and word bounds of Appendix~\ref{sec:resources}, this completes all assertions of Theorem~\ref{thm:main}; Section~\ref{subsec:packed-consequences} derives Corollary~\ref{cor:fixed-powers} from them.

\section{Comparison with earlier bounds}\label{subsec:comparisons}\label{subsec:packed-consequences}
We compare the leading terms of our bounds with B1's general-degree expression~\cite[Section 2, Remark~(iii), and Section S4.9]{B1PublicDraft2026} and the bound of Duan et al.~\cite{DMSY2026}. This section also proves Corollary~\ref{cor:fixed-powers} and the comparisons stated in the Introduction. For the original input, recall $d=m/n$ and $\lambda_0=\log\log n$. Put $L_0\coloneqq\log n$ and write the two benchmarks as
\[
 F_{\rm B}\coloneqq nd\sqrt{L_0}\,[\lambda_0(1+(\log\lambda_0)/d)]^{1/4},\qquad
 F_{\rm P}\coloneqq n\sqrt{L_0}\,(d+\sqrt{d\lambda_0}).
\]
B1 identifies $d=o(\sqrt{\lambda_0})$ as the strict-improvement range of this general-degree calculation. The draft gives the general-degree calculation~\cite[Section S4.9]{B1PublicDraft2026} and explicitly leaves it outside its Theorem B1, citing the absence of frozen pseudocode for that branch~\cite[Section 2, Remark~(iii)]{B1PublicDraft2026}.

For $m=\Theta(n)$, the sparse bound \eqref{eq:main-sparse} removes a factor $(\log\log\log n)^{1/4}$ from B1's stated sparse bound~\cite[Section 2]{B1PublicDraft2026}. When $d=O(\lambda_0^{-1/2})$, both $d$ and $\sqrt{d\lambda_0}$ are $O(\lambda_0^{1/4})$, so $F_{\rm P}$ is already $O(n\sqrt{\log n}\,\lambda_0^{1/4})$. B1's sparse bound can also be smaller than \eqref{eq:main-sparse} at smaller $m$.

To prove Corollary~\ref{cor:fixed-powers}, fix $0\leq a\leq b<1$ and $c_1,c_2>0$, and let $c_1\lambda_0^\alpha\leq d\leq c_2\lambda_0^\alpha$ with $\alpha\in[a,b]$. In the subcase $d\geq3$, the uniform intermediate-density assertion of Theorem~\ref{thm:main} applies eventually and
\[
 \frac{d\log(2+d)}{\sqrt{d\lambda_0}}
       =O\!\left(\lambda_0^{-(1-b)/2}\log\lambda_0\right)=o(1).
\]
The term $d+1$ is also $o(\sqrt{d\lambda_0})$, so $h(d,\lambda_0)=\Theta(\sqrt{d\lambda_0})$ uniformly. The leading term is therefore
\[
 n\sqrt{\log n}\,d^{3/4}\lambda_0^{1/4}
       =O\!\left(n\sqrt{\log n}\,\lambda_0^{(3\alpha+1)/4}\right).
\]
If $d<3$, the sparse assertion of Theorem~\ref{thm:main} gives exponent $1/4$, which is at most $(3\alpha+1)/4$ because $\alpha\geq0$. These two uniform subcases prove the corollary.

For $d\geq3$ with $d=o(\sqrt{\lambda_0})$, the part of B1's stated improvement range covered by our intermediate-density bound, we have $h(d,\lambda_0)=\Theta(\sqrt{d\lambda_0})$. The ratio of the new leading term to $F_{\rm B}$ is
\[
 O\!\left((d+\log\lambda_0)^{-1/4}\right)=o(1).
\]

For $d\geq3$, the leading term in \eqref{eq:main} is $n\sqrt{d\log n\,h(d,\lambda_0)}$, since $h\geq d$ absorbs $n+m$. The expression $F_{\rm P}$ lies within a factor $\sqrt2$ of $n\sqrt{d\log n\,(d+\lambda_0)}$. The squared ratio of the bounds is thus of order $h(d,\lambda_0)/(d+\lambda_0)$. If $d\log(2+d)=o(\lambda_0)$, then $d=o(\lambda_0)$ and $h(d,\lambda_0)=o(\lambda_0)$, so the ratio tends to zero. This includes $d=\Theta(\sqrt{\lambda_0})$, where B1's stated range ends, and density functions crossing that scale.

Conversely, if $d\log(2+d)\geq c\lambda_0$ for some $c>0$ along a subsequence, then along that subsequence $h(d,\lambda_0)\geq\max\{c\lambda_0,d\}\geq\min\{c,1\}(d+\lambda_0)/2$. The ratio therefore tends to zero exactly in \eqref{eq:gainwindow}. Where $d\log(2+d)=\Theta(\lambda_0)$, we have $d=o(\lambda_0)$ and $h(d,\lambda_0)=\Theta(\lambda_0)$, giving bounds of the same order.

For fixed $0<\alpha<1/2$, the exponent of $F_{\rm B}$ exceeds $(3\alpha+1)/4$ by $\alpha/4$. For $1/2\leq\alpha<1$, the exponent of $F_{\rm P}$ exceeds it by $(1-\alpha)/4$. In particular, at $\alpha=1/2$ the exponent in \eqref{eq:main-packed} is $5/8$, compared with $3/4$ for $F_{\rm P}$.

At $d=\Theta(\lambda_0)$, $d\log(2+d)$ dominates, so our stated upper-bound expression exceeds that of Duan et al.\ by a factor $\sqrt{\log\lambda_0}$. This term bounds the $D\log(2+D)$ part of the local coefficient, using $D\leq d$. That part pays for the logarithmic floor in the inner recursion scale and the logarithmic cost of base-heap operations (Section~\ref{sec:packed-cost}).

The public C-HD candidate~\cite{CHDPublic2026} states the bound
\[
 O\!\left(n+m+m\log(m/n)+m^{1/3}(n\log n)^{2/3}\right).
\]
For $3\leq d\leq C\lambda_0$, the C-HD expression is at least $n(\log n)^{2/3}$. On the same range, \eqref{eq:main} is
\[
 O_C\!\left(n\sqrt{\log n}\,\lambda_0\sqrt{\log(2+\lambda_0)}\right).
\]
Thus our bound is uniformly of smaller order than C-HD's displayed bound on this range.

\section{Conclusion}\label{sec:conclusion}
One deterministic algorithm gives the sparse bound \eqref{eq:main-sparse} on each fixed class $m\leq Cn$ and the intermediate-density bound \eqref{eq:main} for $3\leq d\leq C\log\log n$. Within the latter range, the bound improves on Duan et al.\ when $d\log(2+d)=o(\log\log n)$. At $d=\Theta(\log\log n)$ the displayed bound is larger by a factor $\sqrt{\log\log\log n}$. The algorithm uses an exact fallback if its parameter tests fail.

Storage is bounded by a constant multiple of the corresponding running-time bound. The parameter tests have very large thresholds. Below the necessary size threshold, an input reaching those tests falls back to exact Dijkstra (Appendix~\ref{app:finite-thresholds}). A speedup over Duan et~al.\ for $3\leq d\leq C\log\log n$ outside the window~\eqref{eq:gainwindow} and the smallest storage needed for these running times remain open.

The analysis of the lazy worklist uses the operation specifications, amortized work bounds and block-count transitions proved in the C-HD proof package~\cite{CHDGeneric2026}. The cited Lean formalization covers these statements, which Appendix~\ref{app:lazy} names together with their hypotheses; their applicability to our recursion is proved in Section~\ref{sec:framework} and Appendices~\ref{app:lazy} and~\ref{app:return}.

\appendix
\section{Chronological witnesses and cap feasibility}\label{app:witness}
The runtime guard \eqref{eq:guard} compares constant-size signatures. When signatures tie, we use the history of accepted updates to justify the corresponding comparison of full witnesses. The four-field witness order follows Duan et al.~\cite[Section 2.3]{DMSY2026}. The four parts below prove Lemma~\ref{lem:semantics} and the signature agreement stated after \eqref{eq:guard}.

\emph{Nonincrease.} Induct over the entire history of accepted updates to the common array, made in every context. A strict signature improvement is a strict full improvement. If signatures tie, the old head witness has the same final edge, because the graph is simple and the predecessor fields agree. Its old tail prefix was created earlier; by induction the current tail is no larger. Appending their common edge proves full nonincrease. The initialized zero-hop source cannot tie a nonempty witness.

\emph{Cap tests in signature form.} Freeze a legitimate snapshot $C$. For an equal-signature cap test on $(u,v)$, let $\omega_u$ be the historical prefix that created $C(v)$. The prefix $\omega_u$ was the label at $u$ when that update created $C(v)$. Subsequent accepted labels are nonincreasing in full order, so $C(u)\leq \omega_u$. Inductively a finite local witness at $u$ is at most $C(u)$, so its extension is at most $C(v)$ in full order. Strict cap tests and infinite caps are immediate. The same chronological argument resolves current-label ties. Thus the one signature guard agrees with the full cap and current conditions. This agreement completes the induction above, which shows that accepted full witnesses are feasible: each accepted update extends a feasible active tail witness and passes the full cap at its new arrival. This argument also applies when a comparison run restricts future continuations to an induced graph. On each surviving edge the same ambient historical-prefix inequality holds, even if the old witnesses leave that graph. Updates on surviving edges continue a legitimate ambient history with omitted vertices unchanged; the caps need not have been generated inside the induced graph.

\emph{Simple and canonical witnesses.} Closing a cycle at $v$ appends nonnegative length and positive hops to an earlier historical prefix at $v$. The present label is no larger than that prefix, so the local-current guard rejects the closure. A local route that returns to a vertex on its inherited seed path fails the snapshot cap there for the same reason. Hence accepted witnesses are simple. Canonical witnesses can be defined among all cap-feasible routing walks: deleting a routing cycle improves the witness and preserves all later caps by common-suffix monotonicity. There are finitely many simple routes, so the minimum exists and is unique. Replacing a nonoptimal routing prefix by its optimum would improve the full witness, proving prefix consistency. While every write is accepted in the cap context of $C$, every current active witness is at least its optimum, so nonincrease makes equality persist. If an accepted update first attains $\pi_C(v)$ along an edge from $u$, the update used the prefix $\pi_C(u)$ by prefix consistency. Applying this argument to each edge back to the seed shows that all canonical prefixes were attained and remain complete.

\emph{Agreement for a nested capped search.} Consider either a capped invocation or an inner forest search interpreted with its proof-local entry snapshot, starting at a complete vertex $y$ of the containing context. If $C_{\rm out}$ is the containing cap and $C$ the child's entry snapshot, then $C(v)\leq C_{\rm out}(v)$ everywhere and $\pi_{C_{\rm out}}(v)\leq C(v)$ at each containing-feasible vertex. These inequalities hold at active vertices by feasibility and at inactive ones because the common label is their containing cap. A containing canonical suffix from $y$ therefore respects every child cap and begins with exactly the inherited witness $\pi_{C_{\rm out}}(y)$. It is a feasible child route. Conversely, every child route begins with that containing-feasible seed and respects $C\leq C_{\rm out}$, so is containing-feasible. The two minima agree on these suffix targets. The inherited prefix is preserved even if a proof-only restricted routing graph omits some of its vertices.

\section{Ordered blocks and tree pieces}\label{sec:structures}
The recursive search needs a worklist that returns the smallest $M$ stored keys without sorting the whole worklist; in this appendix $M$ is a block quota. A child returns only keys smaller than the unpulled parent keys. We exploit this ordering in the merge operation. The interface and block organization are adapted from Duan et al.~\cite[Lemma 6 and Section 4.2]{DMSY2026} and the related realization in B1~\cite[Appendix F]{B1PublicDraft2026}. We give the full construction to establish the empty and duplicate cases and aggregate work from zero initial potential.

\subsection{Operations and representation}
A structure $\mathcal D$ has an integer block quota $M\geq256$, an exclusive input bound $B$, and a fixed maximum record count $R_{\rm B}$. At outer level $j$, the caller uses $M=M_j$ and $R_{\rm B}=R_j$. Put
\[
 \lambda_{\rm B}\coloneqq\max\{1,\log_2(2+R_{\rm B}/M)\},\qquad
 a_{\rm B}\coloneqq\lfloor M/3\rfloor,\quad
 c_{\rm B}\coloneqq\lceil3M/8\rceil,\quad d_{\rm B}\coloneqq\lfloor7M/8\rfloor,
\]
and assume $\lambda_{\rm B}\leq c_0M$ for one fixed constant $c_0$. Keys are constant-size tuples, and different identifiers have different keys. Each structure has its own direct-address array giving constant-time access to a record by vertex identifier. Each entry carries a structure token, and a lookup uses its pointer only when that token matches the current structure. Initialization assigns a fresh token, so a reused, previously allocated array is logically empty without being cleared. Deletion and pull invalidate removed entries; merge writes recipient entries with the recipient token and retires the consumed source's token. Allocation of these arrays, their initialization and the finite widths of their epochs are caller costs, handled in Appendix~\ref{sec:resources}. No operation allocates or clears an array of length $M$.

The operations have the following meanings.
\begin{itemize}
\item $\Init$ makes an empty structure with the supplied parameter record and bound.
\item $\Insert(v,z)$ keeps the smaller stored key for $v$; all inserted keys satisfy $z<B$.
\item $\Pull$ on a nonempty structure removes its smallest $\min\{M,|\mathcal D|\}$ pairs. If it removes everything, the returned cut is $B$; otherwise it is the minimum remaining key $b$. Every removed key is strictly below $b$.
\item $\Merge$ consumes a distinct source structure once. All incoming keys are below $B$ and strictly smaller than every original recipient key. The latter condition is vacuous when the recipient is empty; the former still applies. A nonbase source has parameter $M'<M/3$ and supplies ordered blocks of size at most $M'$. Alternatively an externally budgeted base search tree supplies singleton blocks in inorder. Duplicates keep their smaller incoming key. Write $q_{\rm in}$ for the number of incoming pairs before duplicate removal.
\end{itemize}

\begin{proposition}\label{prop:ds}
Starting from empty structures, take any finite sequence of the operations above that satisfies their stated conditions and keeps each structure's physical record count at most its own $R_{\rm B}$. The total actual work is
\[
 O\!\left(\#\Init+
      \sum_{\Insert}\lambda_{\rm B}+
      \sum_{\Merge}q_{\rm in}+
      \sum_{\Pull}\#\text{removed pairs}\right).
\]
The constant depends only on $c_0$ and the fixed primitive constants. Empty merges are charged to their consumed source initializations, or to the separate base-call budget.
\end{proposition}

The next three subsections bound insertion, pull and merge; the proof of Proposition~\ref{prop:ds} at the end of the merge analysis sums these bounds. Keep a balanced search tree on an ordered doubly linked list of blocks. A block contains an unordered doubly linked record list, a size counter, and copied interval separators. Records have direct links to their blocks. The intervals are right-closed, $(s_{i-1},s_i]$, with final separator $B$. They are historical separators, so deleting a record that once supplied a separator does not require a search for a new extremum. An insertion finds the first upper separator at least its key.

At operation boundaries an empty structure has only a header. A sole block has size between $1$ and $M$. In a multiblock structure every block has size in $[a_{\rm B},M]$. Call these operation-boundary occupancies \emph{stable}. Thus there are $O(1+R_{\rm B}/M)$ blocks, and an interval search costs $O(\lambda_{\rm B})$. Call a block \emph{centered} if its size lies in $[c_{\rm B},d_{\rm B}]$. A reset creates centered blocks, except that a sole block may be smaller than $c_{\rm B}$.

To \emph{reset} $q_{\rm reg}\geq c_{\rm B}$ records in a consecutive region of $O(M)$ size, take $n_{\rm part}\coloneqq\lceil q_{\rm reg}/d_{\rm B}\rceil$ nearly equal integer part sizes. Compute $n_{\rm part}$ and these sizes by binary long division: a constant number of divisions costs $O(\log(M+2))$ charged operations, included in the $O(M)$ reset work below. If $n_{\rm part}\geq2$, then
\[
 q_{\rm reg}\geq(n_{\rm part}-1)d_{\rm B}+1\geq n_{\rm part}c_{\rm B},
\]
using $d_{\rm B}\geq2c_{\rm B}-1$. Hence every part can have size in $[c_{\rm B},d_{\rm B}]$; the case $n_{\rm part}=1$ follows from $q_{\rm reg}\geq c_{\rm B}$. Deterministic selection divides the keys into this constant number of ordered parts while leaving each part internally unsorted. Retain the old outer endpoints and use each nonlast part's maximum as its upper separator. Record scans and a constant number of search-tree edits cost $O(M+\lambda_{\rm B})=O(M)$. A region of size below $c_{\rm B}$ is permitted only when it becomes the sole block.

\subsection{Local normalization and potential}
Deposit a fixed $\kappa$ credits for every ordinary single-record membership insertion or deletion, counting each member of a bulk append separately. Credits are used only in the proof. After a reset the new blocks start with zero credits. Let $\operatorname{ch}(X)$ count membership changes since a block's birth or reset. For a block of size $s$ maintain
\[
 \operatorname{ch}(X)\geq s-d_{\rm B},
 \qquad
 \operatorname{ch}(X)\geq c_{\rm B}-s\quad\text{if the block is nonsole}.
\]
New centered blocks satisfy these inequalities. A small sole block has no underflow obligation. Each membership change increments $\operatorname{ch}(X)$, preserving the inequalities. The merge procedure below ensures that a sole block becomes nonsole without a reset only when its old size is at least $c_{\rm B}$.

An overflow at size $s\geq M+1$ therefore has at least $\kappa(s-d_{\rm B})\geq \kappa s/8$ credits. An underflow at $a_{\rm B}-1$ has at least $\kappa(c_{\rm B}-a_{\rm B}+1)\geq \kappa M/24$ credits. These pay the reset and the tree edits for a sufficiently large universal $\kappa$. In particular, a centered block needs at least $M/32$ changes before either trigger. Discarding a neighbor's or source's nonnegative credits is harmless.

An ordinary insertion overflow has $M+1$ records and splits into two centered blocks. A bulk append of fewer than $M$ records is completed before normalization and leaves fewer than $2M$ records. This order allows incoming records to remain unsorted. For an underflow caused by one deletion, join a neighboring stable block, choosing the left neighbor when possible. The joined size lies in
\[
 [2a_{\rm B}-1,M+a_{\rm B}-1],
\]
which is at least $c_{\rm B}$ for $M\geq256$ and is $O(M)$. Reset this region. Other blocks do not change membership, so there is no cascade. A sole block may remain small; an empty sole block is replaced by an empty header.

To insert a strictly improved key, delete its old record through the direct link, normalize that deletion, locate the new interval and insert the new record. A key no smaller than the stored key is ignored. The first record in an empty structure creates the interval $(-\infty,B]$. Each successful update deposits only constant credit, and every rebuilding scan updates the direct record links. This proves amortized $O(\lambda_{\rm B})$ insertion cost.

\subsection{Pulling a prefix}
If the total size $q_{\rm cur}$ is at most $M$, transfer or copy all pairs to an output list, invalidate their identifier entries, release actual nodes by traversal, and return $B$. This costs $O(q_{\rm cur})$, with no sequence of logarithmic node deletions. Transferred pairs never refer to freed storage.

If $q_{\rm cur}>M$, remove consecutive smallest whole blocks until their total first exceeds $M$. There are only constantly many such blocks, because every nonsole block has at least $a_{\rm B}$ records. The extracted region has at most $2M$ records. Select its $(M+1)$st smallest key $b$, output exactly the $M$ smaller pairs, and retain the record with key $b$. Every untouched block is larger, so $b$ is the minimum remaining key.

Reset the retained extracted prefix. If it has fewer than $c_{\rm B}$ records and a block follows, include that block; if necessary include one more. Two stable following blocks suffice since $2a_{\rm B}\geq c_{\rm B}$. The touched region remains $O(M)$ and involves constantly many tree-node edits. A sole remainder below $c_{\rm B}$ is allowed; a sole remainder near $M$ is centered by a reset. All this work, including selection and direct-link updates, is paid by the $M$ removed records. Old credits may be discarded. This proves an actual $O(\#\text{removed})$ bound for each pull.

\subsection{Consuming an ordered source}
Merge has two phases. First traverse every incoming record and delete all its recipient duplicates, with immediate ordinary normalization after each deletion. The incoming ordering condition makes each duplicate incoming key strictly smaller. Keep the source-block boundaries in an owned temporary list. Only after these deletions classify the recipient $\mathcal D_0$ as empty, sole or multiblock, and record its size if sole. A recipient that was multiblock before the deletions may now be a tiny sole block or empty.

The source is consumed wholesale: its lists are traversed and freed without rebalancing a source block after each detachment. Its credits may be discarded. Source and recipient arrays remain distinct, recipient entries are updated on every transfer, and the source level is reused only after consumption. Traversal and release cost $O(q_{\rm in}+1)$.

When $q_{\rm in}<M$, append the whole incoming list to the first block of $\mathcal D_0$, or create the sole block if $\mathcal D_0$ is empty. All incoming values fit that first interval because they precede every remaining old value. Deposit ordinary membership credits and normalize only after the append. Fewer than $2M$ records are touched by any reset, and its triggering credits pay its cost.

When $q_{\rm in}\geq M$, pack consecutive whole source blocks into chunks, ending each chunk as soon as its size reaches $c_{\rm B}$. Since each source block has size at most $M'<M/3$, completed chunks have size at least $c_{\rm B}$ and less than $c_{\rm B}+M/3\leq d_{\rm B}$. They are centered, with zero initial credits. There are $O(q_{\rm in}/M)$ chunks, and inserting their nodes costs
\[
 O((q_{\rm in}/M)\lambda_{\rm B})=O(q_{\rm in}).
\]
Find chunk maxima during the already paid traversal and use them as upper separators. Whole source blocks establish interchunk order without sorting their interiors. A base tree provides singleton source blocks in inorder. The last uncompleted tail has fewer than $c_{\rm B}$ records.

The final attachment has three cases.
\begin{enumerate}
\item If $\mathcal D_0$ is sole and has size below $c_{\rm B}$, combine it with the last completed chunk and the tail, and reset their contiguous interval, whose last endpoint is $B$. There is a completed chunk because $q_{\rm in}\geq M$. The region has $O(M)$ records and at least $c_{\rm B}$, so this work is paid directly by $q_{\rm in}$. This case prevents a small sole block from being left underfull behind the new chunks.
\item If $\mathcal D_0$ is nonempty in any other case, append the whole tail to its first block, deposit membership credits and normalize as necessary. An old multiblock first block keeps its earlier credits. An old sole block has size at least $c_{\rm B}$ before it becomes nonsole, which establishes the potential invariant for that transition.
\item If $\mathcal D_0$ is empty, let the last completed chunk inherit the upper endpoint $B$, including when the tail is empty. Append the tail there and normalize if necessary. The endpoint edit and any $O(M)$ final reset are paid by $q_{\rm in}\geq M$. No separate small tail block is created.
\end{enumerate}
New chunks restrict the old first block's lower interval endpoint without removing any old record: all incoming values are smaller. A tail reset preserves the order of the preceding completed chunks. These operations preserve the minimum-per-identifier semantics and stable occupancy, with no uncharged extremum search.

\begin{proof}[Proof of Proposition~\ref{prop:ds}]
Let $\Phi$ be the sum of all live-block credits, measured in actual operation units with one common deposit constant. For each completed operation the preceding analysis gives
\[
 \text{actual work}+\Phi_{\rm after}-\Phi_{\rm before}
 \leq C_{\rm op}\,\text{operation budget},
\]
where the budgets are $\lambda_{\rm B}$ for insertion, $q_{\rm in}$ for a nonempty merge, the number removed for a pull, and constant cost for initialization. An empty merge is charged to its consumed source initialization, or to the external base-call budget when its source is a base tree. Small sole births receive no $M$-sized reserve. Large-merge and pull resets are directly paid, and ordinary triggers release sufficient credits. All source-credit discards are nonnegative.

Sum over any completed-operation prefix. Initially $\Phi=0$ and finally $\Phi\geq0$, so the summed right-hand sides bound the total actual work of the prefix. External base-tree creation, its nodes and any empty-source merge or control are charged to the base call. Each outer base call returns a nonempty output, and these outputs are disjoint, so the number of such constant base controls is bounded by the returned-vertex count used in Lemma~\ref{lem:outer-cost}.
\end{proof}

\subsection{Partitioning a tree}
The forest construction groups sources by small connected pieces. Pieces may share boundary vertices, so both the number of pieces and their total listed vertex occurrences must be bounded. The partition statement is given by Duan et al.~\cite[Section 4.1, Lemma 12]{DMSY2026}. We include an explicit implementation to record residual-edge ownership, linear total listing, unique source assignment and deterministic choices on actual records.

\begin{lemma}\label{lem:partition}
For a tree with $n_T\geq k\geq2$ vertices, there is a deterministic linear-time partition of its edges into connected pieces, each with at least $k$ and fewer than $3k$ vertices. All edges and vertices of all pieces can be listed in $O(n_T)$ total time and space. Each specified source vertex can be assigned to exactly one piece containing it.
\end{lemma}
\begin{proof}
Root the tree and process children in fixed order. A child returns a connected unassigned residual containing its root and having at most $k-2$ edges. At a parent, begin with the zero-edge residual. Attach a child residual and its parent edge, a branch of at most $k-1$ edges. Before attachment the accumulated residual has at most $k-2$ edges. When the union reaches $k-1$ edges, emit the entire union and reset to the zero-edge residual. Every emitted piece has between $k-1$ and $2k-3$ edges, hence between $k$ and $2k-2$ vertices. The residual invariant follows by induction; every edge is assigned once.

At the root the remaining residual has $r_{\rm res}\leq k-2$ edges. If $r_{\rm res}>0$, some emitted piece shares a vertex with it: the full tree has at least $k-1$ edges and is connected. Join the residual to one such piece. Connected edge-disjoint subtrees share at most one vertex, since two common vertices would force the unique path between them to belong to both. The joined piece has at most $(2k-3)+(k-2)+1=3k-4<3k$ vertices. For $k=2$ a nonempty root remainder never occurs. Edge coverage also covers every vertex because $n_T\geq k\geq2$.

Store a residual as a linked edge list with head, tail and count. Attaching and concatenating lists takes constant time by splicing; emission transfers ownership. Thus residuals are not recopied along ancestors. Scan emitted edges once to record an incident piece at each touched endpoint. Scanning the root residual finds an incident emitted piece for the last merge; existing incident identifiers remain valid. To list a piece's vertices, scan its edge endpoints using a piece-identifier stamp in the direct vertex array. If there are $p$ pieces, then
\[
 p(k-1)\leq n_T-1,
 \qquad \sum_{P}|V(P)|=(n_T-1)+p\leq2(n_T-1).
\]
These inequalities pay all endpoint scans and headers. On a specified source's first occurrence, assign it to that incident piece using membership and assigned flags. The choices use actual record order and fixed adjacency order, so the same forest records receive the same partition independently of unused ambient vertices.
\end{proof}

\section{A lazy ordered family with prefix cost bounds}\label{app:lazy}
The inner recursion needs constant amortized merge work even when a child returns many physical entries. The generic lazy-block lemmas of C-HD~\cite{CHDGeneric2026} give this bound for ordered blocks with a shared live map. The recursion must preserve the cut that separates a returning child from its paused parent.

\subsection{Operations and semantic conditions}
A physical entry is $(i,v,z)$, with immutable stored key $z$ for vertex $v$ and a globally fresh entry identifier $i$, allocated in increasing order and never reused within the invocation. One live map $\mathcal L(v)$ specifies its live pair $(i,z)$ or absence. The \emph{view} of a structure, its logical worklist, consists of the pairs $(v,z)$ whose live entry lies in it. Entry identifiers of distinct structures are disjoint. A superseded physical entry is stale; this differs from a live entry whose stored key exceeds its vertex's current key. Each structure has a nonempty list of blocks with strictly increasing separators, first separator $-\infty$ and final bound $B$. A block owns the interval from its separator inclusively to the next separator exclusively, or to $B$ for the last block. Its live entries lie in that interval. Stale entries need not satisfy this interval condition. There is no upper bound on raw block size at insertion; here and below, \emph{raw} quantities include stale entries and empty blocks. Endpoint identity is part of every signature, so distinct live vertices have distinct values.

Initialization creates one empty block, irrespective of $M\geq1$. Here $M$ is the block quota; a level-$j$ structure uses $M=M_j$, including $M_0=1$. Insertion skips precisely when the global live value of its vertex is no larger than the candidate. Otherwise it allocates one fresh entry, invalidates the old pointer, finds the candidate's block by binary search and prepends the new entry. It never splits a block. Deletion clears the live pointer. A circular doubly linked list of live entries per structure gives constant-time emptiness: insertion unlinks a superseded entry from its current list by neighbor links, pull and deletion unlink the corresponding live entries, and merge splices live lists without relabeling entry owners.

Pull prepares the front blocks while the number of passed live entries is at most $M$. A front block with more than $2M+1$ live entries is split at its live median into the values strictly below the median and those at least the median, discarding stale entries encountered in the scan. Repeat on the current front until its live size is at most $2M+1$, then prepare the next block if necessary. Collect front blocks until more than $M$ live entries are present or the list ends. Select the smallest $M$ live values when nonexhausting, returning the selected vertex identifiers and the smallest remaining stored key $b$; keep the remaining collected live entries as the first block and the untouched suffix in order. An exhausting pull returns all represented vertex identifiers and $b=B$, leaving an empty first block. In both cases, clear $\mathcal L(v)$ for every returned vertex $v$ and unlink its live-list entry. Copy the returned identifiers before reclaiming their removed physical entries; discarded stale entries are reclaimed during the already charged scans of those blocks. Untouched stale entries are not traversed. This operation preserves the interval invariant and leaves every nonfirst raw separator strictly above $b$.

To merge a child structure of parameter $M'$ into a parent of parameter $M\geq3M'$, pack consecutive child blocks into chunks by counting all physical entries, live or stale. Close a chunk as soon as it reaches $\lfloor M/3\rfloor$ entries; a final smaller chunk is allowed. The closing test is equivalent to three times the accumulated count plus two being at least $M$; a shift and additions implement it in constant work per visited block. Splice entry lists in constant time per old block. Prepend these chunks of blocks and re-separate the old parent's first block at the child's final bound $b$; drop that old first block when its resulting interval is empty. This requires
\begin{equation}\label{eq:mergepremises}
 b\leq B,\qquad b\leq\text{every live parent value},\qquad
 b<\text{every nonfirst raw parent separator}.
\end{equation}
The last premise concerns stale or empty blocks too. It is supplied by the preceding parent pull and its pause, not merely by live-value separation.

The global skip must refer to a live entry in the active structure, and the candidate must be strictly below its final bound. In the proof package~\cite{CHDGeneric2026}, \texttt{insertL\_spec} uses these conditions to give the keep-min update on the active view. In a pull, \texttt{pullL\_spec} and \texttt{pullL\_post} give the cut, the removal of exactly the returned keys from the live map, and the strict lower bound on the remaining nonfirst raw separators. For two well-formed structures satisfying \eqref{eq:mergepremises}, \texttt{merge\_spec} gives the union of the two views under the current live map. \texttt{deleteKeys\_spec} removes the specified keys from every such view and preserves its block invariant.

The comparison with an ordinary keep-min union uses the parent view before the child ran. Each live pointer changed by the child is either cleared or assigned a fresh identifier. For a changed vertex outside the child output, a live entry of no larger key must survive in the child's returned structure. These are the fresh-change and survival hypotheses of \texttt{lazy\_merge\_eq\_literal}~\cite{CHDGeneric2026}. With globally unique entry identifiers, the resulting live view equals the keep-min union of the original parent view and the returned child view, after deleting the child output. The next subsection checks these conditions for our recursion.

\subsection{Use by the inner recursion}\label{app:lazy-use}
While a child runs, all paused ancestors' remaining live keys are at least the child cut. Every active insertion is strictly below its input bound. A global skip therefore finds its no-larger live key in the active structure. The parent's pull left every nonfirst raw separator strictly above the cut. The parent remains paused and the child's final bound is that cut, so both separation conditions for ordered merge in \eqref{eq:mergepremises} hold.

Ancestor invalidation means that a smaller insertion replaces the global live pointer to an ancestor's physical entry, making that entry stale. A second condition prevents this from losing an unreturned key: every vertex placed in a child subtree is, on return, either in its output or represented by a no-larger live key in its returned structure. Pull sends residual vertex identifiers to the child; source restoration puts every unreturned input back at its final current key; merge preserves those replacements; and deletion removes only completed outputs. This induction, given with the recursive proof in Appendix~\ref{app:return}, establishes the condition for all actual placements, including base conversions. Merge followed by deletion of child output consequently restores the parent's logical worklist after removal of completed vertices.

\subsection{The imported operation bounds}\label{app:lazy-primitives}
For $M\geq1$ put
\[
 \eta_M(i)\coloneqq\left\lfloor\log_2\left\lfloor i/M\right\rfloor\right\rfloor
 \quad\text{with value $0$ when $i<2M$},\qquad
 S_M(s)\coloneqq\sum_{i=1}^s\eta_M(i),
\]
and let $E_{\mathcal D}$ be a structure's total physical-entry count, including live and stale entries. Here $|Y|$ is the number of physical entries in block $Y$, including stale entries. Write $(x)_+\coloneqq\max\{x,0\}$. The nonnegative potential is
\begin{equation}\label{eq:lazyphi}
 \Phi_M(\mathcal D)\coloneqq2\operatorname{stale}(\mathcal D)
 +\sum_{\text{blocks }Y}\bigl[3+420(|Y|-M)_++210S_M(|Y|)\bigr]
 +\left\lfloor12E_{\mathcal D}/M\right\rfloor.
\end{equation}
This is the potential \texttt{potM} of the proof package~\cite{CHDGeneric2026}. Its stale term pays for inspecting stale entries. The excess and logarithmic terms pay for median splits. When a merge moves child blocks to a quota at least three times as large, the drop in the entry term pays the cost of assembling the chunks.

The table gives the cited bounds on abstract work plus the change of potential, using the unit block charge $T=1$ of the proof package; in Section~\ref{subsec:outer-run}, $T$ denotes the outer scale. A skipped insertion costs two units. A placement costs $\operatorname{bsCost}(n_{\rm blk})+3$, where $n_{\rm blk}$ is the block count before insertion and $\operatorname{bsCost}(n_{\rm blk})=1+\lfloor\log_2 n_{\rm blk}\rfloor$. Write $V_{\rm pull}$ for the list of vertex identifiers a pull returns. Pull charges scans of physical records, selection and partition work, and block accesses. Deleting a list $Z$ takes $|Z|+1$ units. Merge charges the number of child blocks, the number of new chunks and four additional units.
\begin{center}
\begin{tabular}{lll}
\toprule
Operation&Charge&Declaration in~\cite{CHDGeneric2026}\\
\midrule
New&constant, including initial potential $3$&\texttt{potM\_new}\\
Insert&$\operatorname{bsCost}(n_{\rm blk})+438+210\eta_M(E_{\mathcal D}+1)$&\texttt{insertM\_amortized}\\
Pull&$735|V_{\rm pull}|+950$&\texttt{pullM\_amortized\_sharp}\\
Delete&$3|Z|+1$ for one structure&\texttt{deleteKeys\_amortizedM}\\
Merge&$14$, replacing both input potentials&\texttt{mergeM\_amortized}\\
\bottomrule
\end{tabular}
\end{center}

The block invariant is the proof package's \texttt{WF}. Uniqueness of the physical identifiers gives \texttt{IdsNodup}. Insertion uses these two conditions, \texttt{FreshOK} for the next identifier, and $M\geq1$. Pull uses \texttt{WF}, \texttt{IdsNodup}, $M\geq1$, and the endpoint projection of a stored key supplies \texttt{KeyInj}. Deletion uses \texttt{IdsNodup} for the structure whose potential changes, even when some deleted keys belong to another structure. Merge uses a nonempty parent block list, $M'\geq1$, and $3M'\leq M$. Initialization has one empty block for every quota.

The operation inequalities involve no reserve proportional to $M$ at initialization. Taking their sum over all owned structures gives a finite-prefix bound. A fresh insertion invalidates at most one old live entry in all other structures combined, adding at most $2$ to their stale potential. Pull changes no other structure's liveness. Each deleted key clears at most one live entry, and a cleared entry in another structure adds $2$ to that structure's stale potential; hence deletion has family charge $5|Z|+1$. Merge consumes both structures, so their potentials are replaced. Beginning with no structures, summation yields
\begin{equation}\label{eq:family}
 \begin{split}
 W_{\rm DS}+\Phi_{\rm final}\leq c_{\rm init}\#\Init
 &+\sum_{\Insert}\bigl(\operatorname{bsCost}(n_{{\rm blk},i})+440+210\eta_{M_i}(E_i+1)\bigr)\\
 &+\sum_{\Pull}(735|V_{{\rm pull},i}|+950)
 +\sum_{\Delete}(5|Z_i|+1)+14\#\Merge.
 \end{split}
\end{equation}
Here $W_{\rm DS}$ is the actual abstract work of the operation prefix and $\Phi_{\rm final}$ is the sum of the remaining structure potentials. Each operation-specific sum uses its own index. At insertion $i$, $n_{{\rm blk},i},E_i,M_i$ are the block count, physical-entry count before insertion, and parameter of the active structure; $V_{{\rm pull},i}$ is the list returned by pull $i$, and $Z_i$ the list passed to deletion $i$. Final potential is nonnegative at every completed-operation boundary, including one followed by an outward stop.

To bound the logarithmic insertion charge by actual placements, use the separate block-count quantity. For a companion exposition of this quantity and its transitions, see the public C-HD candidate~\cite[Section 3.4, Block count]{CHDPublic2026}. Put
\[
 \Psi(\mathcal D)\coloneqq M\#\{Y:|Y|<\max(1,\lfloor M/3\rfloor)\}+2E_{\mathcal D}.
\]
The transitions in the proof package~\cite{CHDGeneric2026} are \texttt{psi\_new} (value $M$), \texttt{psi\_insertL} (increase at most $2$), \texttt{psi\_pullL} (decrease at least $M$ for a nonexhausting pull or reset to $M$ for an exhausting pull), and \texttt{psi\_merge} (increase at most $M+2E_{\mathcal D'}$). Insertion uses a nonempty block list, and pull uses the same hypotheses as the sharp pull bound. The merge transition requires $M\geq3$, which follows from $3M'\leq M$ and $M'\geq1$.

Pair each parent's pull with the merge of its unique child, if any; in the recursion every merge into a structure is so paired (each iteration of the loop in Algorithm~\ref{alg:recursion} merges only the child formed from its own pull), and \eqref{eq:psi} is claimed only under this pairing. If $R_X$ counts all successful placements so far in that call's subtree, every physical child entry has a distinct placement origin during that immediate-child execution, from the parent's pull to the matching merge, including placements in descendants. We show that
\begin{equation}\label{eq:psi}
 \Psi(\mathcal D)\leq2M+2R_X.
\end{equation}
A nonexhausting pull pays the merge's $M$ term; an exhausting pull resets the bound; a pull with no merge has no extra charge. Parent $\Psi$ stays fixed while paused. Direct placements pay their own increases. This proves \eqref{eq:psi} at every reached state.

Since $M/\max(1,\lfloor M/3\rfloor)\leq5$, counting large and small blocks gives
\[
 E_{\mathcal D}\leq M+R_X,\qquad \#\text{blocks}=O(1+R_X/M).
\]
Every insertion charge in \eqref{eq:family} is consequently $O(1+\log(2+R_X/M))$. This counts stale placements, group reselections and base conversions as well as relaxation insertions.

For the concrete implementation, store separators in a geometrically growing random-access stack with the smallest block at the top. Pull changes the front, merge pushes chunks in reverse order, and insertion performs binary search. Keep counts and head and tail links so appending the entries of a whole block takes constant time. Each array starts with constant capacity and its total resizing work is linear in its actual pushes. Gather the live keys for deterministic selection in scratch space proportional to the physical entries scanned. A block that fails the split test is scanned by the ensuing collection, so the live-count test adds only a constant factor. The direct-address arrays and frame records of Appendix~\ref{sec:resources} implement the live map in constant time per access or update.

Pay one extra unit of credit per allocation for eventual destruction, including stale entries in a first parent block dropped by merge. Allocation lists transfer with ownership, and each record is destroyed at most once. These costs, the constant-factor implementation of the charged scans and selections, and the stack resizing are bounded on every prefix by a fixed multiple of \eqref{eq:family}.

\section{The recursive return induction}\label{app:return}
We prove Theorem~\ref{thm:return} and the legality of worklist operations by induction on the recursion level. Each nonbase call maintains (J1)--(J6). The proof applies to the inner and the outer recursion because their representations and scanning rules give the same child target interval.

Accepted extensions remain in the current target by Lemma~\ref{lem:foreign}'s containment argument. Each call reached below satisfies (I1) when its forest is formed: the root call by hypothesis and a child by the argument following \eqref{eq:childtarget}. Hence (F) gives \eqref{eq:forest1}--\eqref{eq:forest3} at that call (Section~\ref{sec:framework}). The contact property of stopped trees is part of (F$'$), verified for each local search in the body, and Lemma~\ref{lem:partition} proves the edge partition.

\noindent\textbf{Base call and inductive invariants.}

At level zero, (I1) ensures that every target has a complete canonical ancestor represented in the queue at its optimal key. Scanning a complete extracted tail preserves this coverage for every target not yet extracted by inserting the required canonical boundary heads. If the minimum were incomplete, a proper canonical prefix would still be represented by a smaller key. Thus extractions are complete and strictly increasing. Exhaustion has precedence after the final scan; otherwise the next minimum separates exactly the returned prefix. As $S\neq\varnothing$ and $H_0\geq1$, the first extraction has key $a_X$, so $a_X<\beta$ and $1\leq|U|\leq H_0<2H_0$; a partial return has $|U|=H_0=t^3\geq t|S|$, since $|S|\leq A_0=t^2$. Every residual input is at its current key, and every required boundary head remains at its optimal key. Vertices represented in the logical worklist originate among the input vertices or an extracted tail's outgoing heads. An inner base call converts these residual keys into a fresh lazy structure before parent resumption. The ancestor-bound argument of Appendix~\ref{app:lazy-use} validates each conversion insertion; the conversion preserves the queue's logical view.

For a nonbase call put $T_P\coloneqq\mathcal T(B,P_S)$, and write $U_g$ for the union of prior immediate-child outputs. Maintain the following invariants at loop boundaries:
\begin{enumerate}[label=(J\arabic*)]
\item $U_g$ is exactly the complete prefix of $T_P$ below $L$.
\item Every group is its original membership minus $U_g$, with a designated representative at its current stored key or a no-larger live replacement in the lazy representation.
\item Vertices represented in the logical worklist lie in $T_P$, with stored keys in $[L,B)$ at least their current keys.
\item Every residual target has a complete canonical ancestor represented in the worklist at its optimal key, or a complete residual group ancestor $y$ whose designated representative's stored key is at most $\sig(\pi(y))$.
\item Vertices represented in the logical worklist lie in $S\cup N^+(U_g)$.
\item Every canonical edge from $U_g$ to a residual target has its complete head represented in the worklist at its optimal key.
\end{enumerate}
At loop exit, (J1) is used for item~1 of Theorem~\ref{thm:return}, (J3) and (J5) for item~3, and (J6) for item~4, while (J2) and (J4) keep every residual target covered by a worklist entry or a group representative. (J1)--(J6) hold initially by \eqref{eq:forest3} and choice of a representative with minimum current key. There is no $T_P$ optimum below the initial $L$.

\noindent\textbf{The child target and ordered return.}

We first show that each child's target is the band of $T_P$ in \eqref{eq:childtarget}. Let $b$ be the cut returned by the parent's pull and $S_c$ the resulting child source set after filtering and group expansion. A pull removes keys below $b$ and leaves keys at least $b$. A pulled vertex identifier with current key below $L$ is already in $U_g$ by feasibility and (J1). In the lazy representation it was deleted when returned. A record of a vertex already in $U_g$ has stored key at least $L$ by (J3), while its optimum is below $L$ by (J1). Thus it cannot be the optimal-key ancestor guaranteed by (J4). A residual designated representative survives the filter. Expansion of those representatives and (J4) therefore give the inclusion from right to left in
\begin{equation}\label{eq:childtarget}
 \mathcal T(b,S_c)=\{v\in T_P:L\leq\sig(\pi(v))<b\}.
\end{equation}
Conversely, a pulled source lies in $T_P$ by (J3) and is outside $U_g$, since a vertex of $U_g$ is complete below $L$ and is discarded by the filter; expanded group members are outside $U_g$ by (J2). By (J1), every child source has canonical signature at least $L$. Its canonical descendants remain in $T_P$ and have no smaller signature; imposing the child bound $b$ gives the reverse inclusion. Every child source is residual, and each target in this interval has a complete source ancestor. The forest properties, parent (I2) and the already returned prefix give child (I2). At most $M_j$ vertex identifiers are returned by pull; groups have fewer than $3k$ members, so $|S_c|\leq(3k+1)M_j\leq A_{j-1}$. An empty filtered batch removes no required ancestor or representative, creates no entry, and preserves $L$ and all invariants.

A nonempty child returns cut $\beta_c$ and output $U_c$, the complete vertices of its target in $[L,\beta_c)$. Adding $U_c$ to $U_g$ and setting $L=\beta_c$ therefore preserves (J1). Its residual stored keys lie in $[\beta_c,b)$ and unpulled parent keys in $[b,B)$, permitting explicit ordered merge. For the lazy representation, parent pull additionally left all nonfirst raw separators above $b$. A paused parent changes only liveness, so \eqref{eq:mergepremises} holds with the child's final bound $b$. The global skip condition follows because an ancestor key is at least its child cut, while an active candidate is strictly below the active bound. A consumed sibling has no independent remaining structure or logical view; any transferred entries that remain live belong to the parent and obey the same bound separation.

\noindent\textbf{Survival and residual coverage.}

We prove survival simultaneously for every vertex/key placed in a child subtree. A placed vertex is represented by a no-larger live copy until it is returned, superseded by a smaller placement, or pulled. Pull passes a nonreturned vertex identifier into the child's input; a discarded vertex with current key below $L$ is already returned. By induction on the recursion level, the child returns it or restores it at its final current key. Merge preserves every residual replacement. Deletion removes only returned output, and final input-source restoration and insertions from the $W_{\rm new}$ scan leave their residual keys represented. Thus each placed vertex finishes in output or in the returned live structure. Applied to an invalidated ancestor copy, the replacement is no larger by monotonicity. This proves the survival premise in Appendix~\ref{app:lazy}, so merge followed by output deletion restores the parent's logical view minus completed vertices.

Remove returned group members before the parent's scans of the returned child output. A pulled residual representative was a child input source and is restored at its current key. An unpulled representative remains unless it was returned through another source, which marks its group. In the lazy representation survival supplies a no-larger replacement if child activity invalidated the old copy. Dirty reselection restores a current minimum after a representative's return. Hence (J2) persists under the representative rule of Section~\ref{sec:framework}.

For (J4), follow a previous complete canonical ancestor of a residual target. An unpulled fresh key has optimum at least $b$, so was not returned. A pulled residual ancestor is restored by the child. A returned ancestor is replaced along its first canonical boundary edge: below $b$ the child covers the head, and at least $b$ the parent scans it. A previous complete group ancestor $y$ with $\sig(\pi(y))<b$ was expanded into the child source set and is either returned or restored at its optimal key. If $\sig(\pi(y))\geq b$, it remains in the group. An unpulled and unreturned representative keeps its previous stored-key bound, using a no-larger live replacement if necessary; a pulled but unreturned representative is restored by the child at a key below $b\leq\sig(\pi(y))$; and a returned representative is replaced by dirty reselection, whose current minimum is at most $\sig(\widehat d(y))=\sig(\pi(y))$. This preserves (J4).

For (J3), all residual group members have optimum at least $\beta_c$, so reselected current keys are at least $\beta_c$. Child residual keys lie in $[\beta_c,b)$, unpulled parent keys lie in $[b,B)$, and the parent inserts scanned candidates only in the latter band. Thus the lower stored-key bound in (J3) is preserved. Child residuals lie in $\mathcal T(b,S_c)\subseteq T_P$ and reselected representatives lie in $P_S\subseteq T_P$. A head accepted from a returned tail of $T_P$ lies in $\mathcal T(B,S)$ by Lemma~\ref{lem:foreign}'s containment argument; if it were outside $T_P$, \eqref{eq:forest2} would make it complete in $W$, so its accepted witness would be canonical and would pass through that tail, a contradiction.

For (J5), child sources came from vertices previously represented in the logical worklist or from the fixed input-source groups, child residuals add only outgoing heads of its output, and the parent's scans of the returned child output have those same complete tails. Group operations use call input sources.

For (J6), the boundary-edge argument for (J4) applies to every canonical edge from $U_c$ to a residual target, and a still-residual head of an earlier such edge is kept if unpulled and otherwise restored by the child at its optimal key.

The complete-tail cursor scans exactly the bands required by the boundary-edge argument above and Section~\ref{subsec:core}. Each scanned tail is already complete, so its candidates are fixed and increase along its sorted static list. Its first base or $W_{\rm new}$ scan leaves the cursor at the first candidate at least that call's input bound if one exists, and at list end otherwise; successive ancestors scan from there across ascending bands. An initial $W_{\rm new}$ scan observes lower-band edges but makes no insertions there, and none is needed: an infeasible head never accepts, and a feasible one lies below $L$ and is complete by (J1), \eqref{eq:forest2} and (I2). Packed-slot deletions leave the static outgoing lists unchanged. Lemma~\ref{lem:foreign} shows that deleted exploration edges remain ineligible for accepted relaxation in the context and no canonical edge is lost. The outer recursion's full outgoing scans traverse every outgoing edge of a scanned tail, but candidates outside the current band are skipped without relaxation, publication or insertion. Within the band these scans apply the same acceptance test and insert each accepted head.

\noindent\textbf{Output, progress and termination.}

If the structure empties, (J2) and (J4) leave no residual target in $T_P$, giving $\beta=B$. Otherwise the size exit has $|U_g|\geq H_j$. A vertex in $W_{\rm new}$ has canonical signature at most its current signature below $\beta$; if it were incomplete, \eqref{eq:forest1} would put it in the already complete $T_P$ prefix. Property \eqref{eq:forest2} supplies every remaining target below $\beta$. Thus the final output is exactly the promised complete prefix. Input-source restoration and final $W_{\rm new}$ insertions leave every residual source at its final current key, including a source lowered by that last scan. Together with (J6) at loop exit and the heads inserted by the final $W_{\rm new}$ scan, this completes item~4 for the call. By (J3) and (J5), followed by source restoration and the final $W_{\rm new}$ scans, the residual support lies in $\mathcal T(B,S)\cap(S\cup N^+(U))$, with stored keys in $[\beta,B)$ at least their current keys. This proves item~3.

Immediately before the last child, $|U_g|<H_j$. That child adds fewer than $2H_{j-1}$ vertices and $|W|\leq kA_j$, so \eqref{eq:capacityguards} yields $|U|<2H_j$. A partial output has $|U|\geq H_j\geq t|S|$. If there are no groups, the complete $W$ covers a nonempty full target. Otherwise the initial minimum representative survives its first pull and gives strict progress. Also $\beta>a_X$: either $\beta=B$, or $\beta$ is at least the first child's returned cut, which exceeds that child's source minimum, itself at least the initial $L\geq a_X$. Every nonempty child returns new vertices; empty batches only consume existing records. The finite forest and legal finite primitive operations therefore terminate.

\noindent\textbf{Worklist legality and handled intervals.}

For explicit worklist legality, at every stage let $U_{\rm fin}$ include completed children, adding the latest child before merge and $W_{\rm new}$ before its final scans. Every physical record, including a record still present for a returned vertex, represents an input source or an outgoing head of $U_{\rm fin}$. These are the only sources of placements by (J5) and the final scan rule. Since the explicit structure has at most one record per vertex and $|U_{\rm fin}|<2H_j$, its record count is at most $A_j+2DH_j=R_j$ from \eqref{eq:outerds}. That guard bounds the logarithmic amortized insertion charge by $2T$ at outer scale $T$. The incoming scale and key separation satisfy Proposition~\ref{prop:ds}, and a base tree provides ordered singleton blocks. For lazy worklists, stale raw placements have distinct entry identifiers, and the placement bound in Appendix~\ref{app:counting} supplies their separate amortized insertion charge.

Line~\ref{line:next-cut} of Algorithm~\ref{alg:recursion} sets $L\gets\beta_c$, so successive children have consecutive, hence disjoint, handled intervals. They lie in the parent's handled interval: the parent's initial $L$ is at least its $a_X$, which is at least the lower end of its own handled interval by \eqref{eq:childtarget} (and by definition at the root call), and its returned cut is its final $L$. Base calls have no children. A full call's $[a_X,B)$ lies in its handled interval. Together with exact prefix output this proves all interval, disjointness and laminarity clauses, completing the joint induction.

\noindent\textbf{The packed engine.} For the packed engine of Section~\ref{sec:packed}, Lemma~\ref{lem:packed-forest} supplies (F), containment of accepted extensions in the current target, and stopped trees that meet $F$ only at their contact endpoint (Section~\ref{subsec:packed-cost}). All other operations in this appendix are unchanged: base conversion precedes parent resumption; a child returns its input bound for merge; completed child output is deleted; parent static scans use the child's input band; only dirty groups are reselected; and the final pass restores call input sources at their current keys. A packed relaxation introduces no recursive source or keyed placement by itself. The live-map skip and raw-separator conditions depend only on paused ancestor cuts and active insertions below their own input bounds. The survival induction tracks each placed vertex until output or a no-larger residual replacement. It therefore continues to prove lazy worklist legality for the packed engine.

\section{Source origins and completed operation counts}\label{app:counting}
We count feasible completed output, retained foreign vertices, and raw placement events separately within one fixed context. These counts prove Lemma~\ref{lem:completed-shared} in Appendix~\ref{app:completed-shared-proof} and supply the outer counts of Lemma~\ref{lem:outer-cost}. The source-origin argument adapts B1~\cite[Section S4.2, Lemmas G2 and G4]{B1PublicDraft2026} and C-HD~\cite[Section 5.5, (G4)]{CHDPublic2026} to our source-restoration and retention rules. The complete-tail and home-color counts use C-HD's arguments~\cite[Sections 4.4 and 5.3]{CHDPublic2026} with the fixed ranks of Section~\ref{subsec:foreign}.

\subsection{Output bands and persistent roots of small full searches}
Consider an ordinary completed execution whose root returns fully, with output size $r$ and root level $\ell$, and let $E$ count outgoing graph arcs at its output vertices. The completed comparison run has this property. Write $S_X$ for the input source set of a call $X$. Outputs at one level are disjoint, so their total size is at most $r$. A full call contains all its sources in its output; a partial call has $t|S|\leq|U|$. Therefore
\begin{equation}\label{eq:sourcesums}
 \sum_X|S_X|\leq(\ell+1)r,\qquad
 \sum_{X\text{ partial}}|S_X|\leq\ell r/t.
\end{equation}
The full root is excluded from the partial-call sum, leaving only $\ell$ proper-descendant levels. Every call returns a nonempty output.

An edge-insertion event of a call is an accepted relaxation followed by an insertion operation in that call's base, returned-child or $W_{\rm new}$ scan. A common-label update performed by a forest search is a different event. For a fixed complete tail, output calls containing it form one ancestor chain. At the deepest call its band is $(-\infty,B_0)$ for a base extraction or $[\beta_0,B_0)$ for $W_{\rm new}$; successive ancestor bands are $[B_0,B_1),[B_1,B_2),\ldots$. The candidate on its fixed outgoing edge is fixed by completeness. Hence each physical edge has at most one edge-insertion event, including equal acceptance, and the total is at most $E$. Static complete-tail cursor advances separately have at most $E$ events. Packed slot-mask preparation and the deletion of low slots and slots with missing edges have the separate $O(Dq)$ charge of Section~\ref{sec:packed-cost}. Repeated sentinel probes cost one per scan header.

Write $Q_X$ for the set $Q$ of call $X$, the roots of its small full searches. To prove the source-origin bound \eqref{eq:qbound}, we show that repeated occurrences of a vertex in the sets $Q_X$ of nested full calls are separated by edge-insertion events with that head. For a full call $X$ and $v\in Q_X$, consider the instant after its forest and before initial representative insertions. The groups exclude $v$, and there is no old worklist. Until an edge-insertion event with head $v$ occurs in this call or a proper descendant, no proper descendant input can contain $v$. This follows by induction over actual operations: pull and deduplication select vertices already represented in a logical worklist; expansion and reselection select fixed group members; child initialization selects its input; the vertices contributed by a merge come from the child's logical view; and source restoration selects a child's input source set after its children. Returning $v$ in output, retaining it in a forest, or improving its common label does not by itself insert the vertex into a worklist. An accepted relaxation to $v$ followed by insertion in a base, returned-child or $W_{\rm new}$ scan is exactly the required edge-insertion event. The full call's own final input-source restoration occurs after all its children and its final band is empty. Deletion and ownership transfer introduce no new vertex into the logical view. This checks every possible source origin, including intervening partial calls.

Full-$Q$ occurrences of one vertex lie on an ancestor chain, since every full occurrence returns that vertex. Between consecutive occurrences, the source-origin argument requires an edge-insertion event with that head during the execution after the outer occurrence's forest and before the inner occurrence's input. Those time intervals are disjoint. Different vertices have different event heads. Grant the first occurrence of each returned vertex for free and inject all repetitions into the at most $E$ edge events. Partial occurrences are bounded by \eqref{eq:sourcesums}. We obtain
\begin{equation}\label{eq:qbound}
 \sum_X|Q_X|\leq r+E+\ell r/t.
\end{equation}

\subsection{Retained-vertex counts, groups and raw placements}
In a completed comparison on $q$ vertices, fixed ranks bound the total number of vertices in the retained forests of full calls by $q$ at each level. First-contact duplication at most doubles it. Each piece has at least $k-1$ edges by Lemma~\ref{lem:partition}; hence the number of full source pieces is $O((\ell+1)q/k)$, including pieces later dropped for empty groups. The small full searches, whose discarded trees are not in these forests, and the searches of partial calls are counted in the proof of Lemma~\ref{lem:completed-shared} (Appendix~\ref{app:completed-shared-proof}).

We adapt C-HD's home-color charging argument~\cite[Section 5.3]{CHDPublic2026} to the fixed cap ranks $\mu(v)$. To pay full-call group reselections by one charge per piece plus at most one charge per routing arc in the context, color every retained tree vertex by the immediate child's handled interval containing $\sig(\mu(v))$, using one auxiliary color for ranks outside every child's handled interval. A group member removed by a child lies in that child's output, whose ranks lie in its handled interval, so it has that child's color. A piece with $r_{\rm col}$ removal colors has at least $r_{\rm col}-1$ cross-color edges, so one charge per piece and those edges pay its reselections. If an edge is cross-colored in a descendant call with a full return, both endpoint ranks lie in the handled interval of the ancestor's child whose subtree contains that descendant, so the edge has one color at that ancestor. Incomparable full calls have disjoint intervals, since the children of one call receive consecutive handled intervals (Algorithm~\ref{alg:recursion}, line~\ref{line:next-cut}) and every child interval nests in its parent's. Thus each physical routing arc has at most one such full-call charge in the context. Retained vertices that belong to no group are colored in the same way; they serve only to locate cross-color edges. The retained-vertex count and the per-piece and per-edge charges pay the full-call forest and reselection terms in Lemmas~\ref{lem:outer-cost} and~\ref{lem:completed-shared}; partial calls are paid by $t|S|\leq|U|$ from Theorem~\ref{thm:return}, and this appendix sums both.

Full dirty-group reselections are charged once per piece plus cross-color tree edges. A tree edge leaves a source or an accepted target of its full call, which that call returns (Theorem~\ref{thm:return}), so the home-color argument above bounds these cross-color edges by $E$. For partial calls, each reselection follows the permanent removal of a distinct representative belonging to $S_X$ from a fixed group, giving at most $|S_X|$ reselections. A group expansion costs $O(k)$ per pulled designated representative. A partial child pays by $t|S_c|\leq|U_c|$ and $k\leq t/16$. A full child returns the representative, so its expansion is paid by the resulting reselection or the group's one permanent emptying. Dirty lists cost only their recorded marks.

Appendix~\ref{app:lazy-primitives} bounds each lazy insertion charge by $O(1+\log(2+R_X/M))$, where $R_X$ counts the successful placements so far in the subtree of the call $X$ that owns the structure; we now bound this count. Consider any ordinarily returned inner subtree rooted at a call $X$ of level $j$, whether $X$ returns fully or partially. Count all successful raw placements, not only surviving live entries. Its source sums are at most $(j+1)|U_X|$; partial proper descendants contribute at most $j|U_X|/t$, with the additional $|S_X|$ if $X$ itself is partial. Initial representatives are paid by these source sums. Input-source restoration occurs only in partial bands and is paid by the partial-source sum just stated. Full reselections use their per-piece counts $p_Y$ plus $O(D|U_X|)$ edge charges. Here $p_Y$ counts the pieces receiving nonempty assigned input-source groups when the forest of $Y$ is formed. Since these groups are disjoint subsets of $S_Y$, $p_Y\leq|S_Y|$; this counts only groups kept after empty groups are dropped. This weaker piece bound suffices here; partial reselections use source sums. Edge placements in complete-tail scans total at most $D|U_X|$. Base conversions are at most the sum of base inputs and their own edge placements, hence at most $(1+D)|U_X|$. Other operations create no fresh entry identifiers for keyed placements. Thus one fixed $C_E$ satisfies
\begin{equation}\label{eq:placements}
 \operatorname{placements}(X)
 \leq C_E\bigl[|S_X|+(D+j+1)|U_X|\bigr].
\end{equation}
Pure foreign retention or accepted publication alone introduces no keyed placement or new membership in a source group or recursive source set. Later worklist insertions and source roles are counted by the origins above, and slot-mask preparation is a separate work cost. Thus retained metadata adds no placement class to \eqref{eq:placements}.

For a nonbase level fix the analytic placement ceiling
\[
 \widehat E_j\coloneqq C_E[A_j+2(D+j+1)H_j].
\]
Since $\widehat E_j/M_j=O(t^2(D+j+1)2^t)$, the amortized insertion charge $O(1+\log(2+R_X/M_j))$ proved in Appendix~\ref{app:lazy}, with $R_X$ at most this ceiling as shown below, is $O(t+\log(t+2)+\log(D+j+1))=O(t)$ under the logarithmic guard in \eqref{eq:packed-innerguards}. At the base, the analytic placement ceiling is $O(t^2+Dt^3)$ and the amortized insertion charge is $O(\log(2+tD))$. These analytic placement ceilings are determined by the fixed parameter record before execution. They are analysis bounds, not allocated capacities or runtime stopping thresholds; lazy operations use their parameter $M$ and actual entries. For an ordinarily returned subtree, \eqref{eq:placements}, $|S_X|\leq A_j$ and $|U_X|<2H_j$ give the ceiling, which also bounds every earlier placement count. For an actual stopped invocation use its same-record completed comparison on the retained induced graph from Section~\ref{subsec:packed-projection}: the coupled prefix extends there to an ordinary completion under Theorem~\ref{thm:return}. This bounds the monotone placement count without borrowing future potential credit or executing any continuation after the actual stop.

\subsection{Proof of Lemma~\ref{lem:completed-shared}}\label{app:completed-shared-proof}

\begin{proof}[Proof of Lemma~\ref{lem:completed-shared}]
\emph{Full return.} Appendix~\ref{app:witness} proves that the ambient history of the inherited and cap witnesses still supplies the cap semantics on the induced graph. The run is the root call $\calR(B_*,\{x\},\ell)$ of Section~\ref{subsec:core} in the invocation's cap context. This single-seed root has $|S|=1\leq A_\ell$ and satisfies (I1)--(I2) (Section~\ref{sec:framework}); the inner forest search supplies (F) and (F$'$); $k\geq2$; \eqref{eq:shared-guards} gives \eqref{eq:capacityguards} and $3M_{j-1}\leq M_j$; and the lazy representation is used with the base, conversion and scan rules of Section~\ref{subsec:core}. Hence Theorem~\ref{thm:return} applies. A partial root return would contain at least $H_\ell>P\geq q$ output vertices, more than the induced graph has, so the root returns fully. Its output is the complete target, which is nonempty by item~1 of Theorem~\ref{thm:return}; thus $1\leq r\leq q$, and since every output vertex has at most $D$ outgoing arcs, $E\leq Dr\leq Dq$.

\emph{Source sums and small full searches.} The run is an ordinary completed execution whose root returns fully, with output size $r$ and root level $\ell$. Because the inner forest search creates no keyed placement, recursive source or group membership by itself, the source sums \eqref{eq:sourcesums} and the source-origin bound \eqref{eq:qbound} proved above apply to it.

\emph{Retained mass.} The $i$th inner forest search retains $r_i$ vertices, including its root, so $1+r_i\leq2r_i$. Fix a level. By (F$'$) the vertices retained by the searches satisfy the hypotheses of Lemma~\ref{lem:foreign}, so by the fixed ranks of Section~\ref{subsec:foreign} the retained forests of the full calls at this level contain at most $q$ vertices in total. A search stopped on contact adds only one previously retained endpoint, so the stopped searches of the full calls at this level retain at most $2q$ vertices, counted with multiplicity. Every other search of a full call exhausts with fewer than $k$ retained vertices and has its root in $Q$; over all levels these searches retain fewer than $k\sum_X|Q_X|$ vertices, even when their discarded trees overlap. A partial call $X$ runs at most $|S_X|$ searches, each retaining at most $k$ vertices, so by \eqref{eq:sourcesums} the searches of partial calls retain at most $k\ell r/t$ vertices in total. Summing over the $\ell+1$ levels and using \eqref{eq:qbound},
\[
 \mathcal M\leq2\sum_ir_i\leq2\bigl[2(\ell+1)q+k(r+E+\ell r/t)+k\ell r/t\bigr],
\]
which gives \eqref{eq:packed-mass}. By \eqref{eq:engine-work}, the inner forest searches cost $O(c_s\mathcal M+W_{\rm extra})$.

\emph{Shared operation classes.} The remaining work does not depend on the inner forest search. Count all successful raw placements, including conversions and stale copies. By \eqref{eq:placements} and the analytic placement ceilings above, which count stale copies as well as live entries, the logarithmic guard in \eqref{eq:shared-guards} bounds the amortized lazy insertion charge by $O(t)$ above the base and by $O(a_0)$ at the base. Full source pieces number $O((\ell+1)q/k)$ (shown above), so initial representatives and full-call per-piece representatives cost $O(t(\ell+1)q/k)$. Full dirty groups add this per-piece charge plus $O(tE)$ for cross-color edges (shown above), including their $O(kE)$ scans since $k\leq t$; partial dirty groups and input-source restoration cost $O(\ell r)$. Nonbase edge-insertion events contribute another $O(tE)$, since there are at most $E$ such events and each has amortized insertion charge $O(t)$. Group expansion is paid as shown above. New, merge, delete and frame/header costs total $O((\ell+1)r)$ under the lazy family inequality \eqref{eq:family}. Every vertex identifier returned by pull belongs to the next child input after group expansion, except an empty filtered batch; in the lazy structure returned outputs were globally deleted. Hence the total of pulled vertex occurrences is paid by source sums and progress. Stale raw inspection is already in the family potential, which is nonnegative at every completed-operation boundary, and consumed children transfer ownership.

\emph{Base, scans and output.} Base heap operations and fresh level-zero conversions cost $O((r+E)a_0)$, because base inputs sum to at most $r$ and successful base edges belong to the edge-insertion events. Static complete-tail traversal, with its scan headers and sentinel probes, costs $O(E+(\ell+1)r)$. Returning completed output lists through the recursive levels costs $O((\ell+1)r)$, and handing the invocation's retained tree to its caller costs $O(q)$. A retention-observer action takes constant work per probe. Each accepted write, including an equal-signature update, publishes in both publication contexts and adds constant work to its probe transaction. Destruction is prepaid by creation, and the concrete allocation of Appendices~\ref{app:lazy} and~\ref{sec:resources} costs at most a fixed multiple of these counts.

Using $r\leq q$ and $E\leq Dq$, the classes of the last two paragraphs total $O(q[(\ell+1)(t/k+1)+D(t+a_0)+a_0+1])$. Adding the cost $O(c_s\mathcal M+W_{\rm extra})$ of the inner forest searches proves \eqref{eq:shared-coefficient}.
\end{proof}

\subsection{Explicit outer records}
For the explicit outer structure of Appendix~\ref{sec:structures}, we count pulled vertex/key records by their origins at each parent. Initial representatives, child merges, insertions from scans of returned child outputs and reselections after a representative's permanent removal are the only origins during its loop. A full child returns no records; a partial child returns at most $|S_c|+D|U_c|$ distinct vertex/key records. Disjoint child outputs and progress bound all incoming occurrences by $O((D+1)|U_X|)$. The same count pays pull occurrences with multiplicity. Final source and $W_{\rm new}$ insertions occur after this loop; when pulled later they are received as child-merge records at the next parent. This establishes the per-level explicit merge and scan terms retained in \eqref{eq:outercost}.

\section{Resource bounds, parameter records and mask tables}\label{sec:resources}
The stopped bound charges actual touches, so each invocation must avoid initialization proportional to $P$ or to the ambient graph size. We describe the reusable state and then prove a finite word bound from coarse event counts. Together these give the storage and word-size assertions of Theorem~\ref{thm:main}. The other subsections list the branch tests, bound the inner record (completing the proof of Theorem~\ref{thm:packed-cap}), verify that the driver tests pass (proving Lemma~\ref{lem:bounded-degree-guards}), derive a necessary size threshold for branch selection, and realize the mask operations of the packed engine.

\subsection{State proportional to actual work}\label{app:state}

After the parameter tests pass, allocate a constant number of arrays indexed by routing vertices and edges once in $O(N+M)$ time and cells for the packed engine. Distinct arrays serve accepted activity, invocation retention, the lazy family live map, complete-tail cursors and top pointers for recursive roles. The outer forest membership and publication arrays are separate. All epoch fields begin invalid. Every capped invocation and nonoverlapping scratch operation uses a fresh token; an invalid token means the logical default, and an old payload is read only after its token is checked.

The persistent inner roles include source, group, forest, $U$, $W$ and assignment records. Each role has a direct top pointer at each vertex. A first membership in frame $f_{\rm frame}$ creates a record containing its frame token, value, previous pointer and link in $f_{\rm frame}$'s actual touch list. Lookup in the active frame checks only that token; it does not search ancestors. On exit, restore the previous pointers by traversing the actual lists. Parents are paused, and descendants are destroyed first, so these restorations recover the correct parent state. On an outward stop, process frames deepest first. Returned lazy structures use their separate global live map and transfer ownership until merge; their entries require no level relabeling.

The packed engine initializes a tail's slot mask on first accepted use, reading the exact entry caps for the exclusions in \eqref{eq:rootlow}. Each mask remains private to this invocation. Retention observation alone performs no acceptance or tail initialization; an unactivated foreign leaf receives no slot mask, and a retained leaf that later accepts is promoted normally. A complete-tail cursor is independently initialized when the tail first enters complete output, as in Section~\ref{subsec:core}, and persists through ancestor bands. State management per capped invocation is therefore
\[
 O(q+D|A_*|+\text{actual inner membership and record events}),\qquad |A_*|\leq q.
\]
Every listed event is already counted in Appendix~\ref{app:counting}, Lemma~\ref{lem:completed-shared} or Section~\ref{sec:packed-cost}, including the per-search initializations and final selections and the repeated retained occurrences in \eqref{eq:packed-mass}. No clearing across all recursion levels is performed. A stopped capped invocation exports its actual retained tree and leaves its accepted common-label publications in place. Ordinary completion supplies the complete accepted output. Its retained tree is handed to the caller for disposal, not added to the retained forest. Tree and output copying use their actual sizes, and neither case copies or traverses the old external forest.

Every dynamic object has one owner and one actual allocation list. Its creation prepays a constant destruction cost. Geometrically growing arrays begin at constant capacity, and their total allocated and copied cells over every prefix are linear in actual pushes; obsolete capacity destruction is prepaid. Scratch space is proportional to inspected records. On stopping, destroy or transfer only actual objects, restore actual touch lists and retire the epoch. The lazy potential is nonnegative at that boundary. Keeping allocations until their owning context ends gives storage at most a constant multiple of actual work. The outer implementation keeps its separate once-allocated per-level arrays, costing $O(N(e+1))$ time and cells, already present in \eqref{eq:outercost} and \eqref{eq:packed-outercost}. Their stamps are initialized once. Fresh structure tokens and constant-time validity checks are charged to initialization and the corresponding record operations; a level is reused only after its preceding structure has been consumed.

For the packed engine, the immutable entry cap is needed to prepare each tail's slot mask. The containing context's entry accepted-membership bit is needed only to initialize the comparison of Appendix~\ref{app:coupling-packed}. The cost bound includes saving both values with one token and touch record. At the running invocation's first accepted common write at $v$, including an equal-signature write, save the old four-field signature and the containing context's logical accepted-membership bit before the common-label write or either accepted publication. Only after both payloads are initialized, stamp the record and append $v$ to the cap-touch list. Later writes leave both saved values unchanged. If the token is absent, no accepted write or publication by this invocation has occurred at $v$, and the containing context is paused, so its current common label and accepted-membership bit are still their entry values. Entry-cap and entry-membership reads return their saved value when stamped and the corresponding current value otherwise, in constant time. The underlying full entry witness remains the proof object of Appendix~\ref{app:witness}. Retention alone never invokes this write operation. Cleanup discards the saved values along the cap-touch list and retires the invocation's cap token, leaving the common labels and accepted membership as published. Initialization and cleanup visit only the seed and touched heads.

For the packed branch, the lookup tables, walk records and child masks have the global preparation cost \eqref{eq:packed-pre}. Endpoint-mask construction reuses one stamped scratch array across roots and clears only its touched list. We charge the setup for two auxiliary vertex-indexed arrays to this bound. Their $O(N)$ storage realizes the membership test and label overlay of the proof-only completed comparison in Appendix~\ref{app:coupling-packed}. No search operation reads them in the actual run. Charging these arrays and the saved entry bits lets us use the same finite-word accounting for both executions, with $q\leq P\leq N$ and the same parameter record in the comparison.

\subsection{Branch tests before allocation}\label{app:branch-tests}
\noindent\textbf{Packed outer record.} Recall $L=\J(N)$ and the outer record $(T,P,e)$ from \eqref{eq:packed-outer}. Line~\ref{alg:driver-outertests} first tests
\begin{align}
 &4096\leq T\leq L,\quad T^2\leq N,\quad2\leq P\leq N,
       \quad D\leq N,\label{eq:packed-outerguards}\\
 &16(3P+1)\leq T,\quad P\leq T/16,
       \quad\J(D+e+2)+6\J(T+2)\leq T/4.\notag
\end{align}
The inequalities $16(3P+1)\leq T$ and $P\leq T/16$ give \eqref{eq:capacityguards} for $(t,k)=(T,P)$. The logarithmic guard pays the outer sorting and base costs, and $T\leq L$ enters the cost bound of Section~\ref{subsec:packed-branch} and the word bounds below.
Line~\ref{alg:driver-outertests} also tests the explicit allowance $2M^o_j+A^o_j+2DH^o_j\leq M^o_j2^{2T}$ at every positive outer level, with the capacities \eqref{eq:capacities} and superscript $o$ denoting the outer record. This is the explicit allowance \eqref{eq:outerds}, whose cancelled scalar form in Appendix~\ref{app:word-bounds} permits testing without allocating capacity tables. The base uses its separate search tree and does not require the nonbase allowance.
The checks $D,P,T,t,k\leq N$ and $M\leq N^2$ of line~\ref{alg:driver-innertests}, together with the listed guards, are the coarse bounds used to derive the capacity and word bounds below; there are no additional resource tests.

\subsection{Finite parameter computations and word bounds}\label{app:word-bounds}

Construct all parameters once using the bounded-word primitives of Section~\ref{sec:model}. Binary long division and shift/add products are charged algorithms on those primitives; integer logarithms and fixed-degree roots are derived searches, and doubling can use addition. A fixed polynomial in $\log(N+M+2)$ bounds the scalar searches and power tests and is $O(N+M+2)$ on every input. A multiplication stops before exceeding its tested bound; a failed parameter test never constructs a large rejected table. Capacity powers are generated only after the elementary scale bounds. The explicit outer allowance can be tested by its cancelled scalar expression, then its passing records constructed within the outer array-setup budget. Explicitly, after $T\leq L$ has passed, \eqref{eq:outerds} can be checked without constructing its tables: cancelling $M_j$ makes it equivalent, at every nonbase level, to
\[
 32+T(1+2DT)2^T\leq16\cdot2^{2T}.
\] The runtime block computations in Appendices~\ref{sec:structures} and~\ref{app:lazy} use these same primitives and are charged there. Integer operations never encode or round real weights.

The packed driver also computes $\lambda_n,D_{\rm lim}$ and the inner $\gamma=\J(A+D+2)$ once. Multiplication of $\gamma$ by $256$ uses a fixed number of doublings; its intermediate values have fixed polynomial size in the original input. The ordered tests in Section~\ref{subsec:packed-driver} check $t\leq A$ before capacity powers or walk enumeration, and every non-scalar initialization follows only after these tests pass. The extra scalar work fits the same bound.

To fix one word width for every input before any optimized density exponent is used, first consider the routing computation and its routing setup, interpreting counters as mathematical integers and using the already proved event inequalities. Recall $R=N+M+2$. In the capacity bounds displayed below, use $(P,t,\ell)=(N,T,e)$ for the outer scale, and for the inner scale use the retention limit $P$ and the record of \eqref{eq:packed-inner}. The driver checks $D,P,T,t,k\leq N$ and $M\leq N^2$; the outer tests \eqref{eq:packed-outerguards} give $T\leq L=\J(N)$ and $P\leq N$, and the inner guards \eqref{eq:packed-innerguards} give $t\leq A=\J(P)$. Thus each scale has $P\leq N$, $t\leq\J(P)$ and $\ell=\lceil\log_2(P)/t\rceil$, where $P=N$ at the outer scale. Hence
\[
 2^{\ell t}\leq4(P+2)^2\leq4R^2,
 \quad A_j\leq4R^4,\quad H_j\leq4R^5,
 \quad M_j\leq64R^3.
\]
The bounds for all successful raw placements in an inner call's subtree, established in Appendix~\ref{app:counting}, are $O(R^6)$ by these capacity bounds. The unoptimized coefficient \eqref{eq:packed-coefficient} is at most $C R^3$. Independently the outer retained-vertex count gives
\[
\sum_{\text{capped invocations}}q
 \leq C'[N(e+1)+P(N+M+Ne/T)]\leq C''R^3.
\]
The global preparation has at most $O(R^2)$ operations and cells by \eqref{eq:packed-globalguards} and \eqref{eq:packed-pre}; its lookup tables and the vertex-indexed comparison arrays are smaller. The source bounds and the retained mass \eqref{eq:packed-mass} bound the actual packed selections before using any optimized density exponent. Thus the routing computation has at most a fixed $O(R^6)$ operations and allocations, absorbing its outer and setup bounds. Every routing epoch, frame and placement identifier, internal record index, actual routing-list length and routing counter generated by this trace is bounded by this fixed polynomial. A candidate hop is at most the number of accepted writes plus one, with at most $k$ additional relative hops in a prepared walk record. For these internal quantities choose a word width representing every integer from $0$ through $C_7R^8$, for a sufficiently large fixed constant $C_7$ covering them and the routing-setup intermediates. They then use $O(\log R)$-bit words without overflow. Proof potentials are not runtime counters.

Original-input scans, counts, edge identifiers, and preprocessing and output bookkeeping use integer values and records bounded by a fixed polynomial in $n+m+2$; the fallback's indices, heap ranks and counters have the same bound. This includes the original-edge identifiers stored with routing arcs. They require $O(\log(n+m+2))$-bit words. For the degree reduction, \eqref{eq:packed-routing} gives $R=O(n+m+2)$ even when $N$ is larger than a constant multiple of $n$. Thus the routing bound also fits this width, as do addresses in the combined storage. On the fallback path the routing setup and the test scalars computed before the failure fit as well: $R\leq2n+5m+2$ by \eqref{eq:packed-routing}, and every value that a scalar test forms is at most a fixed polynomial in $N+M+2$: each product stops before exceeding its tested bound, and the largest tested bound, $16\cdot2^{2T}$ in the cancelled allowance, is formed only after $T\leq L$ has passed and is then below $64(N+2)^2$. Choosing the width to cover these bounds gives the stated $O(\log(n+m+2))$-bit integer operations uniformly. Exact real-weight comparisons and additions remain separate.

When the padding of Section~\ref{subsec:sparse} is used, these bounds apply to the augmented counts $n'=2n$ and $m'=m+8n+1$, where $n,m$ are the unpadded counts. Their sum is $O(n+m)$, so the same fixed polynomial bounds give $O(\log(n+m+2))$-bit words for the whole computation. The added graph, the preserved original input and the original-output pass cost $O(n+m)$ time and cells.

\subsection{Guards and coefficient of the inner record}\label{app:record-guards}

This subsection proves the implications among the inner guards that Section~\ref{subsec:packed-cost} states for the packed parameter record, and the bound on its local coefficient; $A$, $\ell$ and the capacities are those of Section~\ref{sec:invocation-cost}. The algorithm tests every guard, including those that other guards imply.

The packed guards are the shared guards \eqref{eq:shared-guards} with $64\leq k\leq P$ and $H_\ell>P$ added:
\begin{align}
 &1\leq D\leq P,\quad4096\leq t\leq A,\quad t^2\leq P,
       \quad64\leq k\leq P,\label{eq:packed-innerguards}\\
 &16(3k+1)\leq t,\quad k\leq t/16,
       \quad\J(D+\ell+2)+6\J(t+2)\leq t/4,\notag\\
 &H_\ell>P.\notag
\end{align}
These are checked before the global preparation. Here $D\leq P$ bounds the routing degree by the invocation's retention limit, $t^2\leq P$ bounds the base capacity $A_0$, and $k\leq P$ keeps the inner retention limit within it. For this fixed record only $t\leq A$ can fail once the outer test $P\leq N$ has passed, as the packed-record paragraph below shows. The lower bound for $t$ is imposed before $k$ is formed, and $k\geq64$ is tested before any division by $k$. Lemma~\ref{lem:bounded-degree-guards} shows that the packed driver's parameters pass all these guards on all sufficiently large inputs in the degree range stated there.

The packed record \eqref{eq:packed-inner} has $A=\J(P)$, $\ell=\lceil\log_2(P)/t\rceil$, $\gamma=\J(A+D+2)$, $t_{\rm sq}=\min\{s\in\mathbb Z_{>0}:Ds^2\geq A\}$, $t=\max\{4096,256\gamma,t_{\rm sq}\}$ and $k=\lfloor t/64\rfloor$. The base factor of Lemma~\ref{lem:completed-shared} is $a_0=1+\log(2+tD)$.

\noindent\textbf{Packed record.} For this fixed record the three tests $D\leq P$, $t^2\leq P$ and $k\leq P$ are redundant: $t\leq A$ and $t\geq256\gamma$ give $A\geq4096$ and $D<2^{A/256}<P$, while $t^2\leq A^2<P$ and $k\leq t$. Also $H_\ell>P$ follows from the capacities and $\ell t\geq\log_2P$. On every record passing $t\leq A$, one has $\ell\leq A$ and
\begin{equation}\label{eq:packed-logguard}
 \J(D+\ell+2)+6\J(t+2)\leq7\gamma
       \leq7t/256<t/4.
\end{equation}
Thus the logarithmic floor is sufficient for this guard. Also $t\geq4096$ gives $k\geq64$, $64\leq t/k<65$, $16(3k+1)\leq3t/4+16\leq t$ and $k\leq t/16$.

\noindent\textbf{Packed coefficient.} This paragraph completes the proof of Theorem~\ref{thm:packed-cap} by bounding \eqref{eq:packed-coefficient} on every passing record. The elementary inequality $\log_2(1+x)=O(\sqrt x)$ for $x\geq0$, together with
\[
 A+D+4\leq(D+4)(1+A/D),
\]
gives the finite estimate
\begin{equation}\label{eq:packed-floorcost}
 D\gamma=O\!\left(D+D\log(2+D)+D\log(1+A/D)\right)
         =O\!\left(\sqrt{DA}+D\log(2+D)+D\right).
\end{equation}
Since $t_{\rm sq}\leq\sqrt{A/D}+1$ and the maximum defining $t$ is at most the sum of its terms,
\[
 Dt\leq4097D+256D\gamma+\sqrt{DA}
       =O\!\left(\sqrt{DA}+D\log(2+D)+D\right).
\]
Also $t\geq\sqrt{A/D}$ gives $\ell+1\leq A/t+2\leq\sqrt{DA}+2$. The ratios $64\leq t/k<65$, $kD=O(Dt)$, $k\ell/t=O(\ell)$ and $k=O(t)$ bound all nonlogarithmic terms of \eqref{eq:packed-coefficient} by $O(\ell+1+Dt+D+1)$. For the base factor, the tested bound $t\leq A$ gives
\[
 2+tD\leq2+AD\leq(D+2)^2(1+A/D),\qquad
 Da_0=O\!\left(\sqrt{DA}+D\log(2+D)+D\right).
\]
As $D\geq1$, this bounds $a_0$ as well. The entire coefficient is therefore $O(\sqrt{DA}+D\log(2+D)+D+1)$ with an absolute constant on every passing record.

\subsection{Passage of the driver tests}\label{app:test-passage}
This subsection proves that every test of the packed driver passes on all sufficiently large inputs of its density range (Lemma~\ref{lem:bounded-degree-guards}).

\begin{proof}[Proof of Lemma~\ref{lem:bounded-degree-guards}]
Uniformly on the range $3\leq D\leq\lambda/1024$, $M\leq ND$ of the lemma,
\[
 \sqrt{D\lambda}\leq G_p=O(\lambda\log(2+\lambda)),\qquad
 DG_p=O(\lambda^2\log(2+\lambda)).
\]
The least-integer choice in \eqref{eq:packed-outer} gives $T=\Theta(\sqrt{LG_p/D})$, with $\sqrt L\leq T=O(\sqrt{L\lambda\log(2+\lambda)})$. As in Section~\ref{subsec:outer-cost}, before accounting for the rounding down in $P$, observe that
\[
 \frac{T}{G_p}\geq\sqrt{\frac{L}{DG_p}}
       \geq c\frac{\sqrt L}{\lambda\sqrt{\log(2+\lambda)}}\longrightarrow\infty.
\]
Hence, uniformly,
\begin{equation}\label{eq:packed-band-scales}
 \begin{gathered}
 P=\Theta\!\left(\sqrt{L/(DG_p)}\right),\qquad
 c'\frac{\sqrt L}{\lambda\sqrt{\log(2+\lambda)}}\leq P\leq C'\sqrt L,\\
 \log_2P=\tfrac12\log_2L+O(\log(2+\lambda)),\qquad
 A=\Theta(\lambda),\qquad\gamma=\Theta(\log(2+\lambda)),\\
 t=O\!\left(\sqrt{\lambda/D}+\log(2+\lambda)\right),\qquad k\leq t/64.
 \end{gathered}
\end{equation}
The fixed floor $4096$ is absorbed in these estimates for sufficiently large $N$.

The outer tests $4096\leq T\leq L$, $T^2\leq N$ and $2\leq P\leq N$ follow, and $D\leq N$ holds. Since $G_p\geq\sqrt{3\lambda}\to\infty$, one has $P/T\leq1/(1024G_p)\to0$, proving $16(3P+1)\leq T$ and $P\leq T/16$. The left side of the outer logarithmic guard is $O(\lambda)=o(T)$. At every positive level the explicit allowance ratio is
\[
 \frac{A^o_j+2DH^o_j}{M^o_j}
       =\frac{T+2DT^2}{16}\,2^T.
\]
Its logarithm, after adding the constant $2$, is $T+O(\log(TD))<2T$ eventually.

For the inner tests, $D=O(\lambda)$ and $t^2=O(\lambda+(\log(2+\lambda))^2)$ are smaller than the lower bound for $P$ in \eqref{eq:packed-band-scales}; thus $D\leq P$, $t^2\leq P$ and $k\leq P$. Also $t=o(A)$, so $t\leq A$. The fixed floor gives $t\geq4096$ and $k\geq64$, and the inequalities following \eqref{eq:packed-logguard} give $16(3k+1)\leq t$ and $k\leq t/16$. Equation~\eqref{eq:packed-logguard} itself proves the inner logarithmic guard. Finally
\[
 H_\ell=t^3 2^{\ell t}\geq t^3P>P.
\]
The inequalities just proved give $D,P,T,t,k\leq N$, and simplicity gives $M\leq N^2$. These establish the coarse checks; Appendix~\ref{app:word-bounds} derives the capacity and word bounds from them before any optimized running time is used.

For global preparation, $\log_2(D+1)/\sqrt D$ is bounded for $D\geq3$. Therefore \eqref{eq:packed-band-scales} gives
\[
 \log_2S_0=(k+1)\log_2(D+1)
       =O\!\left(\sqrt\lambda+(\log(2+\lambda))^2\right)=o(\lambda).
\]
Consequently
\[
 \log_2U_0=2\log_2S_0+D+4\log_2(D+k+2)
       \leq\lambda/1024+o(\lambda).
\]
Since $\log_2L\geq\lambda-2$ for large $N$, this proves $8\log_2U_0<\log_2L$ with a uniform margin. Also $R\geq N$ gives $\log_2\lfloor\log_2R/8\rfloor\geq\lambda-O(1)$, proving \eqref{eq:walkguard}. Thus both tests in \eqref{eq:packed-globalguards} pass. The event-based word bound of Appendix~\ref{app:word-bounds} then covers all tables from the global preparation, capacities and actual records.
\end{proof}

\subsection{Necessary size threshold for the packed branch}\label{app:finite-thresholds}
The floor $t\geq4096$ makes the size threshold of Lemma~\ref{lem:bounded-degree-guards} large. On a record passing \eqref{eq:packed-innerguards}, $\J(P)\geq t\geq4096$, that is, $P\geq2^{4095}-1$. The definition of $P$ in \eqref{eq:packed-outer} then gives $T\geq1024G_pP\geq(2^{4105}-1024)G_p$, hence $T-1\geq(2^{4105}-1025)G_p$ because $G_p\geq1$. Since $T$ is the least positive integer with $DT^2\geq LG_p$, we have $D(T-1)^2<LG_p$, and $D\geq3$ (Section~\ref{subsec:routing}) gives $L>DG_p(2^{4105}-1025)^2\geq3(2^{4105}-1025)^2>2^{8211}$. As $L=\J(N)$ is an integer and $L-1<\log_2(N+2)$, the packed branch can be selected only when $\log_2(N+2)>2^{8211}$. If this necessary inequality fails, an instance reaching these tests takes the Dijkstra fallback at line~\ref{alg:driver-fallback}.

\subsection{Mask tables and walk data}\label{app:mask-tables}
This subsection realizes the frontier operations of Section~\ref{sec:packed} in the integer model of Section~\ref{sec:model}. We use standard small-universe tabulation~\cite{ADKF1970}, with the following construction. For the mask width $w_0=\lfloor\log_2R/8\rfloor$ of Section~\ref{sec:packed} and integers below $Q_0\coloneqq2^{w_0}$, prepare their halves and parities by enumerating $2a$ and $2a+1$. Fill two $Q_0$ by $Q_0$ tables for intersection and union using
\[
 \operatorname{And}(2a+\epsilon,2b+\eta)=2\operatorname{And}(a,b)+i_{\epsilon\eta},\qquad
 \operatorname{Or}(2a+\epsilon,2b+\eta)=2\operatorname{Or}(a,b)+o_{\epsilon\eta},
\]
where $i_{\epsilon\eta}$ and $o_{\epsilon\eta}$ are the intersection and union bits, respectively, for the four fixed Boolean cases. Smaller half-arguments have already been filled. Decode each mask with the half and parity tables to prepare its complement and its least set position. Difference is intersection with the complement. Successive doubling gives position powers. The value $w_0$ denotes exhaustion and is never used as a bit position. Prepared row offsets make each lookup address an addition. The entire construction takes the time and cells stated in \eqref{eq:mask-setup}. Thus the mask operations of the packed search need only constant-time lookups, additions and comparisons, without a machine bitwise operation.

We construct the endpoint masks in $O(S_0)$ time per root using a reusable vertex-indexed array, one token per tree and a list of touched vertices.

For each prefix $p$ of fewer than $k$ edges, store its child-rank array indexed by slot. Enumerate the at most $2^D$ subsets, remove one least set slot, and add its child bit to the already computed smaller-subset answer. Each answer then takes one lookup during a search. Since $D\leq w_0$, the slot masks fit the same tables. Together with walk generation, exact real additions and sorting by Lemma~\ref{lem:relative-order}, these constructions give \eqref{eq:walk-setup}.

\section{Coupling stopped prefixes with completed comparisons}\label{app:coupling}
This appendix proves Proposition~\ref{prop:packed-projection}; the proof couples an actual prefix with the completed comparison run on the subgraph induced by its retained vertices (Section~\ref{sec:invocation-cost}).

\subsection{Transactions and outward stops}\label{app:transactions}
In each probe transaction, the membership records mark retained and accepted vertices in this invocation and accepted vertices in its containing context. The retained-vertex count determines the size-stop flag; a retained contact with $F_*$ determines the contact-stop flag. If both outward stops fire, keep both flags and the contact endpoint; a fixed tie rule selects only the reported reason. The exported tree is joined at its contact endpoint whenever one is present. Data-structure operations between probes are atomic.

A sentinel probe changes no cap, label, active membership, inner search tree, queue or tail state. The observer uses the invocation's bounds even when a complete-tail scan stops at a smaller bound; consequently a sentinel probe can still trigger an invocation size or contact stop. Thus the invocation counts eligible probed endpoints below $B_*$ even when the current scan makes no relaxation. This is the retained set used in the induced comparison of Section~\ref{subsec:packed-projection}.

\subsection{Coupling for stopped packed invocations}\label{app:coupling-packed}
Freeze an actual packed invocation at an ordinary completion or after the final transaction that triggers an outward stop. Let $K_*$ be its retained set, $q\coloneqq|K_*|\leq P$, and let $\mathcal H\coloneqq G[K_*]$. Proposition~\ref{prop:packed-projection} bounds the actual prefix work by $O(W_H+Dq+q)$, where $W_H$ is the work of the completed comparison run on $\mathcal H$ in the state specified below. 

\noindent\textbf{Sufficiency of the walks prepared on $G$.} Two facts explain why the comparison keeps the walks prepared on $G$ and tests the range before arc membership. (i) If a selected walk record has candidate signature below $B$ and its final slot is not excluded by \eqref{eq:rootlow}, its endpoint $v$ has $\sig(C(v))\geq L_*$. If the head has not yet activated, its common label is $C(v)$; if it has, its accepted witness is at least $L_*$. Since $B\leq B_*$, the actual probe therefore retains the head before its internal tests, including a rejected head, a head low relative to $a_X$, or the endpoint that triggers an outward stop. Its accepted tail is already retained. Every such actual arc belongs to $\mathcal H$. (ii) A least candidate with signature at least $B$ ends the search by the pure range test in line~\ref{alg:packed-range} of Algorithm~\ref{alg:packed-search}, which reads only the selected walk record; it tests no arc membership and reads no label, so a prepared rank that refers to an arc outside $\mathcal H$ is harmless there. Accepted and retained sets remain distinct: (i) retains a selected head before its internal tests, whether or not the probe accepts.

\noindent\textbf{Slot exclusions in the comparison.} Immutable entry caps give the comparison the same initial slot exclusions \eqref{eq:rootlow} as the actual prefix, including at heads outside $\mathcal H$, where no comparison label is stored. Later low-head deletions may delete slots, and the induced comparison may delete a slot whose arc of $G$ is absent from its graph.

\noindent\textbf{Comparison state.} The two auxiliary arrays charged in the global preparation give the proof-only comparison concrete membership and label operations in the same integer-word model as the actual run. Before the comparison starts, its additional setup costs $O(q)$, and $W_H$ measures its execution in that model. Preserve $G$, walk ranks, child masks, the slot identities of arcs and the entire parameter record. The membership array is separate from every evolving retained, active, forest and frame-role array. Stamp it with a fresh comparison token on exactly the $q$ vertices in $K_*$. A selected walk record already certifies an arc $(u,v)$ of $G$, and its active tail is in $\mathcal H$. Since $\mathcal H$ is induced, the arc belongs to it exactly when the stamp at $v$ equals that token. This supplies a constant-time membership test for selected arcs, with $O(q)$ setup and cleanup.

Use a second separate vertex-indexed comparison array as a common-label overlay. Its entries stamped with the comparison token also have two private accepted-membership fields, one for the invocation and one for its containing context. Initialize its $q$ labels with the invocation's \emph{entry} caps $C$, and mark only the seed $x$ active in the invocation, with its inherited witness. Initialize the containing-context field by the entry-membership read of Appendix~\ref{app:state}. These are the exact entry values. The entry-cap and entry-membership reads supply them in $O(q)$ time. Common-label writes and both accepted publications, including equal-signature updates, affect only the overlay's labels and private fields; the frozen label and publication state of the actual run is unchanged. The fields use the fresh comparison token, their initialization and cleanup cost $O(q)$, and publication costs constant work per accepted write.

For a head outside $\mathcal H$, slot-mask preparation reads its immutable entry cap through the frozen invocation's entry-cap read of Appendix~\ref{app:state}; it does not read a comparison label for that head. Such a head was never accepted in the actual prefix, since acceptance implies retention by the invocation's observer. Its unsaved common value is therefore still its entry cap. Each slot-mask preparation uses at most $D$ such reads, within $O(Dq)$. The seed's inherited witness may leave $\mathcal H$, as allowed by Appendix~\ref{app:witness}. Membership in the fixed external forest is read from the paused outer membership array without copying that forest.

\begin{proof}[Proof of Proposition~\ref{prop:packed-projection}]
By (i), every selected below-$B$ arc of the actual prefix belongs to $\mathcal H$. Couple selected packed transactions from the entry state. The accepted prefixes $p_u$, copied candidates, endpoint masks, surviving outgoing slots, local forest and publication records agree. The prepared ranks and child masks are identical. A least candidate with signature at least $B$ ends both searches by the same pure range test, before membership is tested. A first below-$B$ divergence cannot be a missing edge, by (i). The two searches remove the same endpoint occurrences; Lemma~\ref{lem:packed-seen} shows that this removal suppresses no later permissible accepted update, including an equal-signature update. Thus the packed part of the actual prefix matches. Beyond this prefix the comparison remains well defined: a selected missing edge deletes its slot once at its tail, activates no walk prefix, and reads no current label outside $\mathcal H$. Its candidate can be compared with $B$ before this membership test.

For static complete-tail scans, prepare an active tail's induced subsequence from its sorted static outgoing list in $O(D)$ time, once per tail, using the membership array. An out-of-range sentinel whose head is outside $\mathcal H$ is paid by the corresponding scan header. A head outside $\mathcal H$ reached by an in-range candidate must be below $L_*$ in the actual prefix and remains ineligible; the complete-tail cursor advances past it once. Surviving candidates and their ascending bands agree. These omissions and induced-list preparations cost $O(Dq)$ in total over all inner levels.

\emph{Coupled state.} Match the two runs after each corresponding probe transaction or scan end. The comparison has the same labels, accepted histories, active membership and immutable caps on $K_*$. It has the same recursive sources and cuts, groups and outputs, packed search state (accepted prefixes, copied candidates, endpoint masks, surviving outgoing slots and local forest), and logical and raw worklist state, up to physical record renaming. The surviving adjacency order is the restriction of the original order to $K_*$, and cursors agree at the next surviving edge. At a transaction that triggers an outward stop, match the state at the end of the transaction, before every step that the stop skips: at a low head the endpoint marking and the slot deletion; otherwise the recording of line~\ref{alg:packed-record} of Algorithm~\ref{alg:packed-search}, the endpoint marking, the inner stop tests and the child-mask addition of line~\ref{alg:packed-children}; and any pending worklist action. Only the comparison performs those actions there; after every other transaction both runs do. The observer's pure retention membership, count, retained tree and masked outward flags and the allocator order may differ. With outward stops masked, these observer fields do not control relaxation, activation, the inner forest or the worklists. A sentinel probe changes only those fields. Later acceptance promotes a head independently of whether the invocation's observer retained it earlier.

Delete the inert in-range complete-tail probes to heads outside $\mathcal H$ from the transcript. Each remaining in-range probe has the same candidate, guards, accepted write and internal action. Every out-of-range sentinel probe to a head outside $\mathcal H$ corresponds to a comparison sentinel probe or list end with the same effective state, since the induced list keeps the sorted order and every later surviving candidate of that tail is also at least the scan's bound. Induction over these events, the packed selections and scan headers establishes the coupled prefix. The comparison may retain an endpoint earlier, with a different first edge, and may set a masked outward flag earlier. Neither changes the coupled state. At a selected outward stop, the actual prefix ends after the final transaction, and the comparison can perform the deferred actions and continue. For ordinary completion, couple the remaining probe-free control and return operations as well.

Different complete-tail sentinels may give different histories for the invocation's retention observer in the comparison. The invocation's retention metadata, with its outward flags masked, chooses no inner source, group, queue operation or accepted prefix. Coupling therefore tracks effective state and the inner first-contact decisions, while paying the actual invocation's $q$ retention records, retained tree and export separately. That actual retained tree is the one given to the outer forest.

The root cannot return partially because $H_\ell>P\geq q$. Lemma~\ref{lem:packed-forest} supplies the forest premises for Theorem~\ref{thm:return}; the comparison's other operations are the shared static scans and lazy worklist operations. Hence the comparison finishes fully with $1\leq r\leq q$ feasible output vertices. At the transaction triggering an outward stop, observation, any accepted update, both publications and required flags have finished; the actual prefix performs no pending child-mask or worklist action. The coupled operations cost at most a fixed multiple of the comparison's work on the coupled prefix, which is at most $W_H$ because work is nonnegative. Creation has already prepaid every actual cleanup record. The comparison membership array, label overlay and actual retained tree add $O(q)$, and the separately charged work on static outgoing lists and missing or low slots adds $O(Dq)$. A seed-only prefix is included by its initialization, final selections, and scan and frame headers.
\end{proof}

This comparison uses one capped invocation inside a paused outer forest. There is no separate active outer retention observer.

\end{document}